\documentclass[prd,aps,showpacs,epsf,floats,onecolumn]{revtex4}%
\usepackage{amssymb}
\usepackage{amsfonts}
\usepackage{amsmath}
\usepackage{graphicx}%
\providecommand{\U}[1]{\protect\rule{.1in}{.1in}}
\begin{document}
\title{\textbf{Krylov's State Complexity and Information Geometry in Qubit Dynamics}}
\author{\textbf{Carlo Cafaro}$^{1}$, \textbf{Emma Clements}$^{2}$, \textbf{Vishnu
Vardhan Anuboyina}$^{2}$}
\affiliation{$^{1}$ Department of Nanoscale Science and Engineering, University at
Albany-SUNY, Albany, NY 12222, USA}
\affiliation{$^{2}$ Department of Physics, University at Albany-SUNY, Albany, NY 12222, USA}

\begin{abstract}
We compare Krylov's state complexity with an information-geometric (IG)
measure of complexity for the quantum evolution of two-level systems. Focusing
on qubit dynamics on the Bloch sphere, we analyze evolutions generated by
stationary and nonstationary Hamiltonians, corresponding to geodesic and
nongeodesic trajectories. We formulate Krylov complexity in geometric terms,
both instantaneously and in a time-averaged sense, and contrast it with an IG
complexity of quantum evolutions characterized in terms of efficiency and
curvature. We show that the two measures reflect fundamentally different
aspects of quantum dynamics: Krylov's state complexity quantifies the
directional spread of the evolving state relative to the initial state,
whereas the IG complexity captures the effective volume explored along the
trajectory on the Bloch sphere. This geometric distinction explains their
inequivalent behavior and highlights the complementary nature of state-based
and information-geometric notions of complexity in quantum systems.

\end{abstract}

\maketitle

\section{Introduction}

In the domain of quantum physics, various concepts of complexity are present
\cite{baiguera26,felice18}. For instance, one might refer to the complexity
linked to a quantum state, the complexity intrinsic to a quantum circuit, or
the complexity associated with an operator. Although these measures of
complexity possess unique characteristics, they are united by a common
principle: the complexity of a composite entity generally increases in
relation to the number of fundamental components necessary for its
construction. Depending on the context, this relationship can frequently be
expressed through geometrically intuitive concepts such as \emph{lengths},
\emph{spread}, and \emph{volumes}.

In the area of theoretical computer science, Kolmogorov proposed in Ref.
\cite{kolmogorov68} that the complexity of a sequence can be measured by the
length of the shortest Turing machine program that can generate it.
Furthermore, in the field of information theory, Rissanen indicated in Refs.
\cite{rissanen78,rissanen86} that the average minimal code length of a
collection of messages acts as a measure of the complexity of that set.
Quantum circuits are composed of quantum gates that function on quantum
states. Specifically, circuit complexity pertains to the minimum number of
primitive gates necessary to create a circuit that executes a transformation
on a designated quantum state \cite{nielsen,fernando21}. The geometric
characterization of circuit complexity was introduced by Nielsen and his
associates in Refs. \cite{mike06,mike06B,mike08}. Within this geometric
framework, the circuit complexity related to a unitary operator $U$ is
fundamentally continuous and corresponds to the length of the shortest
geodesic that connects the identity operator to $U$ within the unitary group.
The lengths of these geodesic paths provide a lower bound for the minimum
number of quantum gates needed to construct the unitary operator $U$.

In gravity and quantum field theory, Krylov's state complexity \cite{bala22}
characterizes complexity as the least extent of the wavefunction's spread
within the Hilbert space, whereas Krylov's operator complexity \cite{parker19}
seeks to measure the rate at which operators disperse in the entirety of
operator space during their evolution. To establish a cohesive framework for
articulating the complexities of Krylov's operator and state, we recommend
consulting Ref. \cite{banerjee23}. The initial attempts to understand Krylov's
complexity from a geometric perspective were conducted in Refs.
\cite{caputa22,caputa21}. In Ref. \cite{caputa21}, for example, during the
examination of instances of irrational two-dimensional conformal field
theories, it was demonstrated that Krylov's operator complexity is
proportional to the volume in the information geometry defined by the
Fubini-Study metric, which serves as a distance measure between pure quantum
states. In Ref. \cite{craps24}, it was demonstrated that the time average of
Krylov complexity related to state evolution can be represented as a trace of
a particular matrix. This matrix also specifies an upper limit on Nielsen
complexity, which is fundamentally geometric as it is associated with geodesic
flows on curved manifolds, accompanied by a specially designed penalty
schedule tailored to the Krylov basis. In Ref. \cite{rolph24}, it was observed
that the Krylov complexities among three states do not adhere to the triangle
inequality, thus rendering them unsuitable as a measure of distance between
these states: there exists no metric for which Krylov complexity represents
the length of the shortest path to the target state or operator. This is
explicitly illustrated in the simplest case of a single qubit, as well as in a
general context. In Ref. \cite{nath25}, the authors illustrated in the context
of a single qubit that the square root of Krylov's state complexity quantifies
the distance between states that have evolved over time. They achieved this by
explicitly constructing a corresponding parameter space onto which the states
within the Hilbert space are mapped.

In Refs. \cite{leo25,emma25}, we introduced a metric for assessing the
complexity of quantum evolution, defined by the ratio of the difference
between accessible and accessed Bloch-sphere volumes during the transition
from the initial to the final state, relative to the accessible volume for the
specified quantum evolution. Specifically, we examined the complexity
associated with both time-optimal and time sub-optimal quantum Hamiltonian
evolutions that connect arbitrary source and target states on the Bloch
sphere, utilizing the Fubini-Study metric. This analysis was conducted in
several stages. Initially, we characterized each unitary Schr\"{o}dinger
quantum evolution through parameters such as path length, geodesic efficiency
\cite{anandan90}, speed efficiency \cite{uzdin12}, and the curvature
coefficient \cite{dandoloff92,carmel00,dandoloff04,laba17,cafaro24A,cafaro24B}
of the corresponding dynamical trajectory that links the source state to the
target state \cite{leo25b,leo25c}. Subsequently, we began with a classical
probabilistic framework where the concept of information geometric complexity
is applicable to describe the complexity of entropic motion on curved
statistical manifolds, which underpin the physics of systems when only partial
information is available \cite{cafaro07,cafarothesis,cafaro17,cafaro18B}. We
then transitioned to a deterministic quantum framework. In this setting, after
establishing a definition for the complexity of quantum evolution, we
introduced the concept of quantum complexity length scale. We particularly
emphasized the physical relevance of both quantities concerning the accessed
(i.e., partial) and accessible (i.e., total) parametric volumes of the regions
on the Bloch sphere that delineate the quantum mechanical evolution from the
source to the target states. Finally, after computing the complexity measure
and the complexity length scale for each of the two quantum evolutions, we
compared the behavior of our measures with that of path length, geodesic
efficiency, speed efficiency, and curvature coefficient \cite{leo25b,leo25c}.
Our findings indicated that, generally, efficient quantum evolutions exhibit
lower complexity than their inefficient counterparts. Nonetheless, we also
noted that complexity encompasses more than mere length. Indeed, elongated
paths that are adequately curved can demonstrate a behavior that is simpler
than that of shorter paths with a lower curvature coefficient.

\medskip

Driven by the quest for a cohesive geometrical understanding of complexity
measures, we concentrate on Krylov's state complexity alongside our quantum IG
measure of complexity. This approach aims to provide insights into their
similarities and differences through a comparative analysis that utilizes
explicit illustrative examples. Some of the inquiries we aim to explore in
this paper include:

\begin{enumerate}
\item[{[i]}] In the context of two-level quantum systems, can we articulate
Krylov's state complexity using vectors that possess a distinct geometric significance?

\item[{[ii]}] Is there a connection between Krylov's state complexity and the
notions of distance among Bloch vectors, or the Fubini-Study finite volume
element assessed through the spherical angles that parameterize qubit states
on the Bloch sphere?

\item[{[iii]}] What unique features of quantum evolution are represented by
Krylov's state complexity and our quantum IG complexity measure when examining
both geodesic and non-geodesic evolutions governed by stationary or
non-stationary Hamiltonian dynamics?

\item[{[iv]}] Do longer trajectories that link the same initial and final
states on the Bloch sphere inherently exhibit higher levels of Krylov's state
complexity? How do relative phases in quantum states influence the behavior of
Krylov's state complexity and our quantum IG complexity measure?
\end{enumerate}

The significance of addressing these inquiries is twofold. Firstly, similar to
other geometric quantifiers of quantum evolutions, such as geodesic
efficiency, speed efficiency, and the curvature coefficient of quantum
evolutions, it enables the expression of Krylov's state complexity in terms of
straightforward real-valued three-dimensional vectors that have a clear
geometric interpretation. This facilitates a more cohesive understanding of
quantum evolutions from a geometric perspective, utilizing quantifiers that
can all be represented as simple real-valued vectors with an easily
comprehensible geometric meaning. Secondly, it provides an opportunity to
acquire new physical insights regarding the complexity of quantum evolutions
by contrasting two distinct measures, each highlighting specific
characteristics of quantum evolutions, both of which are related to common
concepts such as length, spread, and volume.

\medskip

The remainder of this paper is structured as follows. In Section II, we
provide a concise overview of the physics pertaining to two-level systems,
highlighting the functional forms of the unitary time-propagators across
various \textquotedblleft magnetic\textquotedblright\ field configurations. In
this context, we present the fundamental elements of the concepts of geodesic
efficiency and the curvature coefficient associated with quantum evolutions.
Section III introduces the essential characteristics of Krylov's state
complexity alongside our quantum IG complexity measure for quantum evolutions.
In Section IV, we reformulate both the instantaneous and time-averaged Krylov
state complexity for stationary and nonstationary Hamiltonian evolutions,
representing them through two real-valued vectors that possess clear geometric
significance. In the stationary scenario, these vectors correspond to the
Bloch vector of the initial state and the unit vector that defines the
magnetic field vector incorporated into the Hamiltonian. Conversely, in the
nonstationary scenario, the two vectors represent the Bloch vector of the
initial state and that of the time-evolved state at a subsequent time $t$.
These geometrically significant expressions of Krylov's state complexity are
employed not only to explore configurations that result in maximal and
sub-maximal complexity levels but also to examine the distinctions observed
when comparing stationary and nonstationary qubit dynamics. In Section V, we
conduct a comparative analysis of these two complexity measures by explicitly
examining both geodesic and nongeodesic evolutions characterized by
time-independent Hamiltonians. In Section VI, we follow a similar approach to
Section V, substituting stationary Hamiltonian evolutions with nonstationary
ones. Our concluding remarks are presented in Section VII, while additional
technical details can be found in the appendices.

\section{Efficiency and curvature in qubit dynamics}

In this section, considering their application in the upcoming sections, we
introduce the notions of geodesic efficiency and curvature coefficient
pertaining to quantum evolutions. These notions can be articulated for any
arbitrary finite-dimensional quantum systems in a pure state. Nevertheless,
our focus here is on two-dimensional systems and their geometric
representations.\begin{figure}[t]
\centering
\includegraphics[width=0.5\textwidth] {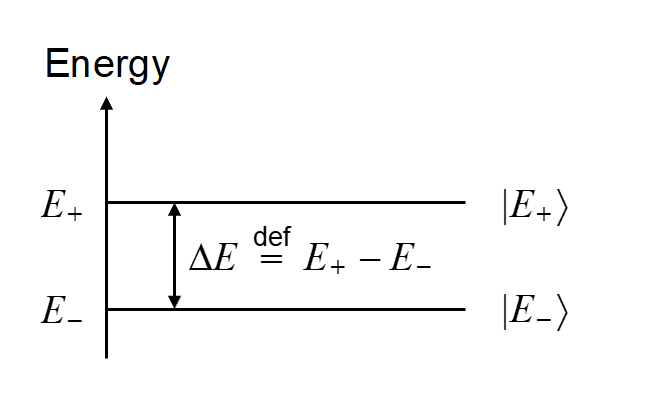}\caption{Sketch of an energy level
diagram for a two-level system. The Hamiltonian \textrm{H} of the system has a
spectral decomposition given by \textrm{H}$=E_{\_}\left\vert E_{\_}%
\right\rangle \left\langle E_{\_}\right\vert +E_{+}\left\vert E_{+}%
\right\rangle \left\langle E_{+}\right\vert $, with energy gap $\Delta
E\overset{\text{def}}{=}E_{+}-E_{\_}\geq0$. In general, $\left\vert
E_{+}\right\rangle =\left\vert e\right\rangle =\left\vert \uparrow
\right\rangle $ denotes the excited state or spin up state (maximum
eigenvalue), while $\left\vert E_{-}\right\rangle =\left\vert g\right\rangle
=\left\vert \downarrow\right\rangle $ denotes the ground state or spin down
state (minimum eigenvalue).}%
\end{figure}

\subsection{Two-level systems}

In this subsection, we present a brief summary of the physics related to
two-level systems, emphasizing the functional representations of the unitary
time-propagators under different \textquotedblleft magnetic\textquotedblright%
\ field arrangements. Two-level systems are significant for at least three
reasons. Firstly, they represent the simplest systems in quantum mechanics as
they encompass a two-dimensional Hilbert space. Secondly, they are quite
prevalent in nature and are extensively utilized in atomic physics. Thirdly, a
two-level system is synonymous with the qubit, which serves as the fundamental
building block of the rapidly advancing field of quantum information and
computation. In Fig. $1$, we display an energy level diagram for a two-level system.

Typical instances of two-level systems include a spin-$1/2$ particle (such as
an electron) in a magnetic field and a two-level atom. The Hamiltonian for the
former system is defined by the term $-\vec{\mu}\mathbf{\cdot B(}t)$ (Zeeman
effect), where $\vec{\mu}=\frac{e}{mc}\mathbf{s=}\frac{e\hslash}%
{2mc}\mathbf{\boldsymbol{\sigma}}=\mu_{\mathrm{B}}\mathbf{\boldsymbol{\sigma}%
}$ represents the magnetic moment operator of the electron in cgs-units,
$\mu_{\mathrm{B}}$ signifies the Bohr magneton, and $\mathbf{B}\left(
t\right)  $ denotes the time-dependent magnetic field vector. The Hamiltonian
for the latter system is characterized by the term $-\mathbf{d\cdot E}\left(
t\right)  $ (Stark effect), where $\mathbf{d}\overset{\text{def}}{\mathbf{=}%
}e\mathbf{r}$ represents the electric dipole moment operator of the electron,
and $\mathbf{E}\left(  t\right)  $ signifies the external electric field. To
elaborate on $\mathbf{d}$, let us consider $\left\vert g\right\rangle $ and
$\left\vert e\right\rangle $ as the ground and excited states, respectively.
Thus, we can express $\mathbf{d}$ as $\mathbf{d}=e\mathbf{r=d}_{eg}\left\vert
e\right\rangle \left\langle g\right\vert +\mathbf{d}_{ge}\left\vert
g\right\rangle \left\langle e\right\vert $, where $\mathbf{d}_{eg}%
\overset{\text{def}}{\mathbf{=}}\left\langle e\left\vert \mathbf{d}\right\vert
g\right\rangle $ and $\mathbf{d}_{ge}\overset{\text{def}}{\mathbf{=}%
}\left\langle g\left\vert \mathbf{d}\right\vert e\right\rangle $. In the case
of a hydrogen atom, the transition from $\left\vert g\right\rangle $ and
$\left\vert e\right\rangle $ is defined by the transition between the $1s$
state (with quantum number $l=0$) and the $2p$ state (with $l=1$). From the
expression of $\mathbf{d}$, it is observed that the off-diagonal terms
facilitate the transitions, whereas the diagonal terms disappear due to
parity. In fact, the dipole operator is odd with respect to parity (i.e.,
$\mathbf{d}\overset{\text{parity}}{\longrightarrow}\mathbf{-d}$), which
necessitates that dipole transitions occur between states of opposite parity.
It is evident that a two-level atom is fundamentally analogous to a spin-$1/2$
particle subjected to a magnetic field.

The fundamental dynamical equations derived from Schr\"{o}dinger's equation
that dictate the evolution of the variables of a two-level atom are
essentially identical to those applicable to spins. In a time-independent
framework, the most comprehensive (traceless and stationary) Hamiltonian model
of a two-level system can be reformulated as%
\begin{equation}
\mathrm{H}=\frac{\hslash}{2}\left(
\begin{array}
[c]{cc}%
\Delta & \Omega^{\ast}\\
\Omega & -\Delta
\end{array}
\right)  =\frac{\hslash}{2}\left(
\begin{array}
[c]{cc}%
\Delta & \Omega_{x}-i\Omega_{y}\\
\Omega_{x}+i\Omega_{y} & -\Delta
\end{array}
\right)  \text{.}%
\end{equation}
In a spin-$1/2$ system, $\Delta$ represents the Zeeman splitting, whereas
$\Omega$ indicates the magnetic field applied in the $xy$-plane. Conversely,
in a laser-driven two-level atom, $\Delta$ signifies the difference between
the laser frequency and the transition frequencies, while $\Omega$ is
proportional to the product of the atomic dipole moment and the amplitude of
the electric field.

In the rest of the paper, we assume to deal with a Hamiltonian which can be
generally non-traceless and nonstationary given by%
\begin{equation}
\mathrm{H}\left(  t\right)  \overset{\text{def}}{=}h_{0}\left(  t\right)
\mathbf{1+h}\left(  t\right)  \mathbf{\cdot\boldsymbol{\sigma}}\text{,}
\label{mille}%
\end{equation}
with $\mathbf{h}\left(  t\right)  \overset{\text{def}}{=}\left(
h_{x}(t)\text{, }h_{y}(t)\text{, }h_{z}(t)\right)  $\textbf{ }being improperly
called here a \textquotedblleft magnetic\textquotedblright\ field vector
(since, from a physical dimension standpoint, it is measured in joules in MKSA
units) and $\mathbf{\boldsymbol{\sigma}}\overset{\text{def}}%
{\mathbf{\boldsymbol{=}}}\left(  \sigma_{x}\text{, }\sigma_{y}\text{, }%
\sigma_{z}\right)  $ denoting the three Pauli spin operators. The quantity
\textquotedblleft$\mathbf{1}$\textquotedblright\ in\textbf{ }Eq. (\ref{mille})
is the identity operator acting on the single-qubit Hilbert space.

We note that when the Hamiltonian remains constant over time, the unitary time
propagator is expressed as $U(t)\overset{\text{def}}{=}e^{-\frac{i}{\hslash
}\mathrm{H}t}$. From a physics perspective, this scenario would represent a
spin-magnetic moment interacting with a static magnetic field. In cases where
the Hamiltonian varies with time but the Hamiltonians at different instances
commute, we have $U(t)\overset{\text{def}}{=}\exp(-\frac{i}{\hslash}\int
_{0}^{t}\mathrm{H}\left(  t^{\prime}\right)  dt^{\prime})$. Physically, this
situation corresponds to a spin-magnetic moment experiencing a magnetic field
that changes in strength over time while maintaining a constant direction.
Lastly, when the Hamiltonian is time-dependent and the Hamiltonians at
different times do not commute, we encounter%
\begin{equation}
U(t)\overset{\text{def}}{=}\mathcal{T}\exp(-\frac{i}{\hslash}\int_{0}%
^{t}\mathrm{H}\left(  t^{\prime}\right)  dt^{\prime})=\mathbf{1+}\sum
_{n=1}^{\infty}\left(  \frac{-i}{\hslash}\right)  ^{n}\int_{0}^{t}dt_{1}%
\int_{0}^{t_{1}}dt_{2}...\int_{0}^{t_{n-1}}dt_{n}\mathrm{H}\left(
t_{1}\right)  \mathrm{H}\left(  t_{2}\right)  ...\mathrm{H}\left(
t_{n}\right)  \text{,} \label{dyson}%
\end{equation}
with $0\leq t_{n}\leq t_{n-1}\leq...\leq t_{2}\leq t_{1}\leq t$. In Eq.
(\ref{dyson}), the symbol \textquotedblleft$\mathcal{T}$\textquotedblright%
\ represents the time-ordering operator, and the series expansion is referred
to as the Dyson series expansion \cite{dyson49}. From a physics perspective,
this unitary time propagator appears when examining a spin-magnetic moment
that is influenced by a magnetic field whose direction varies over time (while
its magnitude may either fluctuate or remain constant). In this series
expansion in Eq. (\ref{dyson}), $\mathbf{1}$ is the zeroth order term,
$(-i/\hslash)\int_{0}^{t}dt_{1}\mathrm{H}\left(  t_{1}\right)  $ is the first
order term, and $(-i/\hslash)^{2}\int_{0}^{t}dt_{1}\int_{0}^{t_{1}}%
dt_{2}\mathrm{H}\left(  t_{1}\right)  \mathrm{H}\left(  t_{2}\right)  $ is the
second order term. We observe that when the Hamiltonian remains constant over
time, the Dyson series in Eq. (\ref{dyson}) simplifies precisely to the
well-known exponential representation of the time-evolution operator%
\begin{equation}
U(t)=\mathbf{1+}\sum_{n=1}^{\infty}\left(  \frac{-i}{\hslash}\right)
^{n}\mathrm{H}^{n}\frac{t^{n}}{n!}=\sum_{n=0}^{\infty}\left(  \frac
{-i}{\hslash}\right)  ^{n}\mathrm{H}^{n}\frac{t^{n}}{n!}=e^{-\frac{i}{\hslash
}\mathrm{H}t}\text{.} \label{dyson2}%
\end{equation}
This occurs due to the commutation of all Hamiltonians at various times,
rendering time ordering unnecessary and transforming the Dyson expansion into
a standard exponential series. Furthermore, although the Dyson integral covers
one ordering of time, if all Hamiltonians commute, all time orderings give the
same contribution, and there are $n!$ of them. Therefore, when $\left[
\mathrm{H}\left(  t_{i}\right)  \text{, }\mathrm{H}\left(  t_{j}\right)
\right]  =0$ for any $t_{i}$ and $t_{j}$, we have
\begin{equation}
U(t)=\mathbf{1+}\sum_{n=1}^{\infty}\left(  \frac{-i}{\hslash}\right)
^{n}\frac{1}{n!}\left(  \int_{0}^{t}\mathrm{H}\left(  t^{\prime}\right)
dt^{\prime}\right)  ^{n}=\sum_{n=0}^{\infty}\left(  \frac{-i}{\hslash}\right)
^{n}\frac{1}{n!}\left(  \int_{0}^{t}\mathrm{H}\left(  t^{\prime}\right)
dt^{\prime}\right)  ^{n}=e^{-\frac{i}{\hslash}\int_{0}^{t}\mathrm{H}\left(
t^{\prime}\right)  dt^{\prime}}\text{.} \label{dyson3}%
\end{equation}
In Fig. $2$, we illustrate a variety of magnetic field vector configurations
$\mathbf{B}\left(  t\right)  $ with $\mathbf{h}\left(  t\right)
\propto\mathbf{B}\left(  t\right)  $ that ultimately lead to one of the time
propagators as presented in Eqs. (\ref{dyson}) (i. e., $\mathbf{B=}B(t)\hat
{B}\left(  t\right)  $ or $\mathbf{B=}B_{0}\hat{B}\left(  t\right)  $),
(\ref{dyson2}) (i. e., $\mathbf{B=}B_{0}\hat{B}_{0}$), and (\ref{dyson3}) (i.
e., $\mathbf{B=}B\left(  t\right)  \hat{B}_{0}$). In this paper, we examine
quantum-mechanical evolutions defined by all magnetic field configurations
presented in (a), (b), (c), and (d) of Fig. $2$. For additional information
regarding two-level systems, we direct the reader to Refs.
\cite{sakurai,allen}.\begin{figure}[t]
\centering
\includegraphics[width=1\textwidth] {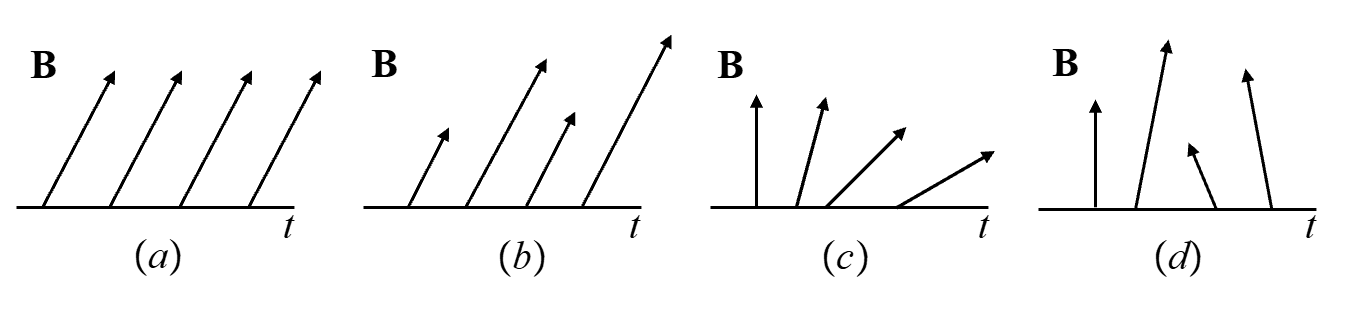}\caption{Sketch of a variety of
magnetic field vector configurations. In (a), $\mathbf{B}\overset{\text{def}%
}{=}B_{0}\hat{B}_{0}$. In (b), $\mathbf{B}\overset{\text{def}}{=}B\left(
t\right)  \hat{B}_{0}$. In (c), $\mathbf{B}\overset{\text{def}}{=}B_{0}\hat
{B}\left(  t\right)  $. Finally, in (d), $\mathbf{B}\overset{\text{def}}%
{=}B\left(  t\right)  \hat{B}\left(  t\right)  $. We note that $\mathbf{B}$ is
the magnetic field vector, $B_{0}$ denotes a constant intensity, and $\hat
{B}_{0}$ is a constant unit vector.}%
\end{figure}

\subsection{Efficiency and curvature}

Concentrating on the dynamics of qubits, we will now outline the fundamental
aspects of geodesic efficiency and the curvature coefficient pertaining to
quantum evolutions limited to two-level systems.

\subsubsection{Geodesic efficiency}

The concept of geodesic efficiency in quantum evolution was first proposed by
Anandan and Aharonov in Ref. \cite{anandan90}. Consider the evolution of a
state vector $\left\vert \psi\left(  t\right)  \right\rangle $ governed by the
time-dependent Schr\"{o}dinger equation, $i\hslash\partial_{t}\left\vert
\psi\left(  t\right)  \right\rangle =\mathrm{H}\left(  t\right)  \left\vert
\psi\left(  t\right)  \right\rangle $, within the interval $t_{A}\leq t\leq
t_{B}$. Consequently, the geodesic efficiency $\eta_{\mathrm{GE}}$ for this
quantum evolution is a time-independent (global) scalar quantity, constrained
by $0\leq\eta_{\mathrm{GE}}\leq1$, as defined in \cite{anandan90}%
\begin{equation}
\eta_{\mathrm{GE}}\overset{\text{def}}{=}\frac{s_{0}}{s}=1-\frac{\Delta s}%
{s}=\frac{2\arccos\left[  \left\vert \left\langle A|B\right\rangle \right\vert
\right]  }{2\int_{t_{A}}^{t_{B}}\frac{\Delta E\left(  t\right)  }{\hslash}%
dt}\text{,} \label{efficiency}%
\end{equation}
with $\Delta s\overset{\text{def}}{=}s-s_{0}$. The term $s_{0}$ represents the
distance along the shortest geodesic path connecting the initial $\left\vert
A\right\rangle \overset{\text{def}}{=}$ $\left\vert \psi\left(  t_{A}\right)
\right\rangle $ and final $\left\vert B\right\rangle \overset{\text{def}}%
{=}\left\vert \psi\left(  t_{B}\right)  \right\rangle $ states within the
complex projective Hilbert space. Furthermore, the variable $s$ in Eq.
(\ref{efficiency}) signifies the distance along the dynamical trajectory
$\gamma\left(  t\right)  :t\mapsto\left\vert \psi\left(  t\right)
\right\rangle $ that corresponds to the evolution of the state vector
$\left\vert \psi\left(  t\right)  \right\rangle $ for $t_{A}\leq t\leq t_{B}$.
It is evident that a geodesic quantum evolution characterized by
$\gamma\left(  t\right)  =\gamma_{\mathrm{geodesic}}\left(  t\right)  $ is
determined by the equation $\eta_{\mathrm{GE}}^{(\gamma_{\mathrm{geodesic})}%
}=1$. Additionally, we observe that by setting \textrm{H}$\left(  t\right)
\overset{\text{def}}{=}h_{0}\left(  t\right)  \mathbf{1+h}\left(  t\right)
\cdot\mathbf{\boldsymbol{\sigma}}$ and $\rho\left(  t\right)  \overset
{\text{def}}{=}(\mathbf{1+a}\left(  t\right)  \cdot\mathbf{\boldsymbol{\sigma
})/}2$ with $t_{A}\leq t\leq t_{B}$, the energy uncertainty $\Delta E\left(
t\right)  \overset{\text{def}}{=}\sqrt{\mathrm{tr}\left(  \rho\mathrm{H}%
^{2}\right)  -\left[  \mathrm{tr}\left(  \rho\mathrm{H}\right)  \right]  ^{2}%
}$ simplifies to $\Delta E\left(  t\right)  =\sqrt{\mathbf{h}^{2}-\left[
\mathbf{a}\left(  t\right)  \cdot\mathbf{h}\right]  ^{2}}$. Ultimately, the
geodesic efficiency expressed in Eq. (\ref{efficiency}) can be reformulated as%
\begin{equation}
\eta_{\mathrm{GE}}=\frac{2\arccos\left(  \sqrt{\frac{1+\mathbf{a}%
\cdot\mathbf{b}}{2}}\right)  }{\int_{t_{A}}^{t_{B}}\frac{2}{\hslash}%
\sqrt{\mathbf{h}^{2}-\left[  \mathbf{a}\left(  t\right)  \cdot\mathbf{h}%
\right]  ^{2}}dt}\text{,} \label{jap}%
\end{equation}
where $\mathbf{a}\left(  t_{A}\right)  \overset{\text{def}}{=}\mathbf{a}$ and
$\mathbf{a}\left(  t_{B}\right)  =\mathbf{b}$ in Eq. (\ref{jap}).
Interestingly, for \textrm{H}$\left(  t\right)  \overset{\text{def}}%
{=}\mathbf{h}\left(  t\right)  \cdot\mathbf{\boldsymbol{\sigma}}$ and by
$\mathbf{h}\overset{\text{def}}{=}\left[  \mathbf{h}\cdot\mathbf{a}\right]
\mathbf{a}+\left[  \mathbf{h}-(\mathbf{h}\cdot\mathbf{a})\mathbf{a}\right]
=\mathbf{h}_{\shortparallel}+\mathbf{h}_{\perp}$, where $\mathbf{a}%
=\mathbf{a}\left(  t\right)  $ in the decomposition of $\mathbf{h}$,
$\eta_{\mathrm{GE}}$ in Eq. (\ref{jap}) reduces to
\begin{equation}
\eta_{\mathrm{GE}}=\frac{\arccos\left(  \sqrt{\frac{1+\mathbf{a}%
\cdot\mathbf{b}}{2}}\right)  }{\int_{t_{A}}^{t_{B}}h_{\bot}(t)dt}\text{.}
\label{goodyo1}%
\end{equation}
Consequently, from Eq. (\ref{goodyo1}), it is evident that $\eta_{\mathrm{GE}%
}$ is solely dependent on the initial Bloch vector $\mathbf{a}$, the final
Bloch vector $\mathbf{b}$, and the magnitude $h_{\bot}(t)\overset{\text{def}%
}{=}\left\Vert \mathbf{h}_{\perp}\right\Vert $ of the vector component of the
magnetic field that is perpendicular to the time-dependent Bloch vector
$\mathbf{a}\left(  t\right)  $. For further information regarding the concept
of geodesic efficiency, we recommend consulting Refs.
\cite{anandan90,leo25b,leo25c}.

\subsubsection{Curvature}

In the realm of quantum-mechanical processes, the curvature coefficient, which
varies with time, is defined by the square of the magnitude of the covariant
derivative of the tangent vector associated with the time-evolving state
vector \cite{cafaro24A,cafaro24B}. Specifically, this curvature coefficient
quantifies the curvature of the quantum trajectory formed by a pure quantum
state that is parallel-transported and evolves unitarily under a nonstationary
Hamiltonian, which governs the Schr\"{o}dinger evolution equation.

For two-level systems, the curvature coefficient $\kappa_{\mathrm{AC}}^{2}$ in
Refs.\cite{cafaro24A,cafaro24B} (with the subscript \textquotedblleft%
\textrm{AC\textquotedblright\ }denoting Alsing and Cafaro) can be entirely
articulated in terms of two distinct real three-dimensional vectors that
possess clear geometric significance. These vectors are the Bloch vector
$\mathbf{a}\left(  t\right)  $ and the magnetic field vector $\mathbf{h}%
\left(  t\right)  $. The former is derived from the density operator
$\rho\left(  t\right)  =$ $\left\vert \psi\left(  t\right)  \right\rangle
\left\langle \psi\left(  t\right)  \right\vert \overset{\text{def}}{=}\left[
\mathbf{1}+\mathbf{a}\left(  t\right)  \cdot\mathbf{\boldsymbol{\sigma}%
}\right]  /2$, while the latter defines the nonstationary Hamiltonian
\textrm{H}$\left(  t\right)  \overset{\text{def}}{=}\mathbf{h}\left(
t\right)  \cdot\mathbf{\boldsymbol{\sigma}}$. Based on the comprehensive
analysis presented in Ref. \cite{cafaro24B}, we obtain%
\begin{equation}
\kappa_{\mathrm{AC}}^{2}\left(  \mathbf{a}\text{, }\mathbf{h}\right)
=4\frac{\left(  \mathbf{a\cdot h}\right)  ^{2}}{\mathbf{h}^{2}-\left(
\mathbf{a\cdot h}\right)  ^{2}}+\frac{\left[  \mathbf{h}^{2}\mathbf{\dot{h}%
}^{2}-\left(  \mathbf{h\cdot\dot{h}}\right)  ^{2}\right]  -\left[  \left(
\mathbf{a\cdot\dot{h}}\right)  \mathbf{h-}\left(  \mathbf{a\cdot h}\right)
\mathbf{\dot{h}}\right]  ^{2}}{\left[  \mathbf{h}^{2}-\left(  \mathbf{a\cdot
h}\right)  ^{2}\right]  ^{3}}+4\frac{\left(  \mathbf{a\cdot h}\right)  \left[
\mathbf{a\cdot}\left(  \mathbf{h\times\dot{h}}\right)  \right]  }{\left[
\mathbf{h}^{2}-\left(  \mathbf{a\cdot h}\right)  ^{2}\right]  ^{2}}\text{,}
\label{XXX}%
\end{equation}
where $\mathbf{\dot{h}}\overset{\text{def}}{\mathbf{=}}d\mathbf{h}/dt$. The
expression of $\kappa_{\mathrm{AC}}^{2}$ in Eq. (\ref{XXX}) is highly
beneficial from a computational perspective for qubit systems and
simultaneously provides a clear geometric interpretation of the curvature of
quantum evolution in relation to the (normalized unitless) Bloch vector a and
the (generally unnormalized, $\left[  \mathbf{h}\right]  _{\mathrm{MKSA}}%
=$\textrm{joules}$=\sec.^{-1}$ when setting $\hslash=1$) magnetic field vector
$\mathbf{h}$. From Eq. (\ref{XXX}), it is observed that for stationary
Hamiltonian evolutions, $\kappa_{\mathrm{AC}}^{2}$ simplifies to a
time-independent quantity as indicated by \cite{cafaro24A}%
\begin{equation}
\kappa_{\mathrm{AC}}^{2}\left(  \mathbf{a}\text{, }\mathbf{h}\right)
=4\frac{\left(  \mathbf{a\cdot h}\right)  ^{2}}{\mathbf{h}^{2}-\left(
\mathbf{a\cdot h}\right)  ^{2}}\text{.} \label{XXXb}%
\end{equation}
Evidently, when $\mathbf{a}$ and $\mathbf{h}$ are orthogonal in Eq.
(\ref{XXXb}), the curvature coefficient $\kappa_{\mathrm{AC}}^{2}\left(
\mathbf{a}\text{, }\mathbf{h}\right)  $ becomes zero. To clarify, although
$\mathbf{a=a}\left(  t\right)  $ in Eq. (\ref{XXX}), it is permissible to
substitute $\mathbf{a}\left(  t\right)  $ with $\mathbf{a}\left(  0\right)  $
in Eq. (\ref{XXXb}). This is fundamentally because when the Hamiltonian
\textrm{H} does not depend on time, $\left\langle \psi\left(  t\right)
\left\vert \mathrm{H}\right\vert \psi\left(  t\right)  \right\rangle
=\left\langle \psi\left(  0\right)  \left\vert \mathrm{H}\right\vert
\psi\left(  0\right)  \right\rangle $ (i.e., $\mathbf{a}\left(  t\right)
\cdot\mathbf{h=a}\left(  0\right)  \cdot\mathbf{h}$). The conservation of
$\left\langle \mathrm{H}\right\rangle $ means that the projection of
$\mathbf{a}\left(  t\right)  $ along $\mathbf{h}$ is constant in time. For
additional information regarding the concept of curvature in quantum
mechanical evolutions, we recommend consulting Refs.
\cite{cafaro24A,cafaro24B,leo25b,leo25c}.

After examining the essential elements of two-level systems and emphasizing
the notions of efficiency and curvature, we are now ready to introduce the two
types of complexity measures employed in our comparative analysis.

\section{Complexity measures}

In this section, we present the fundamental features of Krylov's state
complexity in conjunction with our quantum IG complexity metric for quantum evolutions.

We recall here that the geometry of the Bloch sphere endowed with the
Fubini--Study metric can be interpreted as a form of quantum information
geometry \cite{caves94,cafaro12,cafaro18}. Indeed, when equipped with the
Fubini--Study (FS) metric, the Bloch sphere constitutes a natural quantum
information manifold for pure states. Moreover, the Fubini--Study metric is
directly related to the quantum Fisher information metric, with $g_{\alpha
\beta}^{\mathrm{QFI}}=4g_{\alpha\beta}^{\mathrm{FS}}$ (for details, see
Appendix A). Note that $1\leq\alpha$, $\beta\leq\dim_{%
\mathbb{R}
}%
\mathbb{C}
\mathrm{P}^{1}=2$, where $%
\mathbb{C}
\mathrm{P}^{1}$ is the complex Hilbert space for single-qubit pure states that
is isomorphic to the Bloch sphere $S^{2}$, with $S^{2}\simeq%
\mathbb{C}
\mathrm{P}^{1}$.

Following this series of comments, we are prepared to present Krylov's state
complexity.\begin{figure}[t]
\centering
\includegraphics[width=0.75\textwidth] {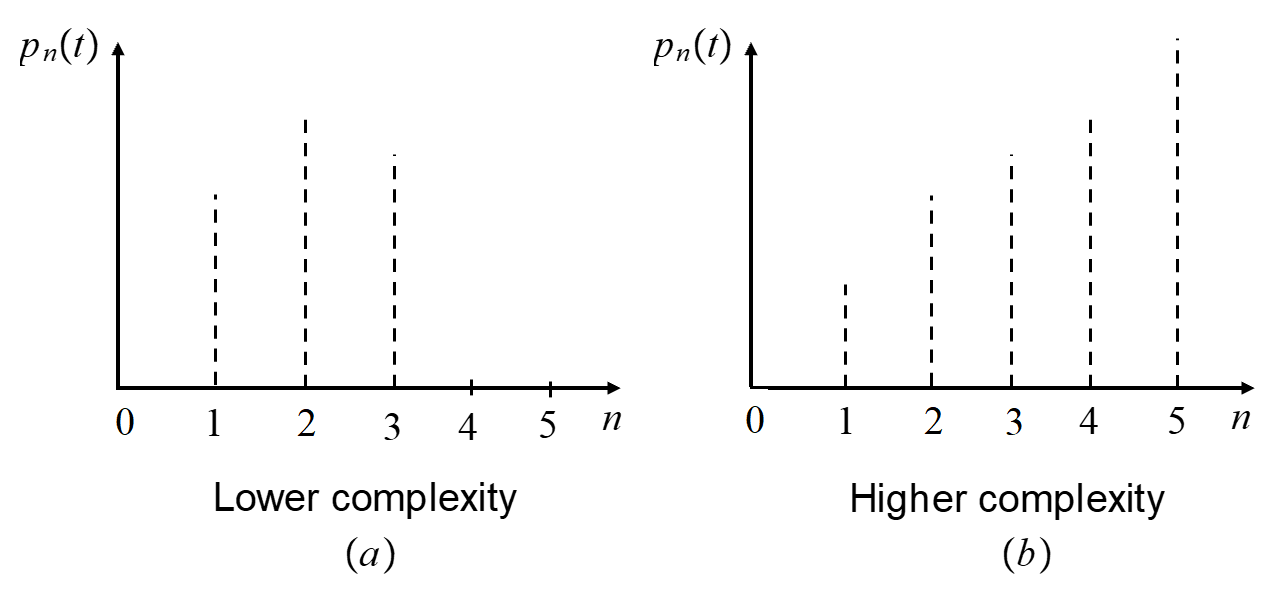}\caption{Illustrations of low (a)
and high (b) scenarios of Krylov state complexity $\mathcal{K}\left(
t\right)  $. In both (a) and (b), $n$ represents the Krylov index, while the
dashed vertical lines indicate the probabilities $p_{n}(t)\overset{\text{def}%
}{=}\left\vert \left\langle K_{n}\left\vert \psi\left(  t\right)  \right.
\right\rangle \right\vert ^{2}$. The Krylov state complexity $\mathcal{K}%
\left(  t\right)  $ is defined as the average position $\left\langle
n\right\rangle $ of this probability distribution $p_{n}(t)$ with $0\leq
p_{n}(t)\leq1$ and $\sum_{n}p_{n}(t)=1$. In scenario (a), the wavefunction
exhibits minimal spreading, resulting in a slow increase in Krylov's state
complexity. Generally, a low complexity scenario is characterized by a narrow
distribution concentrated around $n=0$. Conversely, in scenario (b), the state
investigates numerous Krylov directions, leading to a rapid increase in
Krylov's state complexity. Typically, a high energy scenario is marked by a
wide distribution with its center significantly distant from $n=0$.}%
\end{figure}

\subsection{Krylov's state complexity}

Let us examine a quantum-mechanical system that progresses under a stationary
Hamiltonian \textrm{H}. We assume that the time evolution of a state
$\left\vert \psi\left(  t\right)  \right\rangle $ is governed by the
Schr\"{o}dinger equation $i\hslash\partial_{t}\left\vert \psi\left(  t\right)
\right\rangle $ $=$\textrm{H}$\left\vert \psi\left(  t\right)  \right\rangle $
. Consequently, the solution to the evolution equation can be articulated as%
\begin{equation}
\left\vert \psi\left(  t\right)  \right\rangle =e^{-\frac{i}{\hslash
}\mathrm{H}t}\left\vert \psi\left(  0\right)  \right\rangle =%
{\displaystyle\sum\limits_{n=0}^{+\infty}}
\frac{(-it)^{n}}{\hslash^{n}n!}\left\vert \psi_{n}\right\rangle \text{,}%
\end{equation}
with $\left\vert \psi_{n}\right\rangle \overset{\text{def}}{=}$\textrm{H}%
$^{n}\left\vert \psi\left(  0\right)  \right\rangle $. The Gram-Schmidt
orthonormalization method applied to $\left\{  \left\vert \psi_{n}%
\right\rangle \right\}  =\left\{  \left\vert \psi\left(  0\right)
\right\rangle \text{, }\mathrm{H}\left\vert \psi\left(  0\right)
\right\rangle \text{, }\mathrm{H}^{2}\left\vert \psi\left(  0\right)
\right\rangle \text{,}...\right\}  $ results in an ordered, orthonormal basis
$\left\{  \left\vert K_{n}\right\rangle \right\}  $ for the segment of the
Hilbert space that is investigated by the evolving state
$\vert$%
$\left\vert \psi\left(  0\right)  \right\rangle \equiv\left\vert
K_{0}\right\rangle $. The Krylov basis $\left\{  \left\vert K_{n}\right\rangle
\right\}  $ may contain fewer elements than the dimension of the Hilbert
space, contingent upon the dynamics and the selection of the initial state.
The Krylov basis can be formally represented through the Lanczos recursive
algorithm as \cite{lanczos},%
\begin{equation}
\left\{
\begin{array}
[c]{c}%
\left\vert A_{n+1}\right\rangle =\left(  \mathrm{H}-a_{n}\right)  \left\vert
K_{n}\right\rangle -b_{n}\left\vert K_{n-1}\right\rangle \\
\left\vert K_{n}\right\rangle =b_{n}^{-1}\left\vert A_{n}\right\rangle
\end{array}
\right.  \text{,} \label{918}%
\end{equation}
where $b_{n}\overset{\text{def}}{=}\left\langle A_{n}\left\vert A_{n}\right.
\right\rangle ^{1/2}$, $b_{0}\overset{\text{def}}{=}0$, $\left\vert
K_{0}\right\rangle \overset{\text{def}}{=}\left\vert \psi\left(  0\right)
\right\rangle $, $\left\vert K_{-1}\right\rangle \overset{\text{def}}{=}0$,
and $a_{n}\overset{\text{def}}{=}\left\langle K_{n}\left\vert \mathrm{H}%
\right\vert K_{n}\right\rangle $. The quantities $a_{n}$ and $b_{n}$ in Eq.
(\ref{918}) are known as Lanczos coefficients. Manipulation of the two
relations in Eq. (\ref{918}) yields%
\begin{equation}
\mathrm{H}\left\vert K_{n}\right\rangle =b_{n+1}\left\vert K_{n+1}%
\right\rangle +a_{n}\left\vert K_{n}\right\rangle +b_{n}\left\vert
K_{n-1}\right\rangle \text{.} \label{920}%
\end{equation}
Therefore, given Eq. (\ref{920}), the Hamiltonian $\mathrm{H}$ is tridiagonal
with respect to the Krylov basis. In particular, the tridiagonal elements are
given by the Lanczos coefficients. We remark that the Gram-Schmidt and Lanczos
algorithms are not the same. However, the Lanczos algorithm can be viewed as
an efficient version of the Gram-Schmidt algorithm tailored to Krylov spaces
and Hermitian operators. It is noteworthy that the conventional Krylov
construction utilizing the Lanczos algorithm is applicable solely to
autonomous Hamiltonian systems, which possess a constant Hamiltonian. For an
extension of the Krylov construction to periodically driven systems, also
known as Floquet systems \cite{floquet,shirley65}, where the Lanczos algorithm
is substituted with the Arnoldi iteration \cite{arnoldi51}, we direct the
reader to Refs. \cite{nizami23,takahashi25}. In this periodic setting, we wish
to highlight that the (Hermitian) Hamiltonian operator $\mathrm{H}$ is
substituted with the (unitary) Floquet operator $U_{\mathrm{F}}$ of the system
during the formation of Krylov's basis, with $\left\{  \mathrm{H}%
^{n}\left\vert \psi\left(  0\right)  \right\rangle \right\}  \rightarrow$
$\left\{  U_{\mathrm{F}}^{n}\left\vert \psi\left(  0\right)  \right\rangle
\right\}  $ \cite{nizami23}. We point out that the closest analog of the
Lanczos algorithm for Floquet systems is, in fact, the Szeg\"{o} algorithm
\cite{szego39}. While the Arnoldi iteration provides the most general
framework- applicable to arbitrary time evolution, including dissipative
dynamics- it can be significantly optimized when the evolution operator
possesses additional symmetries. For a recent and detailed account of the
Szeg\"{o} algorithm, its generalizations, and its applications to chaotic
Floquet systems, we refer the reader to Ref. \cite{trunin25}. Furthermore, we
stress that in this work we restrict our attention to evolutions generated by
Hermitian Hamiltonians. For applications of Krylov spread complexity extended
to non-Hermitian systems, we refer the reader to Refs.
\cite{medina25A,medina25B}.

The fundamental concept behind Krylov's state complexity is the expectation
that a more intricate time evolution will distribute $\left\vert \psi\left(
t\right)  \right\rangle $ more extensively across the Hilbert space in
comparison to the initial state $\left\vert \psi\left(  0\right)
\right\rangle $. The degree of this distribution is influenced by the selected
basis. Nevertheless, one can assess the complexity of the evolution by
minimizing the wavefunction's spread across all bases. To measure this spread,
one should consider a cost function \textrm{C}$_{\mathcal{B}}(t)$ in relation
to a complete, orthonormal, ordered basis, $\mathcal{B}=\left\{  \left\vert
B_{n}\right\rangle \right\}  $ for the Hilbert space provided by%
\begin{equation}
\mathrm{C}_{\mathcal{B}}(t)=\sum_{n}c_{n}\left\vert \left\langle \psi\left(
t\right)  \left\vert B_{n}\right.  \right\rangle \right\vert ^{2}=\sum
_{n}c_{n}p_{\mathcal{B}}\left(  n,t\right)  \text{,} \label{943}%
\end{equation}
with $\sum_{n}p_{\mathcal{B}}\left(  n,t\right)  =1$. The quantity $c_{n}$ is
a positive, increasing sequence of real-valued numbers, and $p_{\mathcal{B}%
}\left(  n,t\right)  $ are the probabilities of being in each basis vector. In
general, one takes $c_{n}=n$ so that $\mathrm{C}_{\mathcal{B}}(t)$ in Eq.
(\ref{943})\ measures the average spread of $\left\vert \psi\left(  0\right)
\right\rangle $ in the basis $\mathcal{B}$. Finally, Krylov's state complexity
is defined as \cite{bala22,caputa24,rolph24,nath25}%
\begin{equation}
\mathcal{K}\left(  t\right)  \overset{\text{def}}{=}\underset{\mathcal{B}%
}{\min}\mathrm{C}_{\mathcal{B}}(t)\text{,}%
\end{equation}
that is,%
\begin{align}
\mathcal{K}\left(  t\right)   &  =\sum_{n}c_{n}\left\vert \left\langle
\psi\left(  t\right)  \left\vert K_{n}\right.  \right\rangle \right\vert
^{2}\nonumber\\
&  =\left\langle \psi\left(  t\right)  \left\vert \hat{K}\right\vert
\psi\left(  t\right)  \right\rangle \nonumber\\
&  =\sum_{n}np_{\mathcal{K}}\left(  n,t\right)  \text{,} \label{951}%
\end{align}
where $\hat{K}$ is the $K$-complexity operator. From Eq. (\ref{951}), we note
that $\mathcal{K}\left(  t\right)  $ can be viewed as the expectation value of
the $K$-complexity operator with respect to the state $\left\vert \psi\left(
t\right)  \right\rangle $ or, alternatively, as the average of $n$ in the
probability distribution $p_{\mathcal{K}}\left(  n,t\right)  =\left\vert
\left\langle \psi\left(  t\right)  \left\vert K_{n}\right.  \right\rangle
\right\vert ^{2}$. For simplicity of notation, we use $p_{n}(t)$ instead of
$p_{\mathcal{K}}\left(  n,t\right)  $\textbf{ }in the remaining of the paper.
By construction, $\mathcal{K}\left(  0\right)  =0$ and $\mathcal{K}\left(
t\right)  \geq0$ for $t>0$. Krylov's state complexity $\mathcal{K}\left(
t\right)  $ measures how far the state $\left\vert \psi\left(  0\right)
\right\rangle $ has evolved under \textrm{H} away from the initial site with
$n=0$ ($\leftrightarrow\left\vert K_{0}\right\rangle $) along the $1D$
chain/lattice whose sites $n>0$ label increasingly complex Krylov states
$\left\vert K_{n}\right\rangle $. The selection of the Krylov basis for
measuring the spread is determined by its ability to establish a direction in
which complexity is reduced, particularly near $t=0$. In Fig. $3$, we show a
visual representation that aids in understanding the concept of Krylov's state
complexity in both lower and higher complexity level scenarios.

Krylov's state complexity is constructed from the basis that produces the
least spread, as this is the only scenario in which complexity accurately
represents the inevitable growth dictated by the Hamiltonian, rather than
being a by-product of the chosen basis. While it is possible to utilize a
maximally spreading basis, the outcome is not a significant interpretation of
complexity; instead, it becomes a coordinate-dependent distortion. Complexity
ought to be an attribute of the dynamics, independent of the observer's
coordinate framework. The Krylov basis is determined canonically by the
dynamics through the Lanczos algorithm, eliminating arbitrary basis freedom.
It assesses the intrinsic spread of states related to the Hamiltonian, rather
than the artificial delocalization resulting from an inadequate basis
selection. Furthermore, it establishes a lower bound on the complexity that
the dynamics must exhibit. For further details on the Krylov state complexity,
we suggest Ref. \cite{bala22}.

In the following subsection, we introduce our quantum information geometry
measure of complexity.

\subsection{Quantum information geometry complexity}

\bigskip In furtherance of the research detailed in Refs. \cite{leo25,emma25},
we evaluate the single-qubit quantum dynamics influenced by both stationary
and nonstationary Hamiltonian evolutions across a specified time interval
$\left[  t_{A}\text{, }t_{B}\right]  $, by establishing the complexity
\textrm{C}$\left(  t_{A}\text{, }t_{B}\right)  $ as%
\begin{equation}
\mathrm{C}\left(  t_{A}\text{, }t_{B}\right)  \overset{\text{def}}{=}%
\frac{\mathrm{V}_{\max}\left(  t_{A}\text{, }t_{B}\right)  -\overline
{\mathrm{V}}\left(  t_{A}\text{, }t_{B}\right)  }{\mathrm{V}_{\max}\left(
t_{A}\text{, }t_{B}\right)  }\text{.} \label{QCD}%
\end{equation}
It is important to note from the outset that the proposed IG complexity
measure quantifies the fraction of the accessible volume that remains
unexplored during the evolution, rather than the explored volume itself. The
reasoning for suggesting this expression for the complexity $\mathrm{C}\left(
t_{A}\text{, }t_{B}\right)  $ will be clarified in the following paragraphs.

We begin by defining $\overline{\mathrm{V}}\left(  t_{A}\text{, }t_{B}\right)
$ and $\mathrm{V}_{\max}\left(  t_{A}\text{, }t_{B}\right)  $ as outlined in
Eq. (\ref{QCD}). To clarify the definition of the so-called \emph{accessed
volume} $\overline{\mathrm{V}}\left(  t_{A}\text{, }t_{B}\right)  $, we will
utilize a schematic approach as follows. If possible, analytically integrate
the time-dependent Schr\"{o}dinger evolution equation $i\hslash\partial
_{t}\left\vert \psi(t)\right\rangle =\mathrm{H}\left(  t\right)  \left\vert
\psi(t)\right\rangle $ and express the (normalized) single-qubit state vector
$\left\vert \psi(t)\right\rangle $ at any arbitrary time $t$ in terms of the
computational basis state vectors $\left\{  \left\vert 0\right\rangle \text{,
}\left\vert 1\right\rangle \right\}  $. If this is not achievable, proceed
with a numerical analysis. We derive $\left\vert \psi(t)\right\rangle
=c_{0}(t)\left\vert 0\right\rangle +c_{1}(t)\left\vert 1\right\rangle $, where
$c_{0}(t)$ and $c_{1}(t)$ are reformulated as
\begin{equation}
c_{0}(t)\overset{\text{def}}{=}\left\langle 0\left\vert \psi(t)\right.
\right\rangle =\left\vert c_{0}(t)\right\vert e^{i\phi_{0}(t)}\text{, and
}c_{1}(t)\overset{\text{def}}{=}\left\langle 1\left\vert \psi(t)\right.
\right\rangle =\left\vert c_{1}(t)\right\vert e^{i\phi_{1}(t)}\text{,}
\label{qa}%
\end{equation}
respectively. Moreover, it is crucial to recognize that $\phi_{0}(t)$ and
$\phi_{1}(t)$ represent the real-valued phases of $c_{0}(t)$ and $c_{1}(t)$,
respectively. Following this, by employing the complex quantum amplitudes
$c_{0}(t)$ and $c_{1}(t)$ as outlined in Eq. (\ref{qa}), reformulate the state
$\left\vert \psi(t)\right\rangle $ into a physically equivalent state
articulated in its standard Bloch sphere representation, which is
characterized by the polar angle $\theta\left(  t\right)  \in\left[  0\text{,
}\pi\right]  $ and the azimuthal angle $\varphi\left(  t\right)  \in\left[
0\text{, }2\pi\right)  $. With the temporal variations of the two spherical
angles $\theta\left(  t\right)  $ and $\varphi\left(  t\right)  $ at hand,
ascertain the volume of the parametric region that the quantum-mechanical
system navigates during its evolution from $\left\vert \psi(t_{A}%
)\right\rangle =\left\vert A\right\rangle $ to $\left\vert \psi
(t)\right\rangle $. Lastly, calculate the temporal-average volume of the
parametric region traversed by the quantum-mechanical system as it transitions
from $\left\vert \psi(t_{A})\right\rangle =\left\vert A\right\rangle $ to
$\left\vert \psi(t_{B})\right\rangle =\left\vert B\right\rangle $ with
$t\in\left[  t_{A}\text{, }t_{B}\right]  $.

In alignment with this initial framework, we will now delve into the specifics
of the calculation process for $\overline{\mathrm{V}}\left(  t_{A}\text{,
}t_{B}\right)  $. By employing Eq. (\ref{qa}), we note that $\left\vert
\psi(t)\right\rangle =c_{0}(t)\left\vert 0\right\rangle +c_{1}(t)\left\vert
1\right\rangle $ is physically analogous to the state $\left\vert
c_{0}(t)\right\vert \left\vert 0\right\rangle +\left\vert c_{1}(t)\right\vert
e^{i\left[  \phi_{1}(t)-\phi_{0}(t)\right]  }\left\vert 1\right\rangle $. As a
result, $\left\vert \psi(t)\right\rangle $ can be redefined as%
\begin{equation}
\left\vert \psi(t)\right\rangle =\cos\left[  \frac{\theta\left(  t\right)
}{2}\right]  \left\vert 0\right\rangle +e^{i\varphi\left(  t\right)  }%
\sin\left[  \frac{\theta\left(  t\right)  }{2}\right]  \left\vert
1\right\rangle \text{.} \label{qa2}%
\end{equation}
From a formal standpoint, the polar angle $\theta\left(  t\right)  $ and the
azimuthal angle $\varphi\left(  t\right)  \overset{\text{def}}{=}\phi
_{1}(t)-\phi_{0}(t)=\arg\left[  c_{1}(t)\right]  -\arg\left[  c_{0}(t)\right]
$ as stated in Eq. (\ref{qa2}) can be articulated as%
\begin{equation}
\theta\left(  t\right)  \overset{\text{def}}{=}2\arctan\left(  \frac
{\left\vert c_{1}(t)\right\vert }{\left\vert c_{0}(t)\right\vert }\right)
\text{,} \label{teta}%
\end{equation}
and, assuming that $\operatorname{Re}\left[  c_{1}(t)\right]  >0$ and
$\operatorname{Re}\left[  c_{0}(t)\right]  >0$,%
\begin{equation}
\varphi\left(  t\right)  \overset{\text{def}}{=}\arctan\left\{  \frac
{\operatorname{Im}\left[  c_{1}(t)\right]  }{\operatorname{Re}\left[
c_{1}(t)\right]  }\right\}  -\arctan\left\{  \frac{\operatorname{Im}\left[
c_{0}(t)\right]  }{\operatorname{Re}\left[  c_{0}(t)\right]  }\right\}
\text{,} \label{fi}%
\end{equation}
respectively.\begin{figure}[t]
\centering
\includegraphics[width=0.75\textwidth] {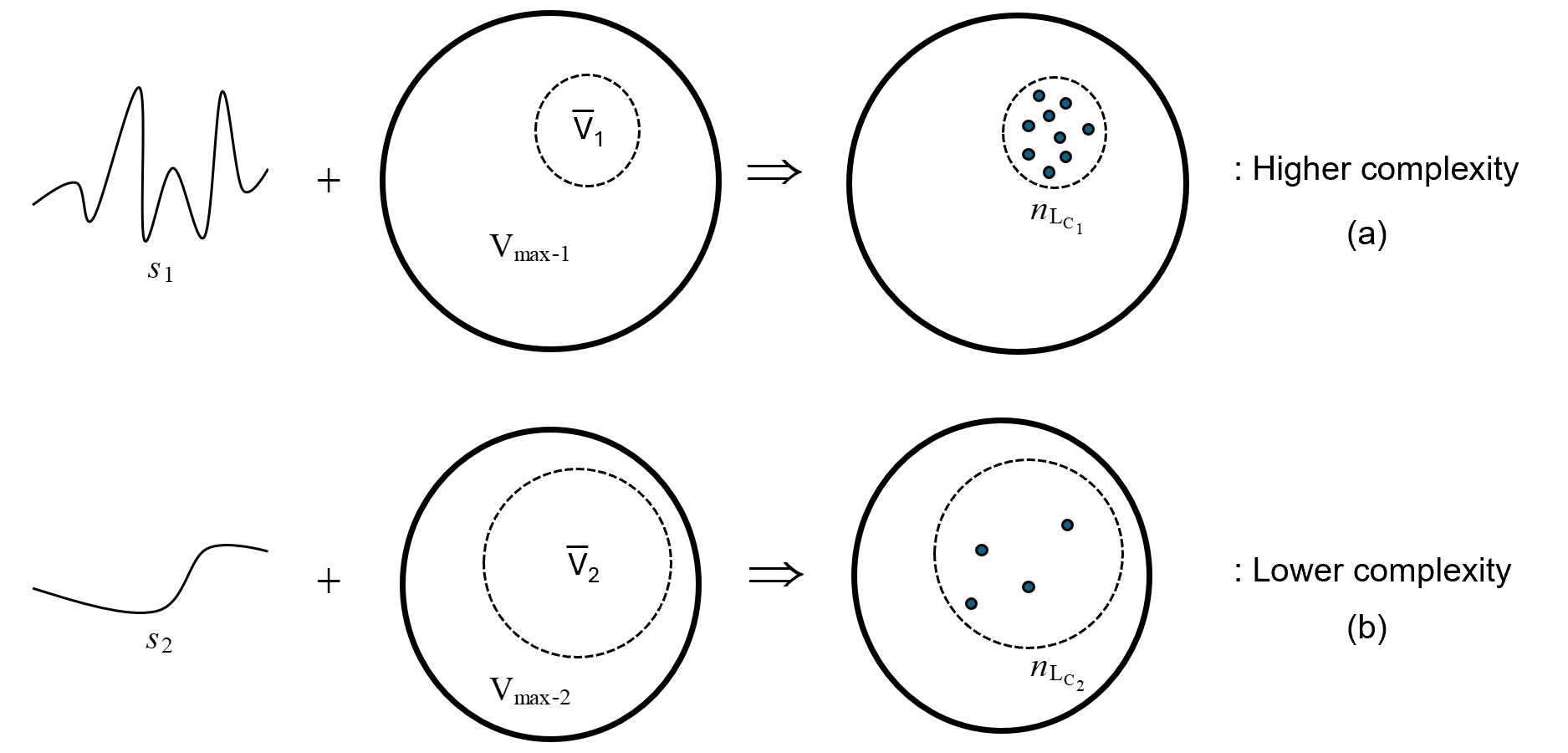}\caption{A visual representation
that aids in understanding the concept of complexity \textrm{C}$\left(
t_{A}\text{, }t_{B}\right)  $ with $t_{A}\leq t\leq t_{B}$. In (a), we
illustrate a complex situation where the accessed volume $\overline
{\mathrm{V}}_{1}$ (the area enclosed by the dashed circular line) constitutes
a minor portion of the total accessible volume \textrm{V}$_{\max\text{-}1}$
(the area enclosed by the thick solid circular line). Conversely, in (b), we
depict a scenario with a reduced level of complexity, as $\overline
{\mathrm{V}}_{2}$ (the area enclosed by the dashed circular line) represents a
larger portion of \textrm{V}$_{\max\text{-}2}$ (the area enclosed by the thick
solid circular line). The length $s$ of a path serves as an estimate for the
maximum number $n_{s}$ of statistically distinguishable states along that
path. Generally, the higher complexity scenario in (a) corresponds to a longer
quantum path of length $s_{1}$, characterized by a small fraction
($\overline{\mathrm{V}}_{1}/\mathrm{V}_{\max\text{-}1}$) of the total
accessible volume \textrm{V}$_{\max\text{-}1}$. In contrast, the lower
complexity scenario in (b) typically corresponds to a shorter quantum path of
length $s_{2}<s_{1}$, described by a larger fraction ($\overline{\mathrm{V}%
}_{2}/\mathrm{V}_{\max\text{-}2}$, with $\overline{\mathrm{V}}_{2}%
/\mathrm{V}_{\max\text{-}2}>\overline{\mathrm{V}}_{1}/\mathrm{V}_{\max
\text{-}1}$) of the total accessible volume \textrm{V}$_{\max\text{-}2}$. The
number $n_{\mathrm{L}_{\mathrm{C}}}$ of points represented in an accessed
volume $\overline{\mathrm{V}}$ is proportional to the square of the length
$s^{2}$ of the path, enhanced by a factor that is the reciprocal of
$\overline{\mathrm{V}}/\mathrm{V}_{\max}$, where $0\leq$ $\overline
{\mathrm{V}}/\mathrm{V}_{\max}\leq1$. These two scenarios in (a) and (b) are
accurately characterized by the concept of complexity length scale
\textrm{L}$_{\mathrm{C}}$, with $n_{\mathrm{L}_{\mathrm{C}}}\overset
{\text{def}}{=}$\textrm{L}$_{\mathrm{C}}^{2}=s^{2}\cdot(\overline{\mathrm{V}%
}/\mathrm{V}_{\max})^{-1}$ \cite{leo25,emma25}.}%
\end{figure}

In general, the functional representation of $\varphi\left(  t\right)  $ as
outlined in Eq. (\ref{fi}) can take on a more intricate expression. This
complexity stems from the need to articulate the phase $\arg\left(  z\right)
$ of a complex number $%
\mathbb{C}
\ni z\overset{\text{def}}{=}x+iy=\left\vert z\right\vert e^{i\arg(z)}$ using
the $2$-\textrm{argument arctangent} function \textrm{atan}$2$ as $\arg(z)=$
\textrm{atan}$2(y$, $x)$. When $x>0$, the function \textrm{atan}$2(y$, $x)$
reduces to $\arctan\left(  y/x\right)  $. For additional mathematical
perspectives on \textrm{atan}$2$, we suggest referring to Ref. \cite{grad00}.
Consequently, with $\theta\left(  t\right)  $ and $\varphi\left(  t\right)  $
established, it can be noted that the unit three-dimensional Bloch vector
$\mathbf{a}\left(  t\right)  $, which corresponds to the state vector
$\left\vert \psi(t)\right\rangle $ as specified in Eq. (\ref{qa2}), is
represented as $\mathbf{a}\left(  t\right)  =(\sin\left[  \theta\left(
t\right)  \right]  \cos\left[  \varphi\left(  t\right)  \right]  $,
$\sin\left[  \theta\left(  t\right)  \right]  \sin\left[  \varphi\left(
t\right)  \right]  $, $\cos\left[  \theta\left(  t\right)  \right]  )$.
Currently, we can define $\overline{\mathrm{V}}\left(  t_{A}\text{, }%
t_{B}\right)  $. Specifically, the accessed volume $\overline{\mathrm{V}%
}\left(  t_{A}\text{, }t_{B}\right)  $, which pertains to the quantum
evolution governed by the Hamiltonian \textrm{H}$\left(  t\right)  $ from
$\left\vert \psi(t_{A})\right\rangle =\left\vert A\right\rangle $ to
$\left\vert \psi(t_{B})\right\rangle =\left\vert B\right\rangle $, where $t$
lies within the interval $\left[  t_{A}\text{, }t_{B}\right]  $, is expressed
as%
\begin{equation}
\overline{\mathrm{V}}\left(  t_{A}\text{, }t_{B}\right)  \overset{\text{def}%
}{=}\frac{1}{t_{B}-t_{A}}\int_{t_{A}}^{t_{B}}V(t)dt\text{.}
\label{avgcomplexity}%
\end{equation}
Eq. (\ref{avgcomplexity}) implies that $\overline{\mathrm{V}}\left(
t_{A}\text{, }t_{B}\right)  $ can be viewed as a mean value of $V(t)$
calculated over the time interval $\left[  t_{A}\text{, }t_{B}\right]  $. The
quantity $V(t)$ in Eq. (\ref{avgcomplexity}) is the instantaneous volume,
which is given by%
\begin{equation}
V(t)=V(\theta(t)\text{, }\varphi\left(  t\right)  )\overset{\text{def}}%
{=}\mathrm{vol}\left[  \mathcal{D}_{\mathrm{accessed}}\left[  \theta(t)\text{,
}\varphi(t)\right]  \right]  \text{,} \label{local-complexity}%
\end{equation}
where $\mathrm{vol}\left[  \mathcal{D}_{\mathrm{accessed}}\left[
\theta(t)\text{, }\varphi(t)\right]  \right]  $ is defined as%
\begin{equation}
\mathrm{vol}\left[  \mathcal{D}_{\mathrm{accessed}}\left[  \theta(t)\text{,
}\varphi(t)\right]  \right]  \overset{\text{def}}{=}\int\int_{\mathcal{D}%
_{\mathrm{accessed}}\left[  \theta(t)\text{, }\varphi(t)\right]  }%
\sqrt{g_{\mathrm{FS}}\left(  \theta\text{, }\varphi\right)  }d\theta
d\varphi\text{.} \label{q3}%
\end{equation}
We stress that $\mathrm{vol}\left[  \mathcal{\cdot}\right]  $
signifies\textbf{ }$\left\vert \mathrm{vol}\left[  \mathcal{\cdot}\right]
\right\vert \geq0$, given that volumes as defined by positive real-valued
numerical values. In Eq. (\ref{q3}), $g_{\mathrm{FS}}\left(  \theta\text{,
}\varphi\right)  \overset{\text{def}}{=}\sqrt{\sin^{2}(\theta)/16}$ denotes
the Fubini-Study area-density factor, while $\sin^{2}(\theta)/16$ is the
determinant of the matrix connected to the Fubini-Study infinitesimal line
element $ds_{\mathrm{FS}}^{2}\overset{\text{def}}{=}(1/4)\left[  d\theta
^{2}+\sin^{2}(\theta)d\varphi^{2}\right]  $. Lastly, $\mathcal{D}%
_{\mathrm{accessed}}\left[  \theta(t)\text{, }\varphi(t)\right]  $ in Eq.
(\ref{q3}) represents the parametric region that the quantum system explores
as it transitions from the initial state $\left\vert \psi(t_{A})\right\rangle
=\left\vert A\right\rangle $ to an intermediate state $\left\vert
\psi(t)\right\rangle $, where $t_{A}$ $\leq t\leq t_{B}$. It is given by%
\begin{equation}
\mathcal{D}_{\mathrm{accessed}}\left[  \theta(t)\text{, }\varphi(t)\right]
\overset{\text{def}}{=}\left[  \theta\left(  t_{A}\right)  \text{, }%
\theta\left(  t\right)  \right]  \times\left[  \varphi\left(  t_{A}\right)
\text{, }\varphi\left(  t\right)  \right]  \subset\left[  0\text{, }%
\pi\right]  _{\theta}\times\left[  0\text{, }2\pi\right)  _{\varphi}\text{.}
\label{j5B}%
\end{equation}
To be precise, we stress that $\mathcal{D}_{\mathrm{accessed}}$ in Eq.
(\ref{j5B}) does not correspond to the exact geometric image of the dynamical
trajectory segment up to time\textbf{ }$t$.\textbf{ }Rather, it serves as a
coordinate-based bounding box- a convenient proxy that parametrizes the
physically relevant two-dimensional region on the Bloch sphere explored during
the quantum evolution. This choice provides a simple and computationally
tractable way to approximate the explored region, capturing the maximal
coordinate spread of the trajectory in the Bloch sphere parametrization, while
avoiding the complexities associated with determining the exact geometric
image of the path\textbf{.} To improve computational efficiency, we note that
the instantaneous volume $V(t)$ in Eq. (\ref{local-complexity}) can be
conveniently reformulated as $V(t)=\left\vert \left(  \cos\left[
\theta\left(  t_{A}\right)  \right]  -\cos\left[  \theta\left(  t\right)
\right]  \right)  \left(  \varphi(t)-\varphi(t_{A})\right)  \right\vert /4$,
where $\theta\left(  t\right)  $ and $\varphi\left(  t\right)  $ are defined
Eqs. (\ref{teta}) and (\ref{fi}), respectively. This general expression of the
instantaneous volume $V(t)$ is valid when both $\theta$ and $\varphi$ are
nonconstant, with $ds_{\mathrm{FS}}^{2}=(1/4)\left[  d\theta^{2}+\sin
^{2}(\theta)d\varphi^{2}\right]  $.\textbf{ }In this general scenario, the
instantaneous volume is a two-dimensional Fubini-Study area. However, if
$\theta$ is constant during the evolution and equals $\theta_{0}$,
$ds_{\mathrm{FS}}^{2}=(1/4)\sin^{2}(\theta_{0})d\varphi^{2}$ and the
instantaneous volume reduces to a one-dimensional integral given by
$V(t)=(1/2)\left\vert \sin\left(  \theta_{0}\right)  \left[  \varphi\left(
t\right)  -\varphi\left(  t_{A}\right)  \right]  \right\vert $\textbf{ }with
$t_{A}\leq t\leq t_{B}$. Similarly, if $\varphi$ is constant during the
quantum evolution and equals $\varphi_{0}$, $ds_{\mathrm{FS}}^{2}%
=(1/4)d\theta^{2}$ and the instantaneous volume reduces to a one-dimensional
integral equal to\textbf{ }$V(t)=(1/2)\left\vert \left[  \theta\left(
t\right)  -\theta\left(  t_{A}\right)  \right]  \right\vert $ with $t_{A}\leq
t\leq t_{B}$. In summary, assuming that the tilde symbol represents the
time-average process, the accessed volume $\overline{\mathrm{V}}\left(
t_{A}\text{, }t_{B}\right)  $ in Eq. (\ref{avgcomplexity}) can be recast as%

\begin{equation}
\overline{\mathrm{V}}\left(  t_{A}\text{, }t_{B}\right)  \overset{\text{def}%
}{=}\widetilde{\mathrm{vol}}\left[  \mathcal{D}_{\mathrm{accessed}}\left[
\theta(t)\text{, }\varphi(t)\right]  \right]  \text{,} \label{cafe1}%
\end{equation}
where $t_{A}$ $\leq t\leq t_{B}$ in Eq. (\ref{cafe1}). Ultimately, in
alignment with the physical justifications outlined in Refs.
\cite{leo25,emma25}, the \emph{accessible volume} $\mathrm{V}_{\max}\left(
t_{A}\text{, }t_{B}\right)  $ in Eq. (\ref{QCD}) is given by%
\begin{equation}
\mathrm{V}_{\max}(t_{A}\text{, }t_{B})\overset{\text{def}}{=}\mathrm{vol}%
\left[  \mathcal{D}_{\text{\textrm{accessible}}}\left(  \theta\text{, }%
\varphi\right)  \right]  =\int\int_{\mathcal{D}_{\text{\textrm{accessible}}%
}\left(  \theta\text{, }\varphi\right)  }\sqrt{g_{\mathrm{FS}}\left(
\theta\text{, }\varphi\right)  }d\theta d\varphi\text{.} \label{j4}%
\end{equation}
The quantity $\mathcal{D}_{\text{\textrm{accessible}}}\left(  \theta\text{,
}\varphi\right)  $ in Eq. (\ref{j4}) is the (local) maximally accessible
two-dimensional parametric region that characterizes the quantum-mechanical
transition from $\left\vert \psi_{A}\left(  \theta_{A}\text{, }\varphi
_{A}\right)  \right\rangle $ to $\left\vert \psi\left(  \theta_{B}\text{,
}\varphi_{B}\right)  \right\rangle $ and is expressed by%
\begin{equation}
\mathcal{D}_{\text{\textrm{accessible}}}\left(  \theta\text{, }\varphi\right)
\overset{\text{def}}{=}\left\{  \left(  \theta\text{, }\varphi\right)
:\theta_{\min}\leq\theta\leq\theta_{\max}\text{, and }\varphi_{\min}%
\leq\varphi\leq\varphi_{\max}\right\}  \text{.} \label{j5}%
\end{equation}
It is relevant to notice that $\theta_{\min}$, $\theta_{\max}$, $\varphi
_{\min}$, and $\varphi_{\max}$ in Eq. (\ref{j5}) are defined as
\begin{equation}
\theta_{\min}\overset{\text{def}}{=}\underset{t_{A}\leq t\leq t_{B}}{\min
}\theta(t)\text{, }\theta_{\max}\overset{\text{def}}{=}\underset{t_{A}\leq
t\leq t_{B}}{\max}\theta(t)\text{, }\varphi_{\min}\overset{\text{def}}%
{=}\underset{t_{A}\leq t\leq t_{B}}{\min}\varphi(t)\text{, and }\varphi_{\max
}\overset{\text{def}}{=}\underset{t_{A}\leq t\leq t_{B}}{\max}\varphi
(t)\text{,} \label{minmax}%
\end{equation}
respectively. We emphasize that because $\varphi$ is an angular variable,
which is usually defined modulo $2\pi$, any interval such as $\left[
\varphi_{\min}\text{, }\varphi_{\max}\right]  $ is fundamentally ambiguous
unless the treatment of periodicity is clearly specified. In our research, we
typically choose the branch cut\textbf{ }$\varphi\in\left[  0\text{, }%
2\pi\right)  $\textbf{ }with the condition that $\varphi_{\min}\leq$
$\varphi_{\max}$ and we do not permit wrap-around intervals. Additionally, the
poles at $\theta=0$ (the north pole) and $\theta=\pi$\textbf{ }(the south
pole) represent coordinate singularities where $\varphi$ is rendered
undefined. In our study, we generally adopt the convention of collapsing
$\varphi$ at these poles, thereby identifying all $\varphi$\textbf{-}%
values\textbf{. }Furthermore, it is important to note that $\mathcal{D}%
_{\text{\textrm{accessed}}}\left(  \theta\text{, }\varphi\right)
\subset\mathcal{D}_{\text{\textrm{accessible}}}\left(  \theta\text{, }%
\varphi\right)  \subset\left[  0,\pi\right]  _{\theta}\times\left[  0\text{,
}2\pi\right)  _{\varphi}$. Finally, given $\overline{\mathrm{V}}\left(
t_{A}\text{, }t_{B}\right)  $ and $\mathrm{V}_{\max}\left(  t_{A}\text{,
}t_{B}\right)  $ as defined in Eqs. (\ref{avgcomplexity}) and (\ref{j4}),
respectively, we can establish our proposed complexity concept \textrm{C}%
$\left(  t_{A}\text{, }t_{B}\right)  $ as outlined in Eq. (\ref{QCD}).

In broad terms, we evaluate the complexity of a quantum evolution
transitioning from an initial state to a final state on the Bloch sphere by
examining the ratio of the unaccessed volume of the sphere that is contained
within the accessible volume of the sphere itself. Specifically, when the
accessible volume $\mathrm{V}_{\max}$ is largely (or minimally) explored, the
complexity \textrm{C}$\overset{\text{def}}{=}\left(  \mathrm{V}_{\max
}-\overline{\mathrm{V}}\right)  /\mathrm{V}_{\max}$\textbf{ }is regarded as
low (or high). We recognize that this convention seems to contradict the
monotonic intuition that some readers might have regarding \textquotedblleft
explored volume\textquotedblright, where a larger explored volume is typically
associated with greater complexity. To alleviate this interpretative
confusion, we emphasize that in our proposal, the complexity measure
\textrm{C}\textbf{ }actually quantifies the unexplored fraction $\left(
\mathrm{V}_{\max}-\overline{\mathrm{V}}\right)  /\mathrm{V}_{\max}$ of the
accessible volume $\mathrm{V}_{\max}$. In simpler terms, we propose that a
greater waste of unaccessed volume correlates with a more complex dynamical
scenario\textbf{.} In Fig. $4$, we present a visual depiction that facilitates
comprehension of our quantum IG measure of complexity across both lower and
higher complexity level scenarios. For further information, we direct you to
Refs. \cite{leo25,emma25}.

To enhance the efficacy of our comparative analysis, we will reframe Krylov's
state complexity in relation to the overlaps of simple three-dimensional
real-valued vectors that define the characteristics of the qubit dynamics
under examination.

\section{Geometry of Krylov's state complexity}

In this section, we rephrase both the instantaneous and time-averaged Krylov
state complexity for stationary and nonstationary Hamiltonian evolutions,
depicting them using two real-valued vectors that have distinct geometric importance.

\subsection{Stationary qubit dynamics}

Here, we calculate the general expression of Krylov's state complexity
$\mathcal{K}\left(  t\right)  $ in the case in which the qubit dynamics are
governed by a stationary Hamiltonian. In particular, the expression is
geometric since it can be fully expressed in terms of the relative geometry of
just two real vectors, the initial Bloch vector of the system $\mathbf{a}%
\left(  t_{i}\right)  $ and the time independent magnetic field vector
$\mathbf{h}$ that specifies the (generally non traceless) Hamiltonian
\textrm{H}$\overset{\text{def}}{=}h_{0}\mathbf{1+h\cdot\boldsymbol{\sigma}}$.

We begin by assuming a Hamiltonian given by \textrm{H}$\overset{\text{def}}%
{=}h_{0}\mathbf{1+h\cdot\boldsymbol{\sigma}}$ and, moreover, we take the
initial state $\left\vert \psi\left(  t_{i}\right)  \right\rangle =\left\vert
\psi\left(  0\right)  \right\rangle $ of the quantum system to be arbitrarily
given by $\left\vert \psi\left(  0\right)  \right\rangle \overset{\text{def}%
}{=}\cos(\theta_{0}/2)\left\vert 0\right\rangle +e^{i\varphi_{0}}\sin
(\theta_{0}/2)\left\vert 1\right\rangle $. Therefore, after some algebra, we
find that the expression of the unitarily time-evolved unit state vector
$\left\vert \psi\left(  t\right)  \right\rangle =U\left(  t\right)  \left\vert
\psi\left(  0\right)  \right\rangle =e^{-\frac{i}{\hslash}\left[
h_{0}\mathbf{1+h\cdot\boldsymbol{\sigma}}\right]  }\left\vert \psi\left(
0\right)  \right\rangle $ with $U(t)\overset{\text{def}}{=}e^{-\frac
{i}{\hslash}h_{0}t}\left[  \cos(\frac{ht}{\hslash})\mathbf{1-}i\sin(\frac
{ht}{\hslash})\mathbf{n\cdot\boldsymbol{\sigma}}\right]  $, $\mathbf{h}%
\overset{\text{def}}{=}h\mathbf{n}$, and $h\overset{\text{def}}{=}\left\Vert
\mathbf{h}\right\Vert $ is given by%
\begin{equation}
\left\vert \psi\left(  t\right)  \right\rangle =e^{-\frac{i}{\hslash}h_{0}%
t}\left[  \cos(\frac{ht}{\hslash})\mathbf{1-}i\sin(\frac{ht}{\hslash
})\mathbf{n\cdot\boldsymbol{\sigma}}\right]  \left\vert \psi\left(  0\right)
\right\rangle \text{.}%
\end{equation}
In the orthonormal Krylov basis $\left\{  \left\vert \phi_{0}\right\rangle
=\left\vert \psi\left(  0\right)  \right\rangle \text{, }\left\vert \phi
_{1}\right\rangle =\left\vert \psi_{\perp}\left(  0\right)  \right\rangle
\right\}  $, we have
\begin{equation}
\left\vert \psi\left(  t\right)  \right\rangle =\left\langle \psi\left(
0\right)  \left\vert \psi\left(  t\right)  \right.  \right\rangle \left\vert
\psi\left(  0\right)  \right\rangle +\left\langle \psi_{\perp}\left(
0\right)  \left\vert \psi\left(  t\right)  \right.  \right\rangle \left\vert
\psi_{\perp}\left(  0\right)  \right\rangle \text{,}%
\end{equation}
where $\left\vert \left\langle \psi\left(  0\right)  \left\vert \psi\left(
t\right)  \right.  \right\rangle \right\vert ^{2}+\left\vert \left\langle
\psi_{\perp}\left(  0\right)  \left\vert \psi\left(  t\right)  \right.
\right\rangle \right\vert ^{2}=1$. Therefore, the expression of Krylov's state
complexity becomes%
\begin{align}
\mathcal{K}\left(  t\right)   &  =\sum_{n=0}^{1}np_{n}(t)=\sum_{n=0}%
^{1}n\left\vert \left\langle \phi_{n}\left\vert \psi\left(  t\right)  \right.
\right\rangle \right\vert ^{2}=\left\vert \left\langle \phi_{1}\left\vert
\psi\left(  t\right)  \right.  \right\rangle \right\vert ^{2}=1-\left\vert
\left\langle \phi_{0}\left\vert \psi\left(  t\right)  \right.  \right\rangle
\right\vert ^{2}\nonumber\\
&  =1-\left\vert \left\langle \psi\left(  0\right)  \left\vert \psi\left(
t\right)  \right.  \right\rangle \right\vert ^{2}=1-\left\vert \left\langle
\psi\left(  0\right)  \left\vert U(t)\right\vert \psi\left(  0\right)
\right\rangle \right\vert ^{2}\text{,}%
\end{align}
that is,%
\begin{equation}
\mathcal{K}\left(  t\right)  =1-\left\vert \left\langle \psi\left(  0\right)
\left\vert U(t)\right\vert \psi\left(  0\right)  \right\rangle \right\vert
^{2}\text{.} \label{now1}%
\end{equation}
Let us then find a convenient form for the quantum overlap $\left\langle
\psi\left(  0\right)  \left\vert U(t)\right\vert \psi\left(  0\right)
\right\rangle $ and, consequently, a geometrically relevant expression for the
probability $\left\vert \left\langle \psi\left(  0\right)  \left\vert
U(t)\right\vert \psi\left(  0\right)  \right\rangle \right\vert ^{2}$. Given
that,%
\begin{equation}
\left\langle \psi\left(  0\right)  \left\vert U(t)\right\vert \psi\left(
0\right)  \right\rangle =e^{-\frac{i}{\hslash}h_{0}t}\left[  \cos(\frac
{ht}{\hslash})-i\sin(\frac{ht}{\hslash})\left\langle \psi\left(  0\right)
\left\vert \mathbf{n\cdot\boldsymbol{\sigma}}\right\vert \psi\left(  0\right)
\right\rangle \right]  \text{,}%
\end{equation}
noting that $\left\langle \psi\left(  0\right)  \left\vert
\mathbf{\boldsymbol{\sigma}}\right\vert \psi\left(  0\right)  \right\rangle
=\mathbf{a}\left(  t_{i}\right)  $ (i.e., it is the Bloch vector of the
initial state $\left\vert \psi\left(  0\right)  \right\rangle $ with $t_{i}%
=0$), we finally arrive at
\begin{equation}
\mathcal{K}\left(  t\right)  =\sin^{2}(\frac{ht}{\hslash})\left\{  1-\left[
\mathbf{n\cdot a}\left(  t_{i}\right)  \right]  ^{2}\right\}  \text{,}
\label{now2}%
\end{equation}
with $\mathbf{a}\left(  t_{i}\right)  =\left(  \sin(\theta_{t_{i}}%
)\cos(\varphi_{t_{i}})\text{, }\sin(\theta_{t_{i}})\sin(\varphi_{t_{i}%
})\text{, }\cos(\theta_{t_{i}})\right)  $. Eq. (\ref{now2}) represents the
geometric expression for $\mathcal{K}\left(  t\right)  $ we were looking for.
Observe that, as expected on physics grounds, $\mathcal{K}\left(  t\right)  $
in Eq. (\ref{now2}) does not depend on $h_{0}$. Interestingly, observe that
when $\mathbf{n}$ and $\mathbf{a}\left(  t_{i}\right)  $ are collinear, the
initial state of the system is an eigenstate of the Hamiltonian \textrm{H}.
Therefore, there is no evolution and $\mathcal{K}\left(  t\right)  $ vanishes.
Moreover, when $\mathbf{n}$ is orthogonal to $\mathbf{a}\left(  t_{i}\right)
$, one achieves the maximal \textquotedblleft instantaneous\textquotedblright%
\ Krylov state complexity with $\mathcal{K}\left(  t\right)  =\sin
^{2}(ht/\hslash)$. Finally, when $\mathbf{n}$ is not orthogonal to
$\mathbf{a}\left(  t_{i}\right)  $, one achieves a sub-maximal
\textquotedblleft instantaneous\textquotedblright\ Krylov state complexity
with $\mathcal{K}\left(  t\right)  \leq\sin^{2}(ht/\hslash)$. It is important
to stress that Eq. (\ref{now2}) is valid only in the stationary case in which
\textrm{H }does not depend on time. It is worth pointing out that using Eq.
(\ref{now2}) along with the mathematical identity%
\begin{equation}
\frac{1}{T}\int_{0}^{T}\sin^{2}(\frac{ht}{\hslash})dt=\frac{1}{2}-\frac
{\sin(\frac{2hT}{\hslash})}{\frac{4hT}{\hslash}}\text{,}%
\end{equation}
we have that in the case of geodesic motion $\mathbf{n\cdot a}\left(
t_{i}\right)  =0$ and $\left\langle \mathcal{K}\right\rangle _{\mathrm{geo}}$
becomes%
\begin{equation}
\left\langle \mathcal{K}\right\rangle _{\mathrm{geo}}=\frac{1}{2}-\frac
{\sin(\frac{2ht_{f}^{\mathrm{geo}}}{\hslash})}{\frac{4ht_{f}^{\mathrm{geo}}%
}{\hslash}}\text{.} \label{io1}%
\end{equation}
Furthermore, in the case of nongeodesic motion, $\mathbf{n\cdot a}\left(
t_{i}\right)  \neq0$ and $\left\langle \mathcal{K}\right\rangle
_{\mathrm{nongeo}}$ reduces to%
\begin{equation}
\left\langle \mathcal{K}\right\rangle _{\mathrm{nongeo}}=\left\{  1-\left[
\mathbf{n\cdot a}\left(  t_{i}\right)  \right]  ^{2}\right\}  \left[  \frac
{1}{2}-\frac{\sin(\frac{2ht_{f}^{\mathrm{nongeo}}}{\hslash})}{\frac
{4ht_{f}^{\mathrm{nongeo}}}{\hslash}}\right]  \text{.} \label{io2}%
\end{equation}
The interested reader can verify that considering the specific proper
expressions for $\mathbf{n}$, $\mathbf{a}\left(  t_{i}\right)  $, $h$,
$t_{f}^{\mathrm{geo}}$, and $t_{f}^{\mathrm{nongeo}}$, $\left\langle
\mathcal{K}\right\rangle _{\mathrm{geo}}$ in Eq. (\ref{io1}) yields Eq.
(\ref{vox3}), while $\left\langle \mathcal{K}\right\rangle _{\mathrm{nongeo}}$
in Eq. (\ref{io2}) leads to Eq. (\ref{vox3b}). With these final remarks, we
end our discussion here.

\subsection{ Nonstationary qubit dynamics}

Here, we find the general expression of Krylov's state complexity
$\mathcal{K}\left(  t\right)  $ when the qubit dynamics are specified\textbf{
}by a nonstationary Hamiltonian. Specifically, the expression is geometric in
nature because it can be fully characterized by means of the relative geometry
of just two real-valued vectors, the initial and the time-evolved Bloch
vectors of the system $\mathbf{a}\left(  t_{i}\right)  $ and $\mathbf{a}%
\left(  t\right)  $, respectively. Without loss of generality, we assume
\textrm{H}$\left(  t\right)  \overset{\text{def}}{=}\mathbf{h}\left(
t\right)  \mathbf{\cdot\boldsymbol{\sigma}}$ with $\mathbf{h}\left(  t\right)
\overset{\text{def}}{=}h\left(  t\right)  \mathbf{n}\left(  t\right)  $ and
$h\left(  t\right)  \overset{\text{def}}{=}\left\Vert \mathbf{h}\left(
t\right)  \right\Vert $ as mentioned in the previous subsection.

The main objective is to derive the analog of $\mathcal{K}\left(  t\right)  $
in Eq. (\ref{now2}) when \textrm{H}$=$\textrm{H}$\left(  t\right)  $. From Eq.
(\ref{now1}), setting $\rho\left(  0\right)  \overset{\text{def}}{=}\left\vert
\psi\left(  0\right)  \right\rangle \left\langle \psi\left(  0\right)
\right\vert =\left[  \mathbf{1+a}(0)\cdot\mathbf{\boldsymbol{\sigma}}\right]
/2$ and $\rho\left(  t\right)  \overset{\text{def}}{=}\left\vert \psi\left(
t\right)  \right\rangle \left\langle \psi\left(  t\right)  \right\vert
=\left[  \mathbf{1+a}(t)\cdot\mathbf{\boldsymbol{\sigma}}\right]  /2$ and
after some standard algebra of Pauli matrices, we get%
\begin{align}
\mathcal{K}\left(  t\right)   &  =1-\left\vert \left\langle \psi\left(
0\right)  \left\vert U(t)\right\vert \psi\left(  0\right)  \right\rangle
\right\vert ^{2}\nonumber\\
&  =1-\left\vert \left\langle \psi\left(  0\right)  \left\vert \psi\left(
t\right)  \right.  \right\rangle \right\vert ^{2}\nonumber\\
&  =1-\mathrm{tr}\left[  \rho\left(  0\right)  \rho\left(  t\right)  \right]
\nonumber\\
&  =1-\mathrm{tr}\left[  \left(  \frac{\mathbf{1+a}(0)\cdot
\mathbf{\boldsymbol{\sigma}}}{2}\right)  \left(  \frac{\mathbf{1+a}%
(t)\cdot\mathbf{\boldsymbol{\sigma}}}{2}\right)  \right] \nonumber\\
&  =1-\frac{1}{2}\left[  1+\mathbf{a}(0)\cdot\mathbf{a}(t)\right] \nonumber\\
&  =\frac{1}{2}\left[  1-\mathbf{a}(0)\cdot\mathbf{a}(t)\right] \nonumber\\
&  =\frac{1}{2}\left[  1-\mathbf{a}(0)\cdot\mathcal{R}\left(  t\right)
\mathbf{a}(0)\right]  \text{,}%
\end{align}
that is,%
\begin{equation}
\mathcal{K}\left(  t\right)  =\frac{1}{2}\left[  1-\mathbf{a}(0)\cdot
\mathcal{R}\left(  t\right)  \mathbf{a}(0)\right]  \text{.} \label{anna}%
\end{equation}
We note that $\left\vert \left\langle \psi\left(  0\right)  \left\vert
U(t)\right\vert \psi\left(  0\right)  \right\rangle \right\vert ^{2}$ is known
as the survival probability (i.e., the probability that the system is still in
the initial state at time $t$), while $1-\left\vert \left\langle \psi\left(
0\right)  \left\vert U(t)\right\vert \psi\left(  0\right)  \right\rangle
\right\vert ^{2}$ is the complement of the survival probability known as decay
probability (or, alternatively, fidelity loss or escape probability). While
earlier sections motivate Krylov state complexity in terms of a Krylov basis
and spreading in the Krylov index, we emphasize that we are not employing a
genuine nonautonomous Krylov/Lanczos construction here. Rather, focusing on
qubit dynamics under nonstationary Hamiltonians, we reinterpret $\mathcal{K}%
\left(  t\right)  $ in Eq. (\ref{anna}) as a fidelity-based proxy for state
complexity, quantifying fidelity loss or, equivalently, the escape
probability.\textbf{ }The emergence of the rotation matrix $\mathcal{R}\left(
t\right)  $ in Eq. (\ref{anna}) is due to the fact that every unitary time
propagator $U\left(  t\right)  \in\mathrm{SU}\left(  2;%
\mathbb{C}
\right)  $ induces a unique rotation matrix $\mathcal{R}\left(  t\right)
\in\mathrm{SO}\left(  3;%
\mathbb{R}
\right)  $ acting on Bloch vectors \cite{morton,frankel,ray22}, such that
$\rho\left(  t\right)  =U(t)\rho\left(  0\right)  U^{\dagger}\left(  t\right)
\leftrightarrow\mathbf{a}(t)=\mathcal{R}\left(  t\right)  \mathbf{a}(0)$. From
Eq. (\ref{anna}), the time averaged Krylov state complexity is formally given
by%
\begin{equation}
\left\langle \mathcal{K}\left(  t\right)  \right\rangle =\frac{1}{2}\left[
1-\left\langle \mathbf{a}(0)\cdot\mathcal{R}\left(  t\right)  \mathbf{a}%
(0)\right\rangle \right]  \text{,} \label{anna1}%
\end{equation}
therefore, $\left\langle \mathcal{K}\left(  t\right)  \right\rangle $ depends
on $\left\langle \mathbf{a}(0)\cdot\mathbf{a}(t)\right\rangle $. Eqs.
(\ref{anna})\ and (\ref{anna1})\ are valid for both stationary and
nonstationary quantum mechanical Hamiltonian evolutions.\begin{figure}[t]
\centering
\includegraphics[width=0.5\textwidth] {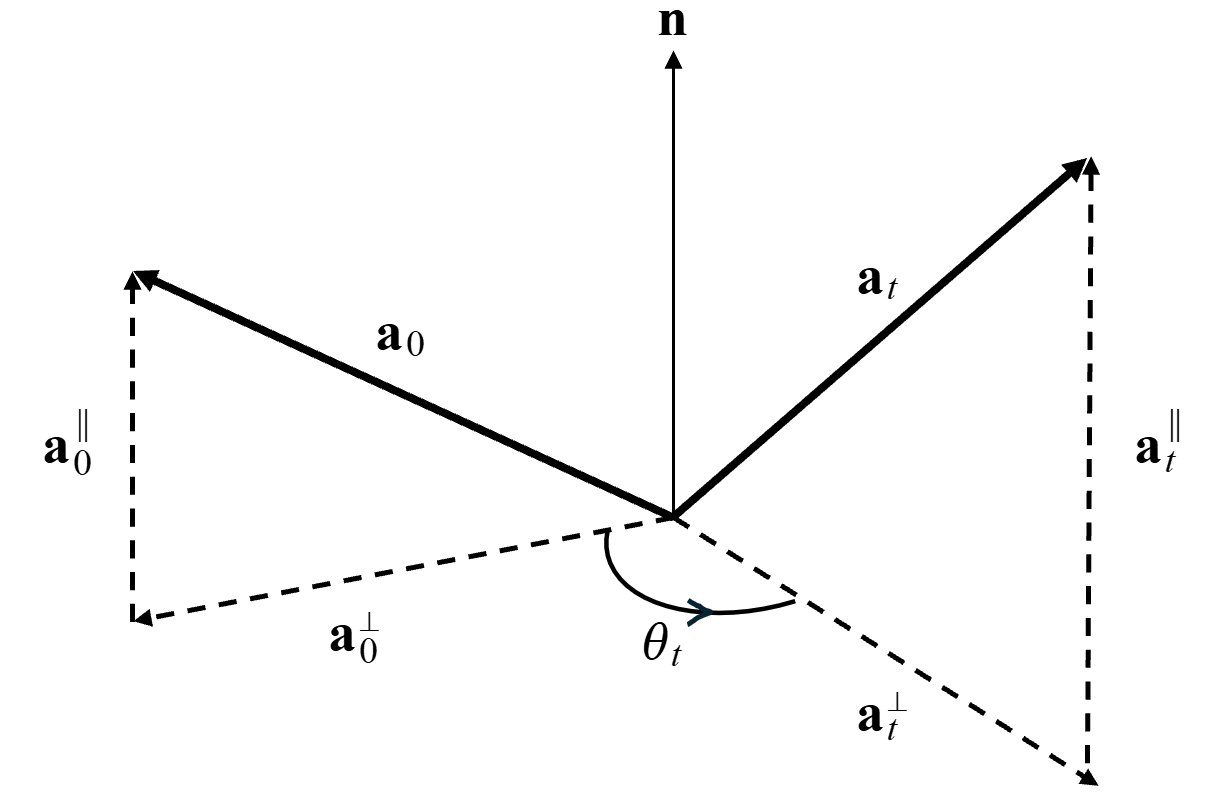}\caption{Geometric sketch for
Rodrigues' rotation formula, $\mathbf{a}_{t}=\mathcal{R}_{\mathbf{n}}\left(
\theta_{t}\right)  \mathbf{a}_{0}$ with $\mathbf{a}_{0}\overset{\text{def}}%
{=}\mathbf{a}_{0}^{\parallel}+\mathbf{a}_{0}^{\perp}$ and $\mathbf{a}%
_{t}\overset{\text{def}}{=}\mathbf{a}_{t}^{\parallel}+\mathbf{a}_{t}^{\perp}$.
We stress that $\mathcal{R}_{\mathbf{n}}\left(  \theta_{t}\right)  $ denotes a
rotation about the axis specified by means of the unit vector $\mathbf{n}$ by
an angle $\theta_{t}$.}%
\end{figure}

Interestingly, one can recover Eq. (\ref{now2}) from Eq. (\ref{anna}) when
\textrm{H} is stationary. Before examining this, let us set $\mathbf{a}%
(0)=\mathbf{a}_{0}$ and $\mathbf{a}(t)=\mathbf{a}_{t}$ for simplicity of
notation. Then, start considering the decomposition of $\mathbf{a}_{0}$ as%
\begin{equation}
\mathbf{a}_{0}=\mathbf{a}_{0}^{\parallel}+\mathbf{a}_{0}^{\perp}\text{,}%
\end{equation}
where $\mathbf{a}_{0}^{\parallel}=\left(  \mathbf{a}_{0}\cdot\mathbf{n}%
\right)  \mathbf{n}$ it the vectorial part of $\mathbf{a}_{0}$ that is
parallel to $\mathbf{n}$ and, in addition, $\mathbf{a}_{0}^{\perp}%
=\mathbf{a}_{0}-\left(  \mathbf{a}_{0}\cdot\mathbf{n}\right)  \mathbf{n}$ is
the vectorial part of $\mathbf{a}_{0}$ that is orthogonal to $\mathbf{n}$. We
stress that $\mathbf{a}_{0}^{\parallel}\cdot\mathbf{a}_{0}^{\perp}=0$ and
$\left\Vert \mathbf{a}_{0}\right\Vert ^{2}=\left\Vert \mathbf{a}%
_{0}^{\parallel}\right\Vert ^{2}+\left\Vert \mathbf{a}_{0}^{\perp}\right\Vert
^{2}=1$. We also remark that $\mathcal{R}\left(  t\right)  $ leaves the
parallel component invariant since $\mathcal{R}\left(  t\right)
\mathbf{a}_{0}^{\parallel}=\mathbf{a}_{0}^{\parallel}$ and, in addition,
\begin{equation}
\mathcal{R}\left(  t\right)  \mathbf{a}_{0}^{\perp}=\mathbf{a}_{0}^{\perp}%
\cos\left(  \frac{2h}{\hslash}t\right)  +\left(  \mathbf{n\times a}_{0}%
^{\perp}\right)  \sin(\frac{2h}{\hslash}t)\text{.} \label{anna2}%
\end{equation}
Eq. (\ref{anna2}) implies that the perpendicular component $\mathbf{a}%
_{0}^{\perp}$ rotates in the plane orthogonal to $\mathbf{n}$ by an angle
$\left(  2h/\hslash\right)  t$. This is an ordinary result of rigid rotations
in $%
\mathbb{R}
^{3}$. For later convenience, we also remark that $\mathbf{a}_{0}%
\in\mathrm{Span}\left\{  \mathbf{n}\text{, }\mathbf{a}_{0}^{\perp}\right\}  $
and, consequently, $\mathbf{a}_{0}\cdot\left(  \mathbf{n}\times\mathbf{a}%
_{0}^{\perp}\right)  =0$. Having presented these properties, we proceed with
the manipulation of $\mathcal{K}\left(  t\right)  $ in Eq. (\ref{anna}). We
have,%
\begin{align}
\mathbf{a}_{0}\cdot\mathcal{R}\left(  t\right)  \mathbf{a}_{0}  &
=(\mathbf{a}_{0}^{\parallel}+\mathbf{a}_{0}^{\perp})\cdot\left[
\mathcal{R}\left(  t\right)  (\mathbf{a}_{0}^{\parallel}+\mathbf{a}_{0}%
^{\perp})\right] \nonumber\\
&  =(\mathbf{a}_{0}^{\parallel}+\mathbf{a}_{0}^{\perp})\cdot\left[
\mathbf{a}_{0}^{\parallel}+\mathbf{a}_{0}^{\perp}\cos\left(  \frac{2h}%
{\hslash}t\right)  +\left(  \mathbf{n\times a}_{0}^{\perp}\right)  \sin
(\frac{2h}{\hslash}t)\right] \nonumber\\
&  =\left\Vert \mathbf{a}_{0}^{\parallel}\right\Vert ^{2}+\left\Vert
\mathbf{a}_{0}^{\perp}\right\Vert ^{2}\cos\left(  \frac{2h}{\hslash}t\right)
\nonumber\\
&  =\left(  \mathbf{a}_{0}\cdot\mathbf{n}\right)  ^{2}+\left[  1-\left(
\mathbf{a}_{0}\cdot\mathbf{n}\right)  ^{2}\right]  \cos\left(  \frac
{2h}{\hslash}t\right)
\end{align}
and, therefore%
\begin{align}
\mathcal{K}\left(  t\right)   &  =\frac{1}{2}\left[  1-\mathbf{a}_{0}%
\cdot\mathcal{R}\left(  t\right)  \mathbf{a}_{0}\right] \nonumber\\
&  =\frac{1}{2}\left\{  1-\left(  \mathbf{a}_{0}\cdot\mathbf{n}\right)
^{2}-\left[  1-\left(  \mathbf{a}_{0}\cdot\mathbf{n}\right)  ^{2}\right]
\cos\left(  \frac{2h}{\hslash}t\right)  \right\} \nonumber\\
&  =\frac{1-\cos\left(  \frac{2h}{\hslash}t\right)  }{2}\left[  1-\left(
\mathbf{a}_{0}\cdot\mathbf{n}\right)  ^{2}\right] \nonumber\\
&  =\sin^{2}\left(  \frac{h}{\hslash}t\right)  \left[  1-\left(
\mathbf{a}_{0}\cdot\mathbf{n}\right)  ^{2}\right]  \text{.} \label{anna3}%
\end{align}
We finally conclude that $\mathcal{K}\left(  t\right)  $ in Eq. (\ref{anna3})
coincides with $\mathcal{K}\left(  t\right)  $ in Eq. (\ref{now2}) and, as a
consequence, Eq. (\ref{now2}) is a special case of Eq. (\ref{anna}). In Fig.
$5$, we provide a visual representation that aids in understanding the
Rodrigues rotation formula. For a step-by-step derivation of the explicit
expression of the rotation matrix $\mathcal{R}\left(  t\right)  $ in Eq.
(\ref{anna3}), we refer to Appendix B. With this final comment, we conclude
this technical subsection here.

\subsection{Difference between stationary and nonstationary qubit dynamics}

Here, we discuss an essential difference in the behavior of Krylov's state
complexity when one transitions from stationary to nonstationary Hamiltonian
evolutions. Specifically, unlike what happens in a time independent setting,
we show that the orthogonality between the unit vectors $\mathbf{n}$ and
$\mathbf{a}_{0}$ (with \textrm{H}$\left(  t\right)  \overset{\text{def}}%
{=}\mathbf{h}\left(  t\right)  \mathbf{\cdot\boldsymbol{\sigma}}$,
$\mathbf{h}\left(  t\right)  \overset{\text{def}}{=}h\left(  t\right)
\mathbf{n}\left(  t\right)  $, and $\rho\left(  0\right)  \overset{\text{def}%
}{=}\left\vert \psi\left(  0\right)  \right\rangle \left\langle \psi\left(
0\right)  \right\vert =\left[  \mathbf{1+a}_{0}\cdot\mathbf{\boldsymbol{\sigma
}}\right]  /2$) does not generally imply that $\mathcal{K}\left(  t\right)  $
achieves its maximal instantaneous value. In other words, it can happen that
there is no time instance in which $\mathcal{K}\left(  t\right)  $ equals one
or, put differently, the evolving state vector $\left\vert \psi\left(
t\right)  \right\rangle $ never gets fully orthogonal to the initial state
$\left\vert \psi\left(  0\right)  \right\rangle $.

\subsubsection{Preliminaries}

Before presenting our physical example where the above mentioned difference
becomes evident, we illustrate some preliminary remarks.

From an historical standpoint, we remark that Rabi \cite{rabi37,rabi54} was a
pioneer in applying the rotating frame concept within quantum mechanics,
specifically for magnetic resonance, during his molecular beam experiments in
the 1930s, which focused on measuring nuclear magnetic moments. His research
demonstrated how a rotating magnetic field can facilitate the comprehension of
transitions in a static field, laying the groundwork for the rotating wave
approximation and proving essential for NMR (nuclear magnetic resonance) and
quantum control. Notably, he proposed the concept of a rotating coordinate
system that aligns with the oscillating magnetic field, thereby significantly
simplifying the visualization and resolution of the time-dependent problem of
spin transitions, particularly in proximity to resonance.

From a practical standpoint, assume that the unitary time propagator $U\left(
t\right)  $ describes the evolution in the laboratory frame of reference and
is associated with the Hamiltonian \textrm{H}$\left(  t\right)  $. In the
laboratory, Schr\"{o}dinger's evolution equation is $i\hslash\partial
_{t}\left\vert \psi\left(  t\right)  \right\rangle =$\textrm{H}$\left(
t\right)  \left\vert \psi\left(  t\right)  \right\rangle $. In the rotating
reference frame, instead, suppose that the unitary time evolution is
characterized by the operator $U_{\mathrm{RF}}\left(  t\right)  $ that
corresponds to the Hamiltonian \textrm{H}$_{\mathrm{RF}}\left(  t\right)  $.
In the rotating frame of reference, Schr\"{o}dinger's evolution equation is
$i\hslash\partial_{t}\left\vert \psi_{\mathrm{RF}}\left(  t\right)
\right\rangle =\mathrm{H}_{\mathrm{RF}}\left(  t\right)  \left\vert
\psi_{\mathrm{RF}}\left(  t\right)  \right\rangle $. Presume that states and
observables in these two frames of reference are connected through the unitary
frame transformation $U_{\mathrm{RF}}\left(  t\right)  $. Specifically, states
and observables are interconnected through the relationships $\left\vert
\psi\left(  t\right)  \right\rangle =U_{\mathrm{RF}}\left(  t\right)
\left\vert \psi_{\mathrm{RF}}\left(  t\right)  \right\rangle $ and $O\left(
t\right)  =U_{\mathrm{RF}}\left(  t\right)  O_{\mathrm{RF}}\left(  t\right)
U_{\mathrm{RF}}^{\dagger}\left(  t\right)  $, respectively \cite{sakurai}. To
find \textrm{H}$_{\mathrm{RF}}\left(  t\right)  $, we will proceed in the
following manner. Observe that, $i\hslash\partial_{t}\left\vert \psi
_{\mathrm{RF}}\left(  t\right)  \right\rangle =\mathrm{H}_{\mathrm{RF}}\left(
t\right)  \left\vert \psi_{\mathrm{RF}}\left(  t\right)  \right\rangle $ where%
\begin{align}
i\hslash\partial_{t}\left\vert \psi_{\mathrm{RF}}\left(  t\right)
\right\rangle  &  =i\partial_{t}\left[  U_{\mathrm{RF}}^{\dagger}\left(
t\right)  \left\vert \psi\left(  t\right)  \right\rangle \right] \nonumber\\
&  =iU_{\mathrm{RF}}^{\dagger}\left(  t\right)  \partial_{t}\left\vert
\psi\left(  t\right)  \right\rangle +i\frac{\partial U_{\mathrm{RF}}^{\dagger
}\left(  t\right)  }{\partial t}\left\vert \psi\left(  t\right)  \right\rangle
\nonumber\\
&  =iU_{\mathrm{RF}}^{\dagger}\left(  t\right)  \left[  \frac{1}{i}%
\mathrm{H}\left(  t\right)  \left\vert \psi\left(  t\right)  \right\rangle
\right]  +i\frac{\partial U_{\mathrm{RF}}^{\dagger}\left(  t\right)
}{\partial t}U_{\mathrm{RF}}\left(  t\right)  \left\vert \psi_{\mathrm{RF}%
}\left(  t\right)  \right\rangle \nonumber\\
&  =\left[  U_{\mathrm{RF}}^{\dagger}\left(  t\right)  \mathrm{H}\left(
t\right)  U_{\mathrm{RF}}\left(  t\right)  +i\frac{\partial U_{\mathrm{RF}%
}^{\dagger}\left(  t\right)  }{\partial t}U_{\mathrm{RF}}\left(  t\right)
\right]  \left\vert \psi_{\mathrm{RF}}\left(  t\right)  \right\rangle \text{.}
\label{way1}%
\end{align}
Therefore, Eq. (\ref{way1}) yields%
\begin{equation}
\mathrm{H}_{\mathrm{RF}}\left(  t\right)  =U_{\mathrm{RF}}^{\dagger}\left(
t\right)  \mathrm{H}\left(  t\right)  U_{\mathrm{RF}}\left(  t\right)
+i\frac{\partial U_{\mathrm{RF}}^{\dagger}\left(  t\right)  }{\partial
t}U_{\mathrm{RF}}\left(  t\right)  \text{.} \label{way2}%
\end{equation}
Given that $U_{\mathrm{RF}}^{\dagger}\left(  t\right)  U_{\mathrm{RF}}\left(
t\right)  =U_{\mathrm{RF}}\left(  t\right)  U_{\mathrm{RF}}^{\dagger}\left(
t\right)  =\mathbf{1}$, Eq. (\ref{way2}) can be recast as%
\begin{equation}
\mathrm{H}_{\mathrm{RF}}\left(  t\right)  =U_{\mathrm{RF}}^{\dagger}\left(
t\right)  \mathrm{H}\left(  t\right)  U_{\mathrm{RF}}\left(  t\right)
-iU_{\mathrm{RF}}^{\dagger}\left(  t\right)  \frac{\partial U_{\mathrm{RF}%
}\left(  t\right)  }{\partial t}\text{.} \label{way3}%
\end{equation}
Given the expression of $\mathrm{H}_{\mathrm{RF}}\left(  t\right)  $ in Eq.
(\ref{way3}), we can now introduce our illustrative example.

\subsubsection{Illustrative example}

In what follows, we study the evolution of a two-level quantum system which is
initially in the state $\left\vert \psi\left(  0\right)  \right\rangle
\overset{\text{def}}{=}\left\vert 0\right\rangle $ and evolves under the time
dependent Hamiltonian
\begin{equation}
\mathrm{H}\left(  t\right)  \overset{\text{def}}{=}\frac{\hslash\omega}%
{2}\left[  \cos\left(  \nu t\right)  \sigma_{x}+\sin\left(  \nu t\right)
\sigma_{y}\right]  \text{.} \label{way4}%
\end{equation}
To clarify, it is important to mention that the quantum evolution defined by
the Hamiltonian in Eq. (\ref{way4}) can be interpreted as an electron situated
within a magnetic field arrangement, similar to that depicted in (c) of Fig.
$2$. Given this initial condition and the Hamiltonian in Eq. (\ref{way4}), one
notices that $\mathbf{n}\left(  t\right)  \mathbf{=}\left(  \cos\left(  \nu
t\right)  \text{, }\sin\left(  \nu t\right)  \text{, }0\right)  $ and
$\mathbf{a}_{0}=\left(  0\text{, }0\text{, }1\right)  $. Therefore,
$\mathbf{n}\left(  t\right)  \cdot\mathbf{a}_{0}=0$ at any instant. However,
as we shall see in this example, there exists no temporal instance $t_{\ast}$
with $t_{i}\leq t_{\ast}\leq t_{f}$ for which $\mathcal{K}\left(  t_{\ast
}\right)  =1$ (unlike what happens in any stationary scenario). Obviously,
$t_{i}$ and $t_{f}$ denote here the initial and final temporal instances,
respectively. To verify the absence of such\textbf{ }$t_{\ast}$, we need to
evaluate Krylov's state complexity $\mathcal{K}\left(  t\right)  =1-\left\vert
\left\langle \psi\left(  0\right)  \left\vert \psi\left(  t\right)  \right.
\right\rangle \right\vert ^{2}$. We therefore need to find the evolving state
$\left\vert \psi\left(  t\right)  \right\rangle $ under $\mathrm{H}\left(
t\right)  $ in\ Eq. (\ref{way4}). To do so, we use the concept of rotating
frame in quantum mechanics. To select the proper rotating frame that
simplifies the integration of the Schr\"{o}dinger equation, we note that%
\begin{equation}
\cos\left(  \nu t\right)  \sigma_{x}+\sin\left(  \nu t\right)  \sigma
_{y}=e^{-i\frac{\nu t}{2}\sigma_{z}}\sigma_{x}e^{i\frac{\nu t}{2}\sigma_{z}%
}\text{.}%
\end{equation}
As a consequence, it happens to be convenient to go into the rotating frame
specified by the unitary $U_{\mathrm{RF}}\left(  t\right)  =e^{-i\frac{\nu
t}{2}\sigma_{z}}$. Substituting this specific expression of $U_{\mathrm{RF}%
}\left(  t\right)  $ in $\mathrm{H}_{\mathrm{RF}}\left(  t\right)  $ in Eq.
(\ref{way3}), we arrive at a time-independent expression for the Hamiltonian
$\mathrm{H}_{\mathrm{RF}}\left(  t\right)  $ given by $\mathrm{H}%
_{\mathrm{RF}}=(1/2)(\omega\sigma_{x}-\nu\sigma_{z})$. Having $\mathrm{H}%
_{\mathrm{RF}}$, we wish to solve the equation $i\partial_{t}\left\vert
\psi_{\mathrm{RF}}\left(  t\right)  \right\rangle =\mathrm{H}_{\mathrm{RF}%
}\left\vert \psi_{\mathrm{RF}}\left(  t\right)  \right\rangle $. Then, given
$\left\vert \psi_{\mathrm{RF}}\left(  t\right)  \right\rangle $, we can find
$\left\vert \psi\left(  t\right)  \right\rangle =U_{\mathrm{RF}}\left(
t\right)  \left\vert \psi_{\mathrm{RF}}\left(  t\right)  \right\rangle $ and
$\mathcal{K}\left(  t\right)  =1-\left\vert \left\langle \psi\left(  0\right)
\left\vert \psi\left(  t\right)  \right.  \right\rangle \right\vert ^{2}$.
Note that $\mathrm{H}_{\mathrm{RF}}=(1/2)(\omega\sigma_{x}-\nu\sigma_{z})$ can
be recast as $\mathrm{H}_{\mathrm{RF}}=(\Omega/2)\mathbf{n}\cdot
\mathbf{\boldsymbol{\sigma}}$, where $\Omega\mathbf{n}\overset{\text{def}%
}{\mathbf{=}}\left(  \omega\text{, }0\text{, }-\nu\right)  $, $\Omega
\overset{\text{def}}{\mathbf{=}}\sqrt{\omega^{2}+\nu^{2}}$, and $\mathbf{n}%
\overset{\text{def}}{\mathbf{=}}\frac{1}{\Omega}\left(  \omega\text{,
}0\text{, }-\nu\right)  $ with $\mathbf{n\cdot n=}1$. Then, setting the
reduced Planck constant $\hslash$ equal to one, we have%
\begin{equation}
\left\vert \psi_{\mathrm{RF}}\left(  t\right)  \right\rangle =e^{-i\mathrm{H}%
_{\mathrm{RF}}t}\left\vert 0\right\rangle =\left[  \cos\left(  \frac{\Omega
t}{2}\right)  \mathbf{1-}i\sin\left(  \frac{\Omega t}{2}\right)
\mathbf{n}\cdot\mathbf{\boldsymbol{\sigma}}\right]  \left\vert 0\right\rangle
\text{,}%
\end{equation}
that is,%
\begin{equation}
\left\vert \psi_{\mathrm{RF}}\left(  t\right)  \right\rangle =\left[
\cos\left(  \frac{\Omega t}{2}\right)  +i\frac{\nu}{\Omega}\sin\left(
\frac{\Omega t}{2}\right)  \right]  \left\vert 0\right\rangle -i\frac{\omega
}{\Omega}\sin\left(  \frac{\Omega t}{2}\right)  \left\vert 1\right\rangle
\text{.} \label{way5}%
\end{equation}
From Eq. (\ref{way5}), we find
\begin{equation}
\left\vert \psi\left(  t\right)  \right\rangle =U_{\mathrm{RF}}\left(
t\right)  \left\vert \psi_{\mathrm{RF}}\left(  t\right)  \right\rangle
=e^{-i\frac{\nu t}{2}\sigma_{z}}\left\{  \left[  \cos\left(  \frac{\Omega
t}{2}\right)  +i\frac{\nu}{\Omega}\sin\left(  \frac{\Omega t}{2}\right)
\right]  \left\vert 0\right\rangle -i\frac{\omega}{\Omega}\sin\left(
\frac{\Omega t}{2}\right)  \left\vert 1\right\rangle \right\}  \text{,}%
\end{equation}
that is,
\begin{equation}
\left\vert \psi\left(  t\right)  \right\rangle =\left[  \cos\left(
\frac{\Omega t}{2}\right)  +i\frac{\nu}{\Omega}\sin\left(  \frac{\Omega t}%
{2}\right)  \right]  e^{-i\frac{\nu t}{2}}\left\vert 0\right\rangle
-i\frac{\omega}{\Omega}\sin\left(  \frac{\Omega t}{2}\right)  e^{i\frac{\nu
t}{2}}\left\vert 1\right\rangle \text{.} \label{way6}%
\end{equation}
The interested reader can explicitly verify that $\left\vert \psi\left(
t\right)  \right\rangle $ in Eq. (\ref{way6}) solves $i\hslash\partial
_{t}\left\vert \psi\left(  t\right)  \right\rangle =$\textrm{H}$\left(
t\right)  \left\vert \psi\left(  t\right)  \right\rangle $ with \textrm{H}%
$\left(  t\right)  $ in Eq. (\ref{way4}) once one sets $\hslash=1$. Finally,
using Eq. (\ref{way6}), $\mathcal{K}\left(  t\right)  =1-\left\vert
\left\langle \psi\left(  0\right)  \left\vert \psi\left(  t\right)  \right.
\right\rangle \right\vert ^{2}$ with $\left\vert \psi\left(  0\right)
\right\rangle =\left\vert 0\right\rangle $ reduces to%
\begin{equation}
\mathcal{K}\left(  t\right)  =\frac{\omega^{2}}{\omega^{2}+\nu^{2}}\sin
^{2}(\frac{\sqrt{\omega^{2}+\nu^{2}}}{2}t)\text{.} \label{way7}%
\end{equation}
From Eq. (\ref{way7}), we make several observations. Firstly, since $\sin
^{2}(\xi t)$ is periodic of $T=\frac{1}{2}\cdot\frac{2\pi}{\xi}=\frac{\pi}%
{\xi}\equiv\frac{2\pi}{\omega_{\mathrm{R}}}$, it oscillates with frequency
$\omega_{\mathrm{R}}=2\xi$. Therefore, in our illustrative example,
$\mathcal{K}\left(  t\right)  $ in Eq. (\ref{way7}) oscillates with frequency
$\Omega=\sqrt{\omega^{2}+\nu^{2}}$. Secondly, as $\nu\rightarrow0$,
\textrm{H}$\rightarrow\frac{\omega}{2}\sigma_{x}$. Then, $\mathcal{K}\left(
t\right)  =\sin^{2}(\frac{\omega}{2}t)\rightarrow1$ as $t\rightarrow\frac{\pi
}{\omega}$. Observe that $e^{-i\frac{\omega}{2}\sigma_{x}t}\left\vert
0\right\rangle \overset{t=\frac{\pi}{\omega}}{=}e^{-i\frac{\pi}{2}\sigma_{x}%
}\left\vert 0\right\rangle =-i\left\vert 1\right\rangle \simeq\left\vert
1\right\rangle $ (with \textquotedblleft$\simeq$\textquotedblright\ denoting
here physical equivalence of quantum states that differ by a global phase
factor). Thirdly, the amplitude of $\mathcal{K}\left(  t\right)  $ is bounded
by $\omega^{2}/(\omega^{2}+\nu^{2})<1$, although $\mathbf{n}\left(  t\right)
\cdot\mathbf{a}_{0}=0$ at any instant. Therefore, we conclude that unlike what
happens in any stationary scenario in which $\mathbf{n}\left(  t\right)
\cdot\mathbf{a}_{0}=0$ for any instant $t$ (where the dynamical trajectory is
a geodesic path on the Bloch sphere given by a fixed great circle), in our
nonstationary scenario $\left\vert \psi\left(  t\right)  \right\rangle $ never
gets fully orthogonal to $\left\vert \psi\left(  0\right)  \right\rangle $.
After detailing these comments, we bring our discussion here a close.

Having outlined the fundamental aspects of Krylov's state complexity and our
quantum IG complexity measure, and having reformulated Krylov's state
complexity in geometric terms, we are now prepared to introduce our
comparative analysis aimed at gaining insights rather than pursuing generality.

\section{Comparative aspects in stationary scenarios}

In this section, we perform a comparative analysis of these two complexity
measures by explicitly investigating both geodesic and non-geodesic evolutions
defined by time-independent Hamiltonians.

\subsection{Geodesic motion}

In this first scenario, we consider the evolution from the state $\left\vert
\psi\left(  t_{i}\right)  \right\rangle =\left\vert 0\right\rangle $ to the
state $\left\vert \psi\left(  t_{f}\right)  \right\rangle =\left\vert
+\right\rangle \overset{\text{def}}{=}\left(  \left\vert 0\right\rangle
+\left\vert 1\right\rangle \right)  /\sqrt{2}$ under the time independent
Hamiltonian $\mathrm{H}\overset{\text{def}}{=}(1/\sqrt{6})\hslash\omega
\sigma_{y}$. Clearly, $t_{i}$ and $t_{f}$ denote the initial and final times,
respectively. We note that the quantum evolution characterized by the
Hamiltonian $\mathrm{H}$ can be understood as an electron positioned within a
configuration of a magnetic field, akin to that illustrated in (a) of Fig. $2$.

\emph{Dynamics}. From a simple calculation, we note that the evolved unit
quantum state vector $\left\vert \psi\left(  t\right)  \right\rangle
=e^{-\frac{i}{\hslash}\mathrm{H}t}\left\vert 0\right\rangle $ with $t_{i}\leq
t\leq t_{f}$ is given by%
\begin{equation}
\left\vert \psi\left(  t\right)  \right\rangle =\cos\left(  \frac{\omega
t}{\sqrt{6}}\right)  \left\vert 0\right\rangle +\sin\left(  \frac{\omega
t}{\sqrt{6}}\right)  \left\vert 1\right\rangle \text{.} \label{voice1}%
\end{equation}
From Eq. (\ref{voice1}), we note that $\left\vert \psi\left(  t_{f}\right)
\right\rangle =\left\vert +\right\rangle $ if and only if $t_{f}=\left[
\sqrt{6}\arccos(1/\sqrt{2})\right]  /\omega=(\pi\sqrt{6})/\left(
4\omega\right)  $. Interestingly, $t_{f}$ coincides with the shortest possible
time $t_{\mathrm{geo}}$ needed to reach the state $\left\vert +\right\rangle
$, when one starts in the state $\left\vert 0\right\rangle $ with a
Hamiltonian $\mathrm{H}$ with energy uncertainty $\Delta E\overset{\text{def}%
}{=}\sqrt{\left\langle 0\left\vert \mathrm{H}^{2}\right\vert 0\right\rangle
-\left\langle 0\left\vert \mathrm{H}\right\vert 0\right\rangle ^{2}}%
=\hslash\omega/\sqrt{6}$. Indeed, in these working conditions,
$t_{\mathrm{geo}}$ becomes%
\begin{equation}
t_{\mathrm{geo}}\overset{\text{def}}{=}\frac{\hslash\arccos\left[  \left\vert
\left\langle 0\left\vert +\right.  \right\rangle \right\vert \right]  }{\Delta
E}=\frac{\sqrt{6}\arccos(1/\sqrt{2})}{\omega}=t_{f}\text{.}%
\end{equation}
Furthermore, it is straightforward to confirm that $\eta_{\mathrm{GE}}$ in Eq.
(\ref{efficiency}) is equal to one for this quantum evolution.

Interestingly, we highlight that our complexity measure \textrm{C}\textbf{
}appears to be independent of the orthonormal basis chosen to represent the
time-evolved quantum state of the system. In this document, we consistently
denote $\left\vert \psi\left(  t\right)  \right\rangle $ (derived by applying
any unitary time propagator $U\left(  t\right)  $ to the initial state
$\left\vert \psi\left(  0\right)  \right\rangle $) using the computational
basis $\left\{  \left\vert 0\right\rangle \text{, }\left\vert 1\right\rangle
\right\}  $. However, this selection is not mandatory, and an alternative
suitable basis may be employed. It is crucial to observe that while
$\overline{\mathrm{V}}$ and \textrm{V}$_{\max}$ are not influenced by the
choice of this basis (and consequently, neither is the complexity \textrm{C}),
the temporal variation of the instantaneous volume $V\left(  t\right)  $ may
indeed be affected by the specific basis selected. In Fig\textbf{.} $6$, we
provide a visual representation that aids in understanding the
basis-independence inherent in our quantum IG complexity measure. We emphasize
that despite not including a formal mathematical proof of such
basis-independence for our complexity measure, all examples examined here
support this assertion. In Appendix C, in particular, these points are
demonstrated through a specific example for the benefit of interested
readers.\begin{figure}[t]
\centering
\includegraphics[width=0.75\textwidth] {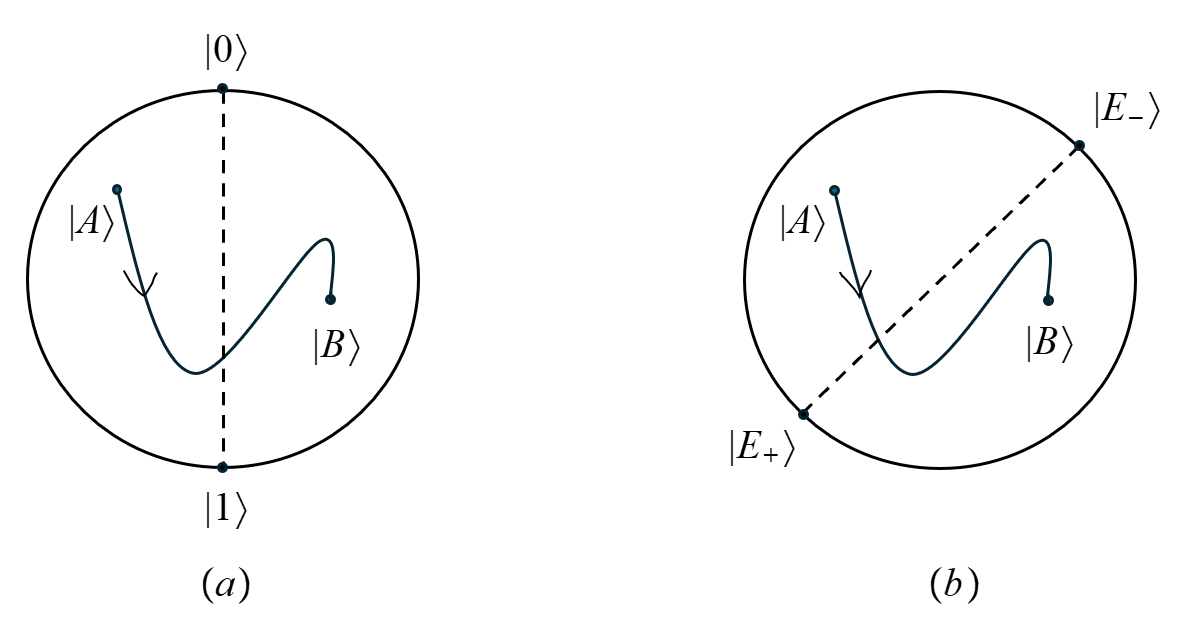}\caption{Sketch to illustrate the
basis-independence of our quantum IG complexity measure. While a rigorous
mathematical proof of basis-independence for our IG complexity measure in
arbitrary physical scenarios is not yet available, our illustrative examples
consistently suggest that this property holds in practice. The essential
argument is that the complexity \textrm{C}$\left(  t_{A}\text{, }t_{B}\right)
$ does not depend on the particular orthonormal basis chosen to decompose the
energy state $\left\vert \psi\left(  t\right)  \right\rangle =\left\langle
0\left\vert \psi\left(  t\right)  \right.  \right\rangle \left\vert
0\right\rangle +\left\langle 1\left\vert \psi\left(  t\right)  \right.
\right\rangle \left\vert 1\right\rangle =\left\langle E_{\_}\left\vert
\psi\left(  t\right)  \right.  \right\rangle \left\vert E_{\_}\right\rangle
+\left\langle E_{+}\left\vert \psi\left(  t\right)  \right.  \right\rangle
\left\vert E_{+}\right\rangle $. Although the parametrizations of the quantum
amplitudes with spherical angles can differ in different bases, the accessed
(i.e., $\overline{\mathrm{V}}$) and accessible (i.e., \textrm{V}$_{\max}$)
volumes used to construct the complexity measure do not change. Here,
$\left\{  \left\vert 0\right\rangle \text{, }\left\vert 1\right\rangle
\right\}  $ denotes the computational basis, while $\left\{  \left\vert
E_{\_}\right\rangle \text{, }\left\vert E_{+}\right\rangle \right\}  $ is the
basis determined by eigenstates of the Hamiltonian of the system. Finally,
$\left\vert A\right\rangle \overset{\text{def}}{=}\left\vert \psi\left(
t_{A}\right)  \right\rangle $ and $\left\vert B\right\rangle \overset
{\text{def}}{=}\left\vert \psi\left(  t_{B}\right)  \right\rangle $ are the
initial and final states, respectively.}%
\end{figure}

\emph{Average Krylov complexity}. We define the average Krylov complexity in
the geodesic case as,%
\begin{equation}
\left\langle \mathcal{K}\right\rangle _{\mathrm{geo}}\overset{\text{def}}%
{=}\frac{1}{t_{f}-t_{i}}\int_{t_{i}}^{t_{f}}\mathcal{K(}t)dt\text{,}
\label{vox2}%
\end{equation}
where $\mathcal{K(}t)\overset{\text{def}}{=}%
{\displaystyle\sum\limits_{n=0}^{1}}
np_{n}(t)$ is the (instantaneous) Krylov complexity and, in addition, the
probabilities $p_{n}(t)$ are specified by $p_{n}(t)\overset{\text{def}}%
{=}\left\vert \left\langle \phi_{n}\left\vert \psi\left(  t\right)  \right.
\right\rangle \right\vert ^{2}$ with $\left\{  \left\vert \phi_{n}%
\right\rangle \right\}  _{0\leq n\leq1}$ being an orthonormal Krylov basis.
Applying the Gram-Schmidt orthonormalization procedure to the set $\left\{
\left\vert 0\right\rangle \text{, }\mathrm{H}\left\vert 0\right\rangle
\right\}  $, one arrives at $\left\{  \left\vert \phi_{n}\right\rangle
\right\}  _{0\leq n\leq1}=\left\{  \left\vert \phi_{0}\right\rangle \text{,
}\left\vert \phi_{1}\right\rangle \right\}  =\left\{  \left\vert
0\right\rangle \text{, }\left\vert 1\right\rangle \right\}  $. Therefore, a
simple calculation yields $\mathcal{K(}t)=\sin^{2}(\omega t/\sqrt{6})$.
Finally, $\left\langle \mathcal{K}\right\rangle _{\mathrm{geo}}$ in Eq.
(\ref{vox2}) reduces to%
\begin{equation}
\left\langle \mathcal{K}\right\rangle _{\mathrm{geo}}=\frac{4\omega}{\pi
\sqrt{6}}\int_{0}^{\frac{\pi\sqrt{6}}{4\omega}}\sin^{2}\left(  \frac{\omega
t}{\sqrt{6}}\right)  dt=\frac{1}{2}-\frac{1}{\pi}\sim0.181\text{.}
\label{vox3}%
\end{equation}
The quantity $\left\langle \mathcal{K}\right\rangle _{\mathrm{geo}}$ in Eq.
(\ref{vox3}) represents the average Krylov complexity of the geodesic
evolution from the state $\left\vert 0\right\rangle $ to the state $\left\vert
+\right\rangle $ under the Hamiltonian $\mathrm{H}\overset{\text{def}}%
{=}(1/\sqrt{6})\hslash\omega\sigma_{y}$.

\textbf{ }Given that $\mathcal{K}\left(  t\right)  $ represents an
instantaneous value, whereas \textrm{C}$\left(  t_{i}\text{, }t_{f}\right)  $
is assessed over a finite time interval\textbf{ }$\left[  t_{i}\text{, }%
t_{f}\right]  $, our analysis solely involves comparative statements regarding
the numerical estimates of\textbf{ }\textrm{C}$\left(  t_{i}\text{, }%
t_{f}\right)  $ in relation to the average $\left\langle \mathcal{K}%
\right\rangle $ of $\mathcal{K}\left(  t\right)  $ across the interval
$\left[  t_{i}\text{, }t_{f}\right]  $.

\emph{Quantum IGC}. In the scenario being considered, the polar and azimuthal
angles that specify the Bloch representation of the state \emph{ }$\left\vert
\psi\left(  t\right)  \right\rangle $ in Eq. (\ref{voice1}) are given by
$\theta\left(  t\right)  =2\omega t/\sqrt{6}$ and $\varphi\left(  t\right)
=0$, respectively, with $0\leq t\leq(\pi\sqrt{6})/(4\omega)$. By means of
simple calculations, we find that the instantaneous volume equals $V\left(
t\right)  \overset{\text{def}}{=}\int_{\theta\left(  0\right)  }%
^{\theta\left(  t\right)  }(1/2)d\theta=\omega t/\sqrt{6}$, the accessible
volume is given by $\overline{\mathrm{V}}\overset{\text{def}}{=}\left(
t_{f}-t_{i}\right)  ^{-1}\int_{t_{i}}^{t_{f}}V(t)dt=\pi/8$ (regardless of the
chosen value of $\omega$) and, lastly, the accessible volume is \textrm{V}%
$_{\max}\overset{\text{def}}{=}\int_{\theta_{\min}}^{\theta_{\max}%
}(1/2)d\theta=\pi/4$ with $\theta_{\min}=0$ and $\theta_{\max}=\pi/2$.
Therefore, the quantum IGC becomes
\begin{equation}
\mathrm{C}_{\mathrm{geo}}\overset{\text{def}}{=}\frac{\mathrm{V}_{\max
}-\overline{\mathrm{V}}}{\mathrm{V}_{\max}}=\frac{1}{2}=0.5\text{.}
\label{vox4}%
\end{equation}
The quantity $\mathrm{C}_{\mathrm{geo}}$ in Eq. (\ref{vox4}) denotes the
quantum IGC of the geodesic evolution from the state $\left\vert
0\right\rangle $ to the state $\left\vert +\right\rangle $ under the
Hamiltonian $\mathrm{H}\overset{\text{def}}{=}(1/\sqrt{6})\hslash\omega
\sigma_{y}$.

The time-dependent Bloch vector is given by
\begin{equation}
\mathbf{a}_{\mathrm{geo}}\left(  t\right)  =\left(  \sin(\frac{2\omega
t}{\sqrt{6}})\text{, }0\text{, }\cos(\frac{2\omega t}{\sqrt{6}})\right)
\text{,} \label{giorgia}%
\end{equation}
with $\mathbf{a}_{\mathrm{geo}}\left(  t_{i}\right)  =\left(  0\text{,
}0\text{, }1\right)  \leftrightarrow\left\vert 0\right\rangle $ and
$\mathbf{a}_{\mathrm{geo}}\left(  t_{f}\right)  =\left(  1\text{, }0\text{,
}0\right)  \leftrightarrow\left\vert +\right\rangle $. Let $\mathbf{n}$ be the
unit vector that specifies the \textquotedblleft magnetic field
vector\textquotedblright\ $\mathbf{h}\overset{\text{def}}{\mathbf{=}%
}h\mathbf{n}$ that enters the Hamiltonian \textrm{H}$=\mathbf{h\cdot
\boldsymbol{\sigma}}$ with\textbf{ }$h\overset{\text{def}}{=}\left\Vert
\mathbf{h}\right\Vert =(1/\sqrt{6})\hslash\omega$). Since\textbf{ }%
$\mathbf{a}_{\mathrm{geo}}\left(  t_{i}\right)  $ is orthogonal to
$\mathbf{n=}\left(  0\text{, }1\text{, }0\right)  $, we have that this
geodesic evolution is characterized by a maximal \textquotedblleft
instantaneous\textquotedblright\ Krylov state complexity given by
$\mathcal{K(}t)=\sin^{2}(\omega t/\sqrt{6})$. Moreover, from Eq. (\ref{XXXb}),
we note that the curvature coefficient $\kappa_{\mathrm{AC}}^{2}$ of this
quantum evolution is zero.

From $\mathbf{a}\left(  t\right)  =\left(  \sin(\frac{2\omega t}{\sqrt{6}%
})\text{, }0\text{, }\cos(\frac{2\omega t}{\sqrt{6}})\right)  $ and
$\mathbf{a}\left(  0\right)  =\left(  0\text{, }0\text{, }1\right)  $, we have
$\left\Vert \mathbf{a}\left(  t\right)  -\mathbf{a}\left(  0\right)
\right\Vert ^{2}=4\sin^{2}(\frac{\omega t}{\sqrt{6}})=4\mathcal{K}\left(
t\right)  $. Therefore, since $\mathcal{K}\left(  t\right)  \propto\left\Vert
\mathbf{a}\left(  t\right)  -\mathbf{a}\left(  0\right)  \right\Vert ^{2}$,
the square root of Krylov's state complexity can be regarded as a measure of
the distance between the initial Bloch vector $\mathbf{a}\left(  0\right)  $
and the time-evolved Bloch vector $\mathbf{a}\left(  t\right)  $ at a later
time $t$. We also note that near $t=0$, $\mathcal{K}\left(  t\right)
\leq\mathrm{C}_{\mathcal{B}}(t)$, with $\mathcal{B}$ being an ordered,
complete, and orthonormal basis for the Krylov subspace defined in terms of
the eigenvectors of the Hamiltonian \textrm{H}$\overset{\text{def}}{=}%
\frac{\hslash\omega}{\sqrt{6}}\sigma_{y}$. Indeed, $\sigma_{y}$ has
eigenvalues $\lambda_{-}\overset{\text{def}}{=}-1$ and $\lambda_{+}%
\overset{\text{def}}{=}+1$, with corresponding eigenspaces $\mathcal{E}%
_{\lambda_{-}}\overset{\text{def}}{=}\mathrm{Span}\left\{  \left\vert
E_{-}\right\rangle \right\}  $ and $\mathcal{E}_{\lambda_{+}}\overset
{\text{def}}{=}\mathrm{Span}\left\{  \left\vert E_{+}\right\rangle \right\}
$. In our case, the orthonormal eigenstates $\left\vert E_{-}\right\rangle $
and $\left\vert E_{+}\right\rangle $ are given by%
\begin{equation}
\left\vert E_{-}\right\rangle \overset{\text{def}}{=}\frac{\left\vert
0\right\rangle -i\left\vert 1\right\rangle }{\sqrt{2}}\text{, and }\left\vert
E_{+}\right\rangle \overset{\text{def}}{=}\frac{\left\vert 0\right\rangle
+i\left\vert 1\right\rangle }{\sqrt{2}}\text{,}%
\end{equation}
respectively. More explicitly, we get $\mathcal{K}\left(  t\right)  =\sin
^{2}(\frac{\omega t}{\sqrt{6}})\leq\frac{1}{2}=\left\vert \left\langle
E_{+}\left\vert \psi\left(  t\right)  \right.  \right\rangle \right\vert
^{2}=\mathrm{C}_{\mathcal{B}}(t)$ for any $0\leq t\leq\frac{\pi\sqrt{6}%
}{4\omega}$. We stress that $\left\vert \left\langle E_{-}\left\vert
\psi\left(  t\right)  \right.  \right\rangle \right\vert ^{2}=\frac{1}{2}$
and, as expected, $\left\vert \left\langle E_{+}\left\vert \psi\left(
t\right)  \right.  \right\rangle \right\vert ^{2}+\left\vert \left\langle
E_{-}\left\vert \psi\left(  t\right)  \right.  \right\rangle \right\vert
^{2}=1$. For additional information regarding the relationships between
$\mathcal{K}\left(  t\right)  $, $\left\Vert \mathbf{a}\left(  t\right)
-\mathbf{a}\left(  0\right)  \right\Vert ^{2}$, and the Fubini-Study finite
volume element $V_{\mathrm{FS}}(t)$, we recommend referring to Appendix D.

\subsection{Nongeodesic motion}

In this second scenario, we consider the evolution from the state $\left\vert
\psi\left(  t_{i}\right)  \right\rangle =\left\vert 0\right\rangle $ to the
state $\left\vert \psi\left(  t_{f}\right)  \right\rangle =\left\vert
+\right\rangle \overset{\text{def}}{=}\left(  \left\vert 0\right\rangle
+\left\vert 1\right\rangle \right)  /\sqrt{2}$ under the time independent
Hamiltonian $\mathrm{H}\overset{\text{def}}{=}(\hslash\omega/2)\mathbf{n\cdot
\boldsymbol{\sigma}}$ with $\mathbf{n}\overset{\text{def}}{=}(1/\sqrt
{3})\left(  1\text{, }1\text{, }1\right)  $ and $\mathbf{\boldsymbol{\sigma}%
}\overset{\text{def}}{=}\left(  \sigma_{x}\text{, }\sigma_{y}\text{, }%
\sigma_{z}\right)  $. For completeness, we observe that the initial and final
states in both geodesic and nongeodesic evolutions are the same. This is
obviously required to have a fair comparison between the different types of
evolutions. We also observe that the quantum evolution defined by the
Hamiltonian \textrm{H} can be interpreted as an electron situated within a
magnetic field configuration, similar to that depicted in (a) of Fig. $2$.

\emph{Dynamics}. From a straightforward calculation, we observe that the
evolved unit quantum state vector $\left\vert \psi\left(  t\right)
\right\rangle =e^{-\frac{i}{\hslash}\mathrm{H}t}\left\vert 0\right\rangle $
with $t_{i}\leq t\leq t_{f}$ reduces to%
\begin{equation}
\left\vert \psi\left(  t\right)  \right\rangle =\left[  \cos\left(
\frac{\omega t}{2}\right)  -i\frac{1}{\sqrt{3}}\sin(\frac{\omega t}%
{2})\right]  \left\vert 0\right\rangle -i\frac{1+i}{\sqrt{3}}\sin\left(
\frac{\omega t}{2}\right)  \left\vert 1\right\rangle \text{.} \label{voice1a}%
\end{equation}
From Eq. (\ref{voice1a})\textbf{,} we note that $\left\vert \psi\left(
t_{f}\right)  \right\rangle $ is physically equivalent to $\left\vert
+\right\rangle $ (i.e., $\left\vert \psi\left(  t_{f}\right)  \right\rangle
=e^{i\phi}\left\vert +\right\rangle $ for some phase $\phi\in%
\mathbb{R}
$) if and only if%
\begin{equation}
\frac{-i\frac{1+i}{\sqrt{3}}\sin\left(  \frac{\omega t_{f}}{2}\right)  }%
{\cos\left(  \frac{\omega t_{f}}{2}\right)  -i\frac{1}{\sqrt{3}}\sin
(\frac{\omega t_{f}}{2})}=1\text{,}%
\end{equation}
that is, if and only if $t_{f}=\left(  2\pi\right)  /(3\omega)$. Observe that
$t_{f}^{\mathrm{geo}}\overset{\text{def}}{=}(\pi\sqrt{6})/\left(
4\omega\right)  \sim1.92/\omega\leq2.09/\omega\sim\left(  2\pi\right)
/(3\omega)\overset{\text{def}}{=}t_{f}^{\mathrm{nongeo}}$. We also notice that
the comparison between $\mathrm{H}\overset{\text{def}}{=}(1/\sqrt{6}%
)\hslash\omega\sigma_{y}$ and $\mathrm{H}\overset{\text{def}}{=}\left[
\left(  \hslash\omega\right)  /(2\sqrt{3})\right]  \left(  \sigma_{x}%
+\sigma_{y}+\sigma_{z}\right)  $ is fair since the energy uncertainty of both
Hamiltonians with respect to the initial state $\left\vert 0\right\rangle $
assumes the same identical value of $\Delta E=\hslash\omega/\sqrt{6}$.
Furthermore, it is simple to verify that $\eta_{\mathrm{GE}}$ in Eq.
(\ref{efficiency}) is stricly less than one for this quantum evolution, since
$\eta_{\mathrm{GE}}=(3\sqrt{6})/8\simeq0.92<1$.

\emph{Average Krylov complexity}. In this case, following the reasoning
outlined for the geodesic evolution scenario, an easy calculation leads to
$\mathcal{K(}t)=(2/3)\sin^{2}(\omega t/2)$. Therefore, the average Krylov
complexity as given in Eq. (\ref{vox2}) becomes%
\begin{equation}
\left\langle \mathcal{K}\right\rangle _{\mathrm{nongeo}}=\frac{3\omega}{2\pi
}\int_{0}^{\frac{2\pi}{3\omega}}\frac{2}{3}\sin^{2}\left(  \frac{\omega t}%
{2}\right)  dt=\frac{1}{3}-\frac{\sqrt{3}}{4}\frac{1}{\pi}\sim0.195\text{.}
\label{vox3b}%
\end{equation}
The quantity$\left\langle \mathcal{K}\right\rangle _{\mathrm{nongeo}}$ in Eq.
(\ref{vox3b}) is the average Krylov complexity of the geodesic evolution from
the state $\left\vert 0\right\rangle $ to the state $\left\vert +\right\rangle
$ under the Hamiltonian $\mathrm{H}\overset{\text{def}}{=}\left[  \left(
\hslash\omega\right)  /(2\sqrt{3})\right]  \left(  \sigma_{x}+\sigma
_{y}+\sigma_{z}\right)  $. Comparing Eqs. (\ref{vox3}) and (\ref{vox3b}), we
note that $\left\langle \mathcal{K}\right\rangle _{\mathrm{nongeo}}%
\geq\left\langle \mathcal{K}\right\rangle _{\mathrm{geo}}$.

\bigskip\emph{Quantum IGC}. In the scenario being considered, we need to find
the polar and azimuthal angles that specify the Bloch representation of the
state \emph{ }$\left\vert \psi\left(  t\right)  \right\rangle $ in Eq.
(\ref{voice1a}), where
\begin{equation}
\left\vert \psi\left(  t\right)  \right\rangle =a(t)\left\vert 0\right\rangle
+b(t)\left\vert 1\right\rangle =\cos\left[  \frac{\theta\left(  t\right)  }%
{2}\right]  \left\vert 0\right\rangle +e^{i\varphi\left(  t\right)  }%
\sin\left[  \frac{\theta\left(  t\right)  }{2}\right]  \left\vert
1\right\rangle \text{,}%
\end{equation}
with $\theta\left(  t\right)  =2\arccos\left[  \left\vert a(t)\right\vert
\right]  $, $\varphi\left(  t\right)  =\arg\left[  b\left(  t\right)  \right]
-\arg\left[  a\left(  t\right)  \right]  $, with complex probability
amplitudes $a(t)$ and $b(t)$ such that $\left\vert a\left(  t\right)
\right\vert ^{2}+\left\vert b(t)\right\vert ^{2}=1$. After some algebraic
manipulations, we arrive at an expression for the polar angle given by
\begin{equation}
\theta\left(  t\right)  =2\arccos\left[  \sqrt{1-\frac{2}{3}\sin^{2}\left(
\frac{\omega t}{2}\right)  }\right]  \text{,} \label{ava1}%
\end{equation}
with $\theta\left(  0\right)  =0$ and $\theta(\frac{2\pi}{3\omega})=\pi/2$.
Moreover, the azimuthal angle becomes
\begin{equation}
\varphi\left(  t\right)  =-\frac{\pi}{4}+\arctan\left[  \frac{1}{\sqrt{3}}%
\tan\left(  \frac{\omega t}{2}\right)  \right]  \text{,} \label{ava2}%
\end{equation}
with $\varphi\left(  0\right)  =-\frac{\pi}{4}$ and $\varphi(\frac{2\pi
}{3\omega})=0$. In this nongeodesic scenario, the instantaneous volume
$V\left(  t\right)  $ is equal to
\begin{align}
V\left(  t\right)   &  =\frac{1}{4}\left\vert \left(  \int_{\theta\left(
0\right)  }^{\theta\left(  t\right)  }\sin(\theta)d\theta\right)  \left(
\int_{\varphi\left(  0\right)  }^{\varphi\left(  t\right)  }d\varphi\right)
\right\vert \nonumber\\
&  =\frac{1}{4}\left\vert \left\{  1-\cos\left[  2\arccos\left(  \sqrt
{1-\frac{2}{3}\sin^{2}(\frac{\omega t}{2})}\right)  \right]  \right\}
\arctan\left[  \frac{1}{\sqrt{3}}\tan(\frac{\omega t}{2})\right]  \right\vert
\text{.} \label{vox5}%
\end{align}
The accessible volume defined as $\overline{\mathrm{V}}\overset{\text{def}}%
{=}\left(  t_{f}-t_{i}\right)  ^{-1}\int_{t_{i}}^{t_{f}}V(t)dt$ can be
estimated numerically using Eq. (\ref{vox5}), once one sets $t_{i}=0$ and
$t_{f}=\left(  2\pi\right)  /\left(  3\omega\right)  $. We have,
$\overline{\mathrm{V}}\sim5.11\times10^{-2}$, irrespective of the specific
value of $\omega$. Finally, the accessible volume is
\begin{equation}
\mathrm{V}_{\max}=\frac{1}{4}\left(  \int_{\theta_{\min}}^{\theta_{\max}}%
\sin(\theta)d\theta\right)  \left(  \int_{\varphi_{\min}}^{\varphi_{\max}%
}d\varphi\right)  =\frac{\pi}{16}\text{,}%
\end{equation}
where $\theta_{\min}=0$, $\theta_{\max}=\pi/2$, $\varphi_{\min}=-\pi/4$, and
$\varphi_{\max}=0$. Therefore, the quantum IGC reduces to
\begin{equation}
\mathrm{C}_{\mathrm{nongeo}}\overset{\text{def}}{=}\frac{\mathrm{V}_{\max
}-\overline{\mathrm{V}}}{\mathrm{V}_{\max}}\sim0.74\text{.} \label{vox4b}%
\end{equation}
The quantity $\mathrm{C}_{\mathrm{nongeo}}$ in Eq. (\ref{vox4b}) specifies the
quantum IGC of the nongeodesic evolution from the state $\left\vert
0\right\rangle $ to the state $\left\vert +\right\rangle $ under the
Hamiltonian $\mathrm{H}\overset{\text{def}}{=}\left[  \left(  \hslash
\omega\right)  /(2\sqrt{3})\right]  \left(  \sigma_{x}+\sigma_{y}+\sigma
_{z}\right)  $. Comparing Eqs. (\ref{vox4}) and (\ref{vox4b}), we conclude
that $\mathrm{C}_{\mathrm{nongeo}}\geq\mathrm{C}_{\mathrm{geo}}$.

For completeness, we remark the time-dependent Bloch vector is given by
\begin{equation}
\mathbf{a}_{\mathrm{nongeo}}\left(  t\right)  =\left(  \sin(\theta_{t}%
)\cos(\varphi_{t})\text{, }\sin(\theta_{t})\sin(\varphi_{t})\text{, }%
\cos(\theta_{t})\right)  \text{,}%
\end{equation}
where $\theta_{t}=\theta\left(  t\right)  $ in Eq. (\ref{ava1}) and
$\varphi_{t}=\varphi\left(  t\right)  $ in Eq. (\ref{ava2}). Lastly, observe
that $\mathbf{a}_{\mathrm{nongeo}}\left(  t_{i}\right)  =\left(  0\text{,
}0\text{, }1\right)  \leftrightarrow\left\vert 0\right\rangle $ and
$\mathbf{a}_{\mathrm{nongeo}}\left(  t_{f}\right)  =\left(  1\text{, }0\text{,
}0\right)  \leftrightarrow\left\vert +\right\rangle $. Since $\mathbf{a}%
_{\mathrm{nongeo}}\left(  t_{i}\right)  $ is not orthogonal to $\mathbf{n=}%
(1/\sqrt{3})\left(  1\text{, }1\text{, }1\right)  $ (i.e., the unit vector
that specifies the \textquotedblleft magnetic field vector\textquotedblright%
\ $\mathbf{h}\overset{\text{def}}{\mathbf{=}}h\mathbf{n}$ that enters the
Hamiltonian \textrm{H}$=\mathbf{h\cdot\boldsymbol{\sigma}}$ with
$h\overset{\text{def}}{=}\left\Vert \mathbf{h}\right\Vert =(\hslash\omega
)/2$), we have that this nongeodesic evolution is not characterized by a
maximal \textquotedblleft instantaneous\textquotedblright\ Krylov state
complexity given that $\mathcal{K(}t)=(2/3)\sin^{2}(\omega t/2)\leq\sin
^{2}(\omega t/2)$.

From $\mathbf{a}\left(  t\right)  =\left(  \sin(\theta_{t})\cos(\varphi
_{t})\text{, }\sin(\theta_{t})\sin(\varphi_{t})\text{, }\cos(\theta
_{t})\right)  $ and $\mathbf{a}\left(  0\right)  =\left(  0\text{, }0\text{,
}1\right)  $, we obtain $\left\Vert \mathbf{a}\left(  t\right)  -\mathbf{a}%
\left(  0\right)  \right\Vert ^{2}=4\sin^{2}(\frac{\theta_{t}}{2})$. Given
that $\theta_{t}=2\arccos(\sqrt{1-\frac{2}{3}\sin^{2}(\frac{\omega t}{2})})$
and $\sin^{2}(\arccos(x))=1-x^{2}$, we arrive at $\left\Vert \mathbf{a}\left(
t\right)  -\mathbf{a}\left(  0\right)  \right\Vert ^{2}=\frac{8}{3}\sin
^{2}(\frac{\omega t}{2})=4\mathcal{K}\left(  t\right)  $. Consequently, given
that $\mathcal{K}\left(  t\right)  \propto\left\Vert \mathbf{a}\left(
t\right)  -\mathbf{a}\left(  0\right)  \right\Vert ^{2}$, the square root of
Krylov's state complexity can be interpreted as an indicator of the distance
separating the initial Bloch vector $\mathbf{a}\left(  0\right)  $ from the
time-evolved Bloch vector $\mathbf{a}\left(  t\right)  $ at a subsequent time
$t$. Furthermore, we notice that the curvature coefficient $\kappa
_{\mathrm{AC}}^{2}$ of this quantum evolution is different from zero since one
can check that $\kappa_{\mathrm{AC}}^{2}=2\neq0$. We also observe that near
$t=0$, $\mathcal{K}\left(  t\right)  \leq\mathrm{C}_{\mathcal{B}}(t)$, with
$\mathcal{B}$ being an ordered, complete, and orthonormal basis for the Krylov
subspace defined by means of the eigenvectors of the Hamiltonian
\textrm{H}$\overset{\text{def}}{=}\frac{\hslash\omega}{2}\frac{\sigma
_{x}+\sigma_{y}+\sigma_{z}}{\sqrt{3}}$. Indeed, $\frac{\sigma_{x}+\sigma
_{y}+\sigma_{z}}{\sqrt{3}}$ possesses eigenvalues $\lambda_{-}\overset
{\text{def}}{=}-1$ and $\lambda_{+}\overset{\text{def}}{=}+1$, with
corresponding eigenspaces $\mathcal{E}_{\lambda_{-}}\overset{\text{def}}%
{=}\mathrm{Span}\left\{  \left\vert E_{-}\right\rangle \right\}  $ and
$\mathcal{E}_{\lambda_{+}}\overset{\text{def}}{=}\mathrm{Span}\left\{
\left\vert E_{+}\right\rangle \right\}  $. In our case, the orthonormal
eigenstates $\left\vert E_{-}\right\rangle $ and $\left\vert E_{+}%
\right\rangle $ are given by%
\begin{equation}
\left\vert E_{-}\right\rangle \overset{\text{def}}{=}\frac{1}{\sqrt{3-\sqrt
{3}}}\left(
\begin{array}
[c]{c}%
-(1-i)\frac{\sqrt{3}}{3+\sqrt{3}}\\
1
\end{array}
\right)  \text{, and }\left\vert E_{+}\right\rangle \overset{\text{def}}%
{=}\frac{1}{\sqrt{\frac{6}{3-\sqrt{3}}}}\left(
\begin{array}
[c]{c}%
(1-i)\frac{\sqrt{3}}{3-\sqrt{3}}\\
1
\end{array}
\right)  \text{,}%
\end{equation}
respectively. More explicitly, we get $\mathcal{K}\left(  t\right)  =\frac
{2}{3}\sin^{2}(\frac{\omega t}{2})\leq\frac{3-\sqrt{3}}{12-6\sqrt{3}%
}=\left\vert \left\langle E_{+}\left\vert \psi\left(  t\right)  \right.
\right\rangle \right\vert ^{2}=\mathrm{C}_{\mathcal{B}}(t)$ for any $t$. For
completeness, we remark that $\left\vert \left\langle E_{-}\left\vert
\psi\left(  t\right)  \right.  \right\rangle \right\vert ^{2}=\frac{1}%
{3+\sqrt{3}}$ and, as expected, $\left\vert \left\langle E_{+}\left\vert
\psi\left(  t\right)  \right.  \right\rangle \right\vert ^{2}+\left\vert
\left\langle E_{-}\left\vert \psi\left(  t\right)  \right.  \right\rangle
\right\vert ^{2}=1$.

After conducting a comparative analysis of these two complexity measures by
examining both geodesic and non-geodesic evolutions defined by stationary
Hamiltonians, we are now prepared to move on to nonstationary Hamiltonian models.

\section{Comparative aspects in nonstationary scenarios}

In this section, we conduct a comparative analysis of these two complexity
measures by explicitly examining both geodesic and non-geodesic evolutions as
defined by time-dependent Hamiltonians.

\subsection{Hamiltonian model}

In this subsection, we introduce a two-parameter family of time-varying
Hamiltonians, which we aim to examine from a geometric viewpoint.
Specifically, our focus is on the time-dependent configurations of the
magnetic fields that define the Hamiltonians. We also will examine the
temporal dynamics of the phase that governs the relative phase factor involved
in the Bloch sphere decomposition of the evolving state by means of the
computational basis state vectors.

In Ref. \cite{uzdin12}, a comprehensive Hermitian nonstationary qubit
Hamiltonian $\mathrm{H}\left(  t\right)  $ is developed in a way that it
produces the identical motion $\pi\left(  \left\vert \psi\left(  t\right)
\right\rangle \right)  $ within the complex projective Hilbert space $%
\mathbb{C}
P^{1}$ (or, correspondingly, on the Bloch sphere $S^{2}\cong%
\mathbb{C}
P^{1}$) as $\left\vert \psi\left(  t\right)  \right\rangle $, where the
projection operator $\pi$ is defined such that $\pi:\mathcal{H}_{2}^{1}%
\ni\left\vert \psi\left(  t\right)  \right\rangle \mapsto\pi\left(  \left\vert
\psi\left(  t\right)  \right\rangle \right)  \in%
\mathbb{C}
P^{1}$. In general, it can be demonstrated that $\mathrm{H}\left(  t\right)  $
can be expressed as%
\begin{equation}
\mathrm{H}\left(  t\right)  =iE\left\vert \partial_{t}m(t)\right\rangle
\left\langle m(t)\right\vert -iE\left\vert m(t)\right\rangle \left\langle
\partial_{t}m(t)\right\vert \text{,} \label{oppio}%
\end{equation}
where, for the sake of simplicity, we define $\left\vert m(t)\right\rangle
=\left\vert m\right\rangle $, $\left\vert \partial_{t}m(t)\right\rangle
=\left\vert \dot{m}\right\rangle $, $E=1$ and, finally, $\hslash=1$. The unit
state vector $\left\vert m\right\rangle $ satisfies the equations $\pi\left(
\left\vert m(t)\right\rangle \right)  =\pi\left(  \left\vert \psi\left(
t\right)  \right\rangle \right)  $ and $i\partial_{t}\left\vert
m(t)\right\rangle =\mathrm{H}\left(  t\right)  \left\vert m(t)\right\rangle $.
The equation $\pi\left(  \left\vert m(t)\right\rangle \right)  =\pi\left(
\left\vert \psi\left(  t\right)  \right\rangle \right)  $ leads us to conclude
that $\left\vert m(t)\right\rangle =c(t)\left\vert \psi\left(  t\right)
\right\rangle $, with $c(t)$ denoting a complex function. By stating that
$\left\langle m\left\vert m\right.  \right\rangle =1$, we infer that
$\left\vert c(t)\right\vert =1$. This condition indicates, therefore, that
$c(t)=e^{i\phi\left(  t\right)  }$ for a particular real-valued phase
$\phi\left(  t\right)  $. Subsequently, by applying the parallel transport
condition $\left\langle m\left\vert \dot{m}\right.  \right\rangle
=\left\langle \dot{m}\left\vert m\right.  \right\rangle =0$, the phase
$\phi\left(  t\right)  $ is determined as $i\int\left\langle \psi\left\vert
\dot{\psi}\right.  \right\rangle dt$. Consequently, $\left\vert
m(t)\right\rangle =\exp(-\int_{0}^{t}\left\langle \psi(t^{\prime})\left\vert
\partial_{t^{\prime}}\psi(t^{\prime})\right.  \right\rangle dt^{\prime
})\left\vert \psi\left(  t\right)  \right\rangle $. It is essential to
emphasize that $\mathrm{H}\left(  t\right)  $ in Eq. (\ref{oppio}) is
inherently traceless, as it consists exclusively of off-diagonal elements
relative to the orthogonal basis \{$\left\{  \left\vert m\right\rangle \text{,
}\left\vert \partial_{t}m\right\rangle \right\}  $. Furthermore,
$\mathrm{H}\left(  t\right)  $ represents a linear combination of traceless
Pauli spin matrices. Finally, the condition $i\partial_{t}\left\vert
m(t)\right\rangle =\mathrm{H}\left(  t\right)  \left\vert m(t)\right\rangle $
indicates that $\left\vert m(t)\right\rangle $ adheres to the Schr\"{o}dinger
evolution equation. As a supplementary note, we wish to highlight that the
unitary time propagator $U\left(  t\right)  =\mathcal{T}\exp(-i\int_{0}%
^{t}\mathrm{H}(s)ds)$, which is associated with the Hamiltonian \textrm{H}%
$\left(  t\right)  $ in Eq. (\ref{oppio}), can be represented as a
path-ordered \textrm{SU}$\left(  2\text{; }%
\mathbb{C}
\right)  $ rotation, formulated as
\begin{equation}
U(t)=\mathcal{T}\exp(-\frac{i}{\hslash}\frac{E}{2}\int_{0}^{t}\left\{  \left[
\mathbf{a}\left(  t^{\prime}\right)  \times\mathbf{\dot{a}}\left(  t^{\prime
}\right)  \right]  \cdot\mathbf{\boldsymbol{\sigma}}\right\}  dt^{\prime
})\text{,}%
\end{equation}
where $\mathcal{T}$ denotes the time-ordering operator. In fact, by defining
the projection operator $P\left(  t\right)  \overset{\text{def}}{=}\left\vert
m(t)\right\rangle \left\langle m(t)\right\vert =\left[  \mathbf{1+a}\left(
t\right)  \cdot\mathbf{\boldsymbol{\sigma}}\right]  /2$ and utilizing the
standard properties of Pauli matrices, one can derive the Hamiltonian
expression \textrm{H}$\left(  t\right)  =iE\left[  \dot{P}\left(  t\right)
\text{, }P\left(  t\right)  \right]  =(E/2)\left[  \mathbf{a}\left(  t\right)
\times\mathbf{\dot{a}}\left(  t\right)  \right]  \cdot
\mathbf{\boldsymbol{\sigma}}$, which is characterized by a \textquotedblleft
magnetic\textquotedblright\ field proportional to $\mathbf{a}\left(  t\right)
\times\mathbf{\dot{a}}\left(  t\right)  $ which is orthogonal to the Bloch
vector $\mathbf{a}\left(  t\right)  $. After providing some critical
preliminary information regarding Uzdin's research in Ref. \cite{uzdin12}, we
are now ready to present our proposed time-dependent Hamiltonian.

To commence, let us examine the normalized state vector $\left\vert
\psi\left(  t\right)  \right\rangle $ defined as,%
\begin{equation}
\left\vert \psi\left(  t\right)  \right\rangle \overset{\text{def}}{=}%
\cos\left[  \alpha\left(  t\right)  \right]  \left\vert 0\right\rangle
+e^{i\beta\left(  t\right)  }\sin\left[  \alpha\left(  t\right)  \right]
\left\vert 1\right\rangle \text{,} \label{do1}%
\end{equation}
where $\alpha\left(  t\right)  $ and $\beta\left(  t\right)  $ are two
generally time-dependent real-valued parameters. From Eq. (\ref{do1}), we
observe that $\left\langle \psi\left(  t\right)  \left\vert \dot{\psi}\left(
t\right)  \right.  \right\rangle =i\dot{\beta}\left(  t\right)  \sin
^{2}\left[  \alpha\left(  t\right)  \right]  \neq0$. Consequently, let us
determine $\left\vert m(t)\right\rangle =e^{-i\phi\left(  t\right)  }$
$\left\vert \psi\left(  t\right)  \right\rangle $, with $\left\langle m\left(
t\right)  \left\vert \dot{m}\left(  t\right)  \right.  \right\rangle =0$. We
note that $\left\langle m\left(  t\right)  \left\vert \dot{m}\left(  t\right)
\right.  \right\rangle =0$ if and only if $\ \dot{\phi}\left(  t\right)
=-i\left\langle \psi\left(  t\right)  \left\vert \dot{\psi}\left(  t\right)
\right.  \right\rangle =\dot{\beta}\left(  t\right)  \sin^{2}(\left[
\alpha\left(  t\right)  \right]  $, which implies%
\begin{equation}
\phi\left(  t\right)  =\int_{0}^{t}\dot{\beta}\left(  t^{\prime}\right)
\sin^{2}\left[  \alpha\left(  t^{\prime}\right)  \right]  dt^{\prime}\text{.}
\label{do4}%
\end{equation}
Given $\left\vert \psi\left(  t\right)  \right\rangle $ and $\phi\left(
t\right)  $ as defined in Eqs. (\ref{do1}) and (\ref{do4}), the state
$\left\vert m(t)\right\rangle $ can be expressed as
\begin{equation}
\left\vert m(t)\right\rangle =e^{-i\int_{0}^{t}\dot{\beta}\left(  t^{\prime
}\right)  \sin^{2}\left[  \alpha\left(  t^{\prime}\right)  \right]
dt^{\prime}}\left\{  \cos\left[  \alpha\left(  t\right)  \right]  \left\vert
0\right\rangle +e^{i\beta\left(  t\right)  }\sin\left[  \alpha\left(
t\right)  \right]  \left\vert 1\right\rangle \right\}  \text{,} \label{do4b}%
\end{equation}
By setting $\left\vert m\right\rangle \left\langle m\right\vert =(1/2)\left(
\mathbf{1+a}\cdot\mathbf{\boldsymbol{\sigma}}\right)  $, the application of
Eq. (\ref{do4b}) results in the formulation of the Bloch vector $\mathbf{a}%
\left(  t\right)  $ in relation to the real-valued time-dependent parameters
$\alpha\left(  t\right)  $ and $\beta\left(  t\right)  $,%
\begin{equation}
\mathbf{a}\overset{\text{def}}{\mathbf{=}}\left(
\begin{array}
[c]{c}%
a_{x}\\
a_{y}\\
a_{z}%
\end{array}
\right)  =\left(
\begin{array}
[c]{c}%
\sin(2\alpha)\cos\left(  \beta\right) \\
\sin(2\alpha)\sin\left(  \beta\right) \\
\cos(2\alpha)
\end{array}
\right)  \text{.} \label{do6}%
\end{equation}
From Eq. (\ref{do6}), we observe that $\mathbf{a\cdot a=}a_{x}^{2}+a_{y}%
^{2}+a_{z}^{2}=1$. Finally, to derive the expression of $\mathrm{H}\left(
t\right)  =i\left\vert \partial_{t}m(t)\right\rangle \left\langle
m(t)\right\vert -i\left\vert m(t)\right\rangle \left\langle \partial
_{t}m(t)\right\vert $ reformulated as $h_{0}\left(  t\right)  \mathbf{1}%
+\mathbf{h}\left(  t\right)  \cdot\mathbf{\boldsymbol{\sigma}}$, it is
necessary to determine the explicit formulas for both $h_{0}\left(  t\right)
$ and the magnetic field vector $\mathbf{h}\left(  t\right)  $. Through a
series of algebraic manipulations, we reach the conclusion that
\begin{align}
\mathrm{H}\left(  t\right)   &  =\left(
\begin{array}
[c]{cc}%
h_{0}+h_{z} & h_{x}-ih_{y}\\
h_{x}+ih_{y} & h_{0}-h_{z}%
\end{array}
\right) \nonumber\\
&  =\left(
\begin{array}
[c]{cc}%
2\dot{\phi}\cos^{2}\left(  \alpha\right)  & e^{-i\beta}\left[  2\dot{\phi}%
\sin\left(  \alpha\right)  \cos\left(  \alpha\right)  -i\dot{\alpha}%
-\dot{\beta}\sin\left(  \alpha\right)  \cos\left(  \alpha\right)  \right] \\
e^{i\beta}\left[  2\dot{\phi}\sin\left(  \alpha\right)  \cos\left(
\alpha\right)  +i\dot{\alpha}-\dot{\beta}\sin\left(  \alpha\right)
\cos(\alpha)\right]  & 2\dot{\phi}\sin^{2}(\alpha)-2\dot{\beta}\sin^{2}\left(
\alpha\right)
\end{array}
\right)  \text{.} \label{dada2}%
\end{align}
From Eq. (\ref{dada2}), we conclude that $h_{0}\left(  t\right)  =0$ (as
anticipated, given that the Hamiltonian is traceless). Furthermore, by
utilizing Eq. (\ref{do4}), we observe from Eq. (\ref{dada2}) that
$\mathbf{h}\left(  t\right)  $ is equivalent to%
\begin{equation}
\mathbf{h}\left(  t\right)  \overset{\text{def}}{=}\left(
\begin{array}
[c]{c}%
h_{x}(t)\\
h_{y}(t)\\
h_{z}(t)
\end{array}
\right)  =\left(
\begin{array}
[c]{c}%
-\frac{\dot{\beta}}{2}\cos(2\alpha)\sin(2\alpha)\cos\left(  \beta\right)
-\dot{\alpha}\sin\left(  \beta\right) \\
-\frac{\dot{\beta}}{2}\cos(2\alpha)\sin(2\alpha)\sin\left(  \beta\right)
+\dot{\alpha}\cos\left(  \beta\right) \\
\frac{\dot{\beta}}{2}\sin^{2}(2\alpha)
\end{array}
\right)  \text{.} \label{magno}%
\end{equation}
By utilizing Eqs. (\ref{do6}) and (\ref{magno}), one can confirm through
straightforward yet tedious algebra that the Bloch vector and the magnetic
vector satisfy the differential equation $\mathbf{\dot{a}=}2\mathbf{h}%
\times\mathbf{a}$ \cite{feynman57}. Notably, it is observed that the Bloch
vector $\mathbf{a}\left(  t\right)  $ in Eq. (\ref{do6}) and the magnetic
field $\mathbf{h}\left(  t\right)  $ in Eq. (\ref{magno}) are orthogonal, as
$\mathbf{a}\cdot\mathbf{h=}0$ at all times. It is important to note that
$\mathbf{\dot{a}=}2\mathbf{h}\times\mathbf{a}$ indicated that $\mathbf{\dot
{a}\cdot h=}0$. Consequently, since $\mathbf{a}\cdot\mathbf{h=}0$ leads to
$\mathbf{\dot{a}}\cdot\mathbf{h+\mathbf{a}\cdot\mathbf{\dot{h}}=}0$, we also
find that $\mathbf{\mathbf{a}\cdot\mathbf{\dot{h}}=}0$. In conclusion, we
summarize the following relationships: (i) $\mathbf{\dot{a}=}2\mathbf{h}%
\times\mathbf{a}$, which indicates $\mathbf{\dot{a}\cdot h=}0$; (ii)
$\mathbf{a}\cdot\mathbf{h=}0$, which leads to $\mathbf{\dot{a}}\cdot
\mathbf{h+\mathbf{a}\cdot\mathbf{\dot{h}}=}0$; (iii) $\mathbf{\mathbf{a}%
\cdot\mathbf{\dot{h}}=}0$, as a result of (i) and (ii). By utilizing the
geometric constraints that exist between the Bloch and magnetic vectors, the
curvature coefficient $\kappa_{\mathrm{AC}}^{2}\left(  \mathbf{a}\text{,
}\mathbf{h}\right)  $ as presented in Eq. (\ref{XXX}) simplifies to%
\begin{equation}
\kappa_{\mathrm{AC}}^{2}\left(  \mathbf{h}\right)  =\frac{\left(
\mathbf{h}^{2}\right)  (\mathbf{\dot{h}}^{2})-\left(  \mathbf{h\cdot\dot{h}%
}\right)  ^{2}}{\mathbf{h}^{6}}\text{.} \label{do20}%
\end{equation}

In the upcoming two subsections, we will focus on the time dependence of the
previously discussed phase $\beta\left(  t\right)  $ to examine two distinct
scenarios: i) no growth; ii) linear growth. These scenarios are characterized
by specific magnetic field configurations $\mathbf{h}\left(  t\right)  $,
which are affected by the generally time-varying real-valued parameters
$\alpha\left(  t\right)  $ and $\beta\left(  t\right)  $. In particular,
$\alpha\left(  t\right)  $ governs the time-dependent behavior of the quantum
amplitudes of the evolving state $\left\vert \psi\left(  t\right)
\right\rangle $ in relation to the computational basis vectors $\left\vert
0\right\rangle $ and $\left\vert 1\right\rangle $, whereas $\beta\left(
t\right)  $ represents the quantum-mechanically observable relative phase
factor that is incorporated into the expression of $\left\vert \psi\left(
t\right)  \right\rangle $. Our primary focus is on quantum evolutions that
transition between orthogonal initial and final states $\left\vert
A\right\rangle \overset{\text{def}}{=}\left\vert \psi\left(  0\right)
\right\rangle =\left\vert 0\right\rangle $ and $\left\vert B\right\rangle
\overset{\text{def}}{=}\left\vert \psi\left(  t_{f}\right)  \right\rangle
\simeq\left\vert 1\right\rangle $. Additionally, to emphasize the alterations
in motion induced by various time-configurations of the relative phase factor
defined by $\beta\left(  t\right)  $, we will establish $\alpha\left(
t\right)  \overset{\text{def}}{=}\omega_{0}t$ with $\omega_{0}\in%
\mathbb{R}
_{+}\backslash\left\{  0\right\}  $, throughout our comparative analysis.
Based on the expression of $\left\vert \psi\left(  t\right)  \right\rangle $,
this assumption regarding $\alpha\left(  t\right)  $ leads us to consider
$t_{f}$ as equal to $\pi/(2\omega_{0})$.\begin{figure}[t]
\centering
\includegraphics[width=0.4\textwidth] {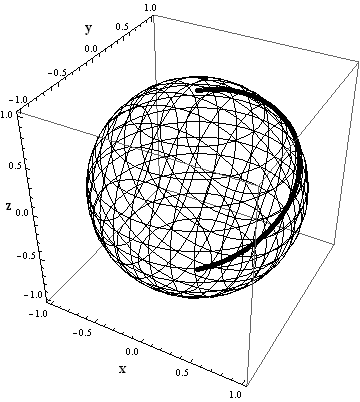}\caption{An illustrative depiction
of the geodesic evolution path (thick solid line) on the Bloch sphere,
generated by the Hamiltonian evolution associated with $\left(  \alpha\left(
t\right)  \text{, }\beta\left(  t\right)  \right)  \overset{\text{def}}%
{=}\left(  \omega_{0}t\text{, }\beta_{0}\right)  $, where $0\leq t\leq\pi/2$.
The evolution shifts from $\left\vert A\right\rangle \overset{\text{def}}%
{=}\left\vert 0\right\rangle $ to $\left\vert B\right\rangle \overset
{\text{def}}{=}\left\vert 1\right\rangle $. For simplicity, we set $\nu
_{0}=\omega_{0}=1$ and $\beta=\pi/4$. Physical units are chosen with
$\hslash=1$.}%
\end{figure}

\subsection{Geodesic motion}

In the first example, we consider that $\mathbf{h}\left(  t\right)  $ is
defined by a phase $\beta\left(  t\right)  $ which remains constant over time.
Specifically, we establish $\beta\left(  t\right)  \overset{\text{def}}%
{=}\beta_{0}\in%
\mathbb{R}
_{+}\backslash\left\{  0\right\}  $ such that $\dot{\beta}=0$. The evolution
of interest transpires from $\left\vert A\right\rangle =\left\vert
0\right\rangle $ to $\left\vert B\right\rangle \simeq\left\vert 1\right\rangle
$ within a time interval $t_{f}=\pi/(2\omega_{0})$, keeping in mind our
selection of $\alpha\left(  t\right)  $ as $\omega_{0}t$. The symbol
\textquotedblleft$\simeq$\textquotedblright\ signifies the physical
equivalence of quantum states, disregarding insignificant global phase factors.

$\emph{Dynamics}$. From the perspective of geodesic efficiency, it is observed
that the geodesic distance $s_{0}$ from $\left\vert A\right\rangle $ to
$\left\vert B\right\rangle $ is given by $s_{0}=\pi$. Additionally, the energy
uncertainty $\Delta E\left(  t\right)  =\sqrt{\left\langle \dot{m}\left\vert
\dot{m}\right.  \right\rangle }$ remains constant and is equal to $\omega_{0}%
$, as $\left\langle \dot{m}\left\vert \dot{m}\right.  \right\rangle
=\dot{\alpha}^{2}+(1/4)\dot{\beta}^{2}\sin^{2}(2\alpha)$. Consequently, this
evolution occurs with a unit geodesic efficiency, $\eta_{\mathrm{GE}}=1$ since
$s=s_{0}=\pi$. The curvature coefficient $\kappa_{\mathrm{AC}}^{2}$ in Eq.
(\ref{do20}) pertaining to this quantum evolution is equal to zero. This
aligns with the observation that the geodesic efficiency is one. From the
perspective of a magnetic field, we have $\mathbf{h}^{2}=\dot{\alpha}^{2}$,
$\mathbf{\dot{h}}^{2}=\ddot{\alpha}^{2}$, and $\left(  \mathbf{h\cdot\dot{h}%
}\right)  ^{2}=\dot{\alpha}^{2}\ddot{\alpha}^{2}$. Consequently, the
nullification of the curvature coefficient arises from the fact that
$\mathbf{h}\left(  t\right)  $ and $\mathbf{\dot{h}}\left(  t\right)  $ are
collinear (i.e., $\partial_{t}\hat{h}\left(  t\right)  =\mathbf{0}$, with
$\mathbf{h}\left(  t\right)  =h(t)\hat{h}(t)$). In other terms, the magnetic
field varies solely in intensity, while its direction remains unchanged. For
the sake of thoroughness, we note that when $\beta\left(  t\right)
\overset{\text{def}}{=}\beta_{0}$, the temporal variation of $\alpha\left(
t\right)  $ does not have to be linear in $t$ to achieve a geodesic evolution
on the Bloch sphere. In fact, let us consider $\beta\left(  t\right)
\overset{\text{def}}{=}\beta_{0}$ and $\alpha\left(  t\right)  \overset
{\text{def}}{=}\omega_{0}^{2}t^{2}$. A simple calculation reveals that
$\mathbf{h}\left(  t\right)  =\dot{\alpha}\left(  t\right)  \left(
-\sin\left(  \beta_{0}\right)  \text{, }\cos(\beta_{0})\text{, }0\right)  $
and $\mathbf{\dot{h}}\left(  t\right)  =\ddot{\alpha}\left(  t\right)  \left(
-\sin\left(  \beta_{0}\right)  \text{, }\cos(\beta_{0})\text{, }0\right)  $ so
that $\mathbf{h}^{2}\left(  t\right)  =4\omega_{0}^{4}t^{2}$ with
$\mathbf{h}\left(  t\right)  $ and $\mathbf{\dot{h}}\left(  t\right)  $ being
collinear. Furthermore, after performing some algebraic manipulations, it can
be confirmed that $\eta_{\mathrm{GE}}=2\arccos(0)/(2\int_{0}^{\frac{1}%
{\omega_{0}}\sqrt{\frac{\pi}{2}}}2\omega_{0}^{2}tdt)=1$ and $\kappa
_{\mathrm{AC}}^{2}=0$. In conclusion, when $\beta\left(  t\right)
\overset{\text{def}}{=}\beta_{0}$, both $\alpha\left(  t\right)
\overset{\text{def}}{=}\omega_{0}t$ and $\alpha\left(  t\right)
\overset{\text{def}}{=}\omega_{0}^{2}t^{2}$ produce a great circle on the
Bloch sphere. The distinction lies in the fact that in the first scenario, the
evolution takes place at a constant speed. Conversely, in the second scenario,
the speed of quantum evolution varies with time, as $v\left(  t\right)
\propto\Delta E\left(  t\right)  =\sqrt{\dot{\alpha}^{2}+(1/4)\dot{\beta}%
^{2}\sin^{2}(2\alpha)}$. We observe that the quantum evolution defined by the
Hamiltonian \textrm{H }associated with $(\alpha\left(  t\right)  $,
$\beta\left(  t\right)  )=(\omega_{0}^{2}t^{2}$, $\beta_{0})$ can be
interpreted as an electron within a magnetic field configuration, similar to
that depicted in (b) of Fig. $2$. In Fig. $7$, we illustrate this geodesic
quantum evolution.

\emph{Average Krylov complexity}. Given that $\left\vert \psi\left(  0\right)
\right\rangle =\left\vert A\right\rangle =\left\vert 0\right\rangle $ and
\begin{equation}
\left\vert \psi\left(  t\right)  \right\rangle _{\mathrm{geo}}=\cos\left(
\omega_{0}t\right)  \left\vert 0\right\rangle +e^{i\beta_{0}}\sin\left(
\omega_{0}t\right)  \left\vert 1\right\rangle \text{,} \label{maduro1}%
\end{equation}
we have that $\mathcal{K}\left(  t\right)  =1-\left\vert \left\langle
\psi\left(  0\right)  \left\vert \psi\left(  t\right)  \right.  \right\rangle
\right\vert ^{2}$ $=\sin^{2}\left(  \omega_{0}t\right)  $. Therefore, the
average Krylov state complexity in this geodesic scenario reduces to
$\left\langle \mathcal{K}\right\rangle _{\mathrm{geo}}$%
\begin{equation}
\left\langle \mathcal{K}\right\rangle _{\mathrm{geo}}=\frac{2\omega_{0}}{\pi
}\int_{0}^{\frac{\pi}{2\omega_{0}}}\sin^{2}(\omega_{0}t)dt=\frac{1}{2}\text{.}
\label{take1}%
\end{equation}

\emph{Quantum IGC}. Ultimately, from the perspective of quantum IGC, we find
that $\theta\left(  t\right)  =2\omega_{0}t$ and $\varphi\left(  t\right)
=\beta_{0}$, with $0\leq\theta\leq\pi$ and $0\leq t\leq\pi/(2\omega_{0})$.
Consequently, a simple calculation provides the formulas for instantaneous,
accessed, and accessible volumes $V(t)=\omega_{0}t$, $\overline{\mathrm{V}%
}=\pi/4$, and $\mathrm{V}_{\max}=\pi/2$, respectively. Thus, the quantum IGC
of this quantum evolution simplifies to
\begin{equation}
\mathrm{C}_{\mathrm{geo}}=\frac{\mathrm{V}_{\max}-\overline{\mathrm{V}}%
}{\mathrm{V}_{\max}}=\frac{\frac{\pi}{2}-\frac{\pi}{4}}{\frac{\pi}{2}}%
=\frac{1}{2} \label{take2}%
\end{equation}
Interestingly, $\left\langle \mathcal{K}\right\rangle _{\mathrm{geo}}$ in Eq.
(\ref{take1}) and $\mathrm{C}_{\mathrm{geo}}$ in Eq. (\ref{take2}) yield
identical numerical values in this particular case.\begin{figure}[t]
\centering
\includegraphics[width=0.4\textwidth] {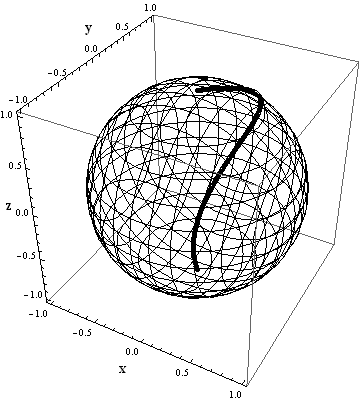}\caption{An illustrative
representation of the nongeodesic evolution trajectory (thick solid line) on
the Bloch sphere, produced by the Hamiltonian evolution linked to $\left(
\alpha\left(  t\right)  \text{, }\beta\left(  t\right)  \right)
\overset{\text{def}}{=}\left(  \omega_{0}t\text{, }\beta_{0}+\nu_{0}t\right)
$, where $0\leq t\leq\pi/2$. The evolution transitions from $\left\vert
A\right\rangle \overset{\text{def}}{=}\left\vert 0\right\rangle $ to
$\left\vert B\right\rangle \overset{\text{def}}{=}\left\vert 1\right\rangle $.
For the sake of simplicity, we establish $\nu_{0}=\omega_{0}=1$ and $\beta
=\pi/4$. Physical units are selected with $\hslash=1$.}%
\end{figure}

\subsection{Nongeodesic motion}

In the second example, we assume that $\mathbf{h}\left(  t\right)  $ is
defined by a phase $\beta\left(  t\right)  $ that increases linearly over
time. Specifically, we express $\beta\left(  t\right)  \overset{\text{def}}%
{=}\beta_{0}+\nu_{0}t$, where $\beta_{0}$, $\nu_{0}\in%
\mathbb{R}
_{+}\backslash\left\{  0\right\}  $, ensuring that $\dot{\beta}=\nu_{0}$. The
evolution of focus occurs from $\left\vert A\right\rangle =\left\vert
0\right\rangle $ to $\left\vert B\right\rangle \simeq\left\vert 1\right\rangle
$ within a time frame $t_{f}=\pi/(2\omega_{0})$, once we recall our selection
indicated by $\alpha\left(  t\right)  =\omega_{0}t$. To ensure thoroughness,
we note that the starting and ending states in both geodesic and non-geodesic
evolutions are identical. This is clearly necessary for a just comparison
among the various types of evolutions.

$\emph{Dynamics}$. From the perspective of geodesic efficiency, we note that
the geodesic distance $s_{0}$ from $\left\vert A\right\rangle $ to $\left\vert
B\right\rangle $ is equal to $s_{0}=\pi$. Furthermore, the energy uncertainty
$\Delta E\left(  t\right)  =\sqrt{\left\langle \dot{m}\left\vert \dot
{m}\right.  \right\rangle }$ is variable, as $\left\langle \dot{m}\left\vert
\dot{m}\right.  \right\rangle =\dot{\alpha}^{2}+(1/4)\dot{\beta}^{2}\sin
^{2}(2\alpha)$ leads to $\Delta E^{2}\left(  t\right)  =\omega_{0}%
^{2}+(1/4)\nu_{0}^{2}\sin^{2}(2\omega_{0}t)$. Consequently, this evolution
takes place with a geodesic efficiency that is not equal to one,
$\eta_{\mathrm{GE}}<1$, since $s>s_{0}=\pi$. Generally, the quantity
$\eta_{\mathrm{GE}}$ depends on the parameters $\omega_{0}$ and $\nu_{0}$ and
can be estimated numerically. For example, when $\omega_{0}=\nu_{0}=1$,
$s\simeq3.33\geq\pi=s_{0}$. We also note that the curvature coefficient
$\kappa_{\mathrm{AC}}^{2}$ in Eq. (\ref{do20}) of this quantum evolution is
different from zero. This agrees with the fact that $\eta_{\mathrm{GE}}<1$.
After some algebra, substituting Eq. (\ref{magno}) into Eq. (\ref{do20}) with
$\alpha\left(  t\right)  \overset{\text{def}}{=}\omega_{0}t$ and $\beta\left(
t\right)  \overset{\text{def}}{=}\beta_{0}+\nu_{0}t$, $\kappa_{\mathrm{AC}%
}^{2}$ becomes%
\begin{equation}
\kappa_{\mathrm{AC}}^{2}\left(  t\text{; }\omega_{0}\text{, }\nu_{0}\right)
=\frac{\mathbf{h}^{2}\left(  t\text{; }\omega_{0}\text{, }\nu_{0}\right)
\mathbf{\dot{h}}^{2}\left(  t\text{; }\omega_{0}\text{, }\nu_{0}\right)
-\left[  \mathbf{h}\left(  t\text{; }\omega_{0}\text{, }\nu_{0}\right)
\mathbf{\cdot\dot{h}}\left(  t\text{; }\omega_{0}\text{, }\nu_{0}\right)
\right]  ^{2}}{\mathbf{h}^{6}\left(  t\text{; }\omega_{0}\text{, }\nu
_{0}\right)  }\text{,} \label{wish1}%
\end{equation}
with $\mathbf{h}^{2}$, $\mathbf{\dot{h}}^{2}$, and $\left(  \mathbf{h\cdot
\dot{h}}\right)  ^{2}$ in Eq. (\ref{wish1}) being given by%
\begin{align}
&  \mathbf{h}^{2}\overset{\text{def}}{=}\frac{1}{8}\nu_{0}^{2}+\omega_{0}%
^{2}-\frac{1}{8}\nu_{0}^{2}\cos\left(  4\omega_{0}t\right)  \text{,
}\nonumber\\
&  \mathbf{\dot{h}}^{2}\overset{\text{def}}{=}\frac{1}{32}\nu_{0}^{4}-\frac
{1}{32}\nu_{0}^{4}\cos\left(  8\omega_{0}t\right)  +2\nu_{0}^{2}\omega_{0}%
^{2}+2\nu_{0}^{2}\omega_{0}^{2}\cos\left(  4\omega_{0}t\right)  \text{,}%
\nonumber\\
&  \text{ }\mathbf{h\cdot\dot{h}}\overset{\text{def}}{=}\frac{1}{4}\nu_{0}%
^{2}\omega_{0}\sin\left(  4\omega_{0}t\right)  \text{,}%
\end{align}
respectively. As a side remark, we note that the short-time limit of
$\kappa_{\mathrm{AC}}^{2}\left(  t\text{; }\omega_{0}\text{, }\nu_{0}\right)
$ is%
\begin{equation}
\kappa_{\mathrm{AC}}^{2}\left(  t\text{; }\omega_{0}\text{, }\nu_{0}\right)
\overset{t\rightarrow0}{\simeq}4\left(  \frac{\nu_{0}}{\omega_{0}}\right)
^{2}-8\left(  \frac{\nu_{0}}{\omega_{0}}\right)  ^{2}\left(  \nu_{0}%
^{2}+2\omega_{0}^{2}\right)  t^{2}+\mathcal{O}\left(  t^{3}\right)  \text{,}%
\end{equation}
where $\kappa_{\mathrm{AC}}^{2}\left(  t\text{; }\omega_{0}\text{, }\nu
_{0}\right)  $ starts at the nonzero value of $4\left(  \nu_{0}/\omega
_{0}\right)  ^{2}$ at $t=0$. Finally, the existence of a non-vanishing
curvature coefficient arises from the fact that $\mathbf{h}\left(  t\right)  $
and $\mathbf{\dot{h}}\left(  t\right)  $ are not collinear (i.e.,
$\partial_{t}\hat{h}\left(  t\right)  \neq\mathbf{0}$, with $\mathbf{h}\left(
t\right)  =h(t)\hat{h}(t)$). Lastly, we remark that the quantum evolution
specified by the Hamiltonian \textrm{H }specified by the pair $(\alpha\left(
t\right)  $, $\beta\left(  t\right)  )=(\omega_{0}t$, $\beta_{0}+\nu_{0}t)$
can be regarded as an electron immersed in a magnetic field configuration,
similar to that illustrated in (d) of Fig. $2$. In Fig. $8$, we display the
nongeodesic quantum-mechanical evolution path on the Bloch sphere.

\emph{Average Krylov complexity}. Given that $\left\vert \psi\left(  0\right)
\right\rangle =\left\vert A\right\rangle =\left\vert 0\right\rangle $ and%
\begin{equation}
\left\vert \psi\left(  t\right)  \right\rangle _{\mathrm{nongeo}}=\cos\left(
\omega_{0}t\right)  \left\vert 0\right\rangle +e^{i\left(  \beta_{0}+\nu
_{0}t\right)  }\sin\left(  \omega_{0}t\right)  \left\vert 1\right\rangle
\text{,} \label{maduro2}%
\end{equation}
we have that $\mathcal{K}\left(  t\right)  =1-\left\vert \left\langle
\psi\left(  0\right)  \left\vert \psi\left(  t\right)  \right.  \right\rangle
\right\vert ^{2}$ $=\sin^{2}\left(  \omega_{0}t\right)  $. Therefore, the
average Krylov state complexity in this nongeodesic scenario reduces to
$\left\langle \mathcal{K}\right\rangle _{\mathrm{nongeo}}$%
\begin{equation}
\left\langle \mathcal{K}\right\rangle _{\mathrm{nongeo}}=\frac{2\omega_{0}%
}{\pi}\int_{0}^{\frac{\pi}{2\omega_{0}}}\sin^{2}(\omega_{0}t)dt=\frac{1}%
{2}\text{.} \label{take3}%
\end{equation}
Interestingly, despite the fact that $\left\vert \psi\left(  t\right)
\right\rangle _{\mathrm{geo}}\neq\left\vert \psi\left(  t\right)
\right\rangle _{\mathrm{nongeo}}$, we have that $\left\langle \mathcal{K}%
\right\rangle _{\mathrm{geo}}$ in Eq. (\ref{take1}) is equal to $\left\langle
\mathcal{K}\right\rangle _{\mathrm{nongeo}}$ in Eq. (\ref{take3}).

\emph{Quantum IGC}. Ultimately, from the perspective of quantum IGC, we find
that $\theta\left(  t\right)  =2\omega_{0}t$ and $\varphi\left(  t\right)
=\beta_{0}+\nu_{0}t$, with $0\leq\theta\leq\pi$, $\beta_{0}\leq\varphi
\leq\beta_{0}+\left(  \pi/2\right)  \left(  \nu_{0}/\omega_{0}\right)  $, and
$0\leq t\leq\pi/(2\omega_{0})$. Consequently, a simple calculation provides
the formulas for instantaneous, accessed, and accessible volumes%
\begin{equation}
V(t)=\frac{\nu_{0}}{4}\left[  1-\cos\left(  2\omega_{0}t\right)  \right]
t\text{, }\overline{\mathrm{V}}=\left(  \frac{1}{4\pi}+\frac{\pi}{16}\right)
\frac{\nu_{0}}{\omega_{0}}\text{, and }\mathrm{V}_{\max}=\frac{\pi}{4}%
\frac{\nu_{0}}{\omega_{0}}\text{,} \label{moment}%
\end{equation}
respectively. Thus, applying Eq. (\ref{moment}), the quantum IGC of this
quantum evolution simplifies to
\begin{equation}
\mathrm{C}_{\mathrm{nongeo}}=\frac{\mathrm{V}_{\max}-\overline{\mathrm{V}}%
}{\mathrm{V}_{\max}}=\frac{3\pi^{2}-4}{4\pi^{2}}\sim0.65\text{.} \label{com2}%
\end{equation}
Interestingly, and in contrast to Krylov's state complexity, we observe that
$\mathrm{C}_{\mathrm{nongeo}}$ in Eq. (\ref{com2}) is approximately $0.65$,
which exceeds $\mathrm{C}_{\mathrm{geo}}=1/2$. It is important to notice that
in the scenario of linear growth, the quantum IGC in Eq. (\ref{com2}) attains
a constant value that remains unaffected by the parameters $\omega_{0}$ and
$\nu_{0}$, stemming from the fact that both $\overline{\mathrm{V}}$ and
$\mathrm{V}_{\max}$ are proportional to the ratio $\left(  \nu_{0}/\omega
_{0}\right)  $.

From our investigation, we note that the phase $\beta\left(  t\right)  $
alters the trajectory length from $\left\vert \psi\left(  0\right)
\right\rangle \overset{\text{def}}{=}\left\vert 0\right\rangle $ to
$\left\vert \psi\left(  t_{f}\right)  \right\rangle \overset{\text{def}}%
{=}\left\vert 1\right\rangle $. Phase motion adds curvature to the path,
increasing its length. However, as we have seen, a change in the path length
does not imply a spread in the Krylov basis. Krylov's state complexity is
blind to phase motion. It depends only on how far the state $\left\vert
\psi\left(  t\right)  \right\rangle $ with $0\leq t\leq t_{f}$ has traveled
from the reference state $\left\vert 0\right\rangle $ in a predetermined
direction, rather than how it navigates around the Bloch sphere. This
predetermined direction is represented by the orthonormal Krylov chain
$\left\{  \left\vert \phi_{n}\right\rangle \right\}  _{n\geq0}$, which is
formed through the repeated application of the Hamiltonian on the reference
state and constructed using Lanczos recursion. The unitary time evolution from
the reference state is restricted to movement along this chain, and Krylov's
state complexity corresponds to the anticipated position on that chain. In
contrast to $\beta\left(  t\right)  $, which modifies the longitude by
rotating the state around the $z$-axis, the parameter $\alpha\left(  t\right)
$ adjusts the latitude of the state, indicating the extent of transition from
$\left\vert 0\right\rangle $ to $\left\vert 1\right\rangle $. Changes in
$\beta\left(  t\right)  $ are identified exclusively by the quantum IGC,
whereas variations in $\alpha\left(  t\right)  $ are acknowledged by both the
quantum IGC and Krylov's state complexity.

We are now prepared to present our summary of results and final observations.

\section{Final Remarks}

In this paper, we conducted a comparative analysis of Krylov's state
complexity $\mathcal{K}\left(  t\right)  $ (Eq. (\ref{951})) in conjunction
with a quantum information geometric (IG) measure of complexity \textrm{C}%
$\left(  t_{A}\text{, }t_{B}\right)  $ with $t_{A}\leq t\leq t_{B}$
(Eq.(\ref{QCD})) that pertains to the quantum evolutions of two-level quantum
systems. This analysis is based on geometric principles such as length,
spread, and volumes, particularly concerning qubit dynamics on the Bloch
sphere. In this context, the quantum mechanical evolutions are characterized
by either stationary or nonstationary Hamiltonians, which can yield either
geodesic (Eqs. (\ref{voice1}) and (\ref{maduro1})) or nongeodesic (Eqs.
(\ref{way6}), (\ref{voice1a}), and (\ref{maduro2})) dynamical trajectories.
After presenting a geometric formulation of Krylov's state complexity in both
its instantaneous and time-averaged forms, we applied the concepts of geodesic
efficiency (i.e., $\eta_{\mathrm{geo}}$ in Eq. (\ref{efficiency})) and
curvature coefficient (i.e., $\kappa_{\mathrm{AC}}^{2}$ in Eq. (\ref{XXX})) of
quantum evolutions to extract physical insights into the various aspects
represented by the two distinct measures of complexity. Ultimately, we
illustrated that this differentiation is fundamentally due to the
non-overlapping quantities utilized in the formulation of these measures,
specifically the spread of the evolving quantum state in relation to a defined
direction determined by the initial state, as well as the volume covered
during the evolution on the Bloch sphere.

\subsection{Summary of results}

Our primary findings can be summarized as follows:

\begin{enumerate}
\item[{[i]}] We demonstrated that Krylov's state complexity can be expressed,
in both stationary (Eqs. (\ref{now1}), (\ref{io1}), and (\ref{io2})) and
nonstationary (Eqs. (\ref{anna}) and (\ref{anna1})) contexts, utilizing
vectors that have a unique geometric significance. In particular, we used
Bloch vectors (i.e., $\mathbf{a}_{0}$ and $\mathbf{a}_{t}$) and magnetic field
vectors (i.e., $\mathbf{h}\left(  t\right)  $), which define the Hamiltonians
that govern the quantum mechanical evolutions.

\item[{[ii]}] We illustrated that, in contrast to any stationary situation
where $\mathbf{n}\left(  t\right)  \cdot\mathbf{a}_{0}=0$ for every moment
$t$, with \textrm{H}$\left(  t\right)  \overset{\text{def}}{=}\mathbf{h}%
\left(  t\right)  \mathbf{\cdot\boldsymbol{\sigma}}$, $\mathbf{h}\left(
t\right)  \overset{\text{def}}{=}h\left(  t\right)  \mathbf{n}\left(
t\right)  $, and $\rho\left(  0\right)  \overset{\text{def}}{=}\left\vert
\psi\left(  0\right)  \right\rangle \left\langle \psi\left(  0\right)
\right\vert =\left[  \mathbf{1+a}_{0}\cdot\mathbf{\boldsymbol{\sigma}}\right]
/2$, and where the dynamic trajectory follows a geodesic path on the Bloch
sphere defined by a constant great circle), in the nonstationary scenario we
examined (\textrm{H}$\left(  t\right)  $ in Eq. (\ref{way4})), $\left\vert
\psi\left(  t\right)  \right\rangle $ does not become completely orthogonal to
$\left\vert \psi\left(  0\right)  \right\rangle $. Consequently, there is no
moment $t$ at which $\mathcal{K}\left(  t\right)  $ reaches its maximum value
of one (Eq. (\ref{way7})).

\item[{[iii]}] We showed that geodesic trajectories on the Bloch sphere,
produced by stationary Hamiltonians, are typically defined by reduced levels
of \emph{time-averaged} Krylov state complexity (Eqs. (\ref{vox3}) and
(\ref{vox3b})). Nevertheless, these geodesic evolutions not only proceed at a
faster speed (i.e., as shorter travel time), but they also manifest with
increased levels of \emph{instantaneous} Krylov state complexity.

\item[{[iv]}] We established that, similar to our quantum IG complexity
measure, Krylov's state complexity exceeds mere length. In particular, by
employing nonstationary Hamiltonian evolutions, we illustrated that different
trajectories on the Bloch sphere that connect the same initial and final
states can possess varying lengths while maintaining identical values of
Krylov's state complexity (Eqs. (\ref{take1}) and (\ref{take3})).
Nevertheless, while $\mathcal{K}\left(  t\right)  $ is dependent on the idea
of \textquotedblleft spread\textquotedblright, \textrm{C}$\left(  t_{A}\text{,
}t_{B}\right)  $ (Eqs. (\ref{take2}) and (\ref{com2})) is founded on the
concept of \ \textquotedblleft volume\textquotedblright. Both spread and
volume are distinct from the concept of length. Furthermore, we also
confirmed, in stationary Hamiltonian contexts and for suitable temporal
regimes, the relationships among $\mathcal{K}\left(  t\right)  $, $\left\Vert
\mathbf{a}_{t}-\mathbf{a}_{0}\right\Vert ^{2}$, and $V_{\mathrm{FS}}\left(
t\right)  $ (Appendix D).

\item[{[v]}] We verified that, in contrast to our quantum IG measure of
complexity (\ref{take2}) and (\ref{com2})), the existence of relative phases
(i.e., $e^{i\beta\left(  t\right)  }$ in Eq. (\ref{do1})) within quantum
states $\left\vert \psi\left(  t\right)  \right\rangle $ does not affect the
behavior of Krylov's state complexity $\mathcal{K}\left(  t\right)  $ (Eqs.
(\ref{take1}) and (\ref{take3})). The phase $\beta\left(  t\right)  $ modifies
the trajectory length from $\left\vert \psi\left(  0\right)  \right\rangle
\overset{\text{def}}{=}\left\vert 0\right\rangle $ to $\left\vert \psi\left(
t_{f}\right)  \right\rangle \overset{\text{def}}{=}\left\vert 1\right\rangle
$. The motion of the phase introduces curvature to the path, thereby
increasing its length. Nevertheless, an alteration in the path length does not
signify a dispersion in the Krylov basis. Krylov's state complexity remains
unaffected by phase motion. It is solely dependent on the distance the state
$\left\vert \psi\left(  t\right)  \right\rangle $ with $0\leq t\leq t_{f}$ has
moved from the reference state $\left\vert 0\right\rangle $ in a specified
direction, rather than the manner in which it traverses the Bloch sphere.
\end{enumerate}

\subsection{Limitations and outlook}

Our research has two significant limitations. First, we have limited our
analysis to quantum evolutions governed by Hamiltonians within the context of
two-level quantum systems. Second, we have not explored the implications of
our findings for understanding complexity in quantum-mechanical systems that
are in mixed quantum states. Although it is theoretically possible to extend
our study to arbitrary time-varying Hamiltonian evolutions, the computational
challenge lies in obtaining accurate analytical solutions to the
time-dependent Schr\"{o}dinger equation for determining probability
amplitudes. This task poses substantial difficulties, even in the domain of
two-level quantum systems
\cite{landau32,zener32,rabi37,rabi54,barnes12,barnes13,messina14,grimaudo18,cafaroijqi,castanos,grimaudo23,elena20}%
. Furthermore, the transition from pure to mixed states presents various
challenges, both computationally and conceptually. It is widely acknowledged
that there exists an infinite variety of distinguishability metrics applicable
to mixed quantum states. This variety results in interpretations of essential
geometric quantities, such as the complexity and volume of quantum states,
which depend on the chosen metric \cite{carluccio1,carluccio2}. In particular,
the non-uniqueness of these distinguishability measures requires a
comprehensive understanding of the physical implications tied to the selection
of a specific metric, a subject of considerable conceptual and practical
importance \cite{silva21,mera22,luongo24,chien24}. Since these complexities
surpass the boundaries of our current study, we intend to tackle certain
aspects of these challenges in our future research efforts. In addition, we
can expand our comparative analysis to incorporate nonstationary Hamiltonian
evolutions defined by significant time-dependent magnetic field configurations
\cite{messina14,grimaudo18,cafaroijqi,castanos,grimaudo23}. Moreover, our
analysis is limited to a two-state system, which currently hinders our ability
to investigate the connection between our complexity measure and quantum
entanglement. However, as a preliminary step towards this objective, we are in
the process of extending our methodology to include $d$-level quantum systems
with $d>2$, within larger finite-dimensional Hilbert spaces
\cite{jakob01,kimura03,krammer08,kurzy11,xie20,siewert21,sharma24,morelli24,sen25}%
.

\medskip

Before presenting our concluding paragraph, we highlight an additional avenue
for future work. Building on the consistency behavior observed for our
proposed IG complexity measure in Ref. \cite{emma25}, an important next step
is to establish a rigorous mathematical proof that this measure is
basis-independent. At present, our support for this property is purely
empirical and limited to a narrow class of cases, namely quantum-mechanical
evolutions governed by stationary Hamiltonians. We also plan to explore
alternative notions of complexity defined through intrinsically
rotation-invariant quantities, such as the areas of surface regions
characterized by solid angles. Preliminary observations in this direction are
outlined in Appendix C.

\medskip

Notwithstanding its constraints, we assert that our research signifies a
considerable endeavor to forge a link between Krylov's state complexity and
our quantum IG measure of complexity. Specifically, it seeks to geometrically
interpret Krylov's state complexity by investigating the relationships among
the notions of \emph{length} (i.e., a metric of the distance a state has
traversed), \emph{spread} (i.e., a metric of the number of independent
directions occupied by the state), and \emph{volume} (i.e., a metric of the
extent of the multi-dimensional region explored). Overall, these findings
underscore that no single scalar quantity can fully capture quantum
complexity; rather, a synthesis of complementary geometric and dynamical
perspectives is essential to describe the full richness of quantum evolution.

\begin{acknowledgments}
C.C. (Principal Investigator) and E.C. (Undergraduate Research Aide)
acknowledge the partial funding received from the University at Albany-SUNY
through the FRAP-B (Faculty Research Award Programs) grant number 102106.
Moreover, V.V.A. (Graduate Research Aide) expresses gratitude to the
University at Albany-SUNY for providing a Research Internship opportunity as
an international student. Finally, the opinions, findings, conclusions, or
recommendations presented in this material are solely those of the author(s)
and do not necessarily represent the perspectives of their respective
Institutions. The authors gratefully acknowledge Dmitrii Trunin (Princeton
University) for insightful comments on the Szeg\"{o} algorithm, and Edward
Medina-Guerra (Weizmann Institute of Science) for valuable remarks on recent
applications of the Krylov state complexity to non-Hermitian systems. They
also thank an anonymous referee for constructive feedback that significantly
improved the manuscript.
\end{acknowledgments}

\pagebreak

\appendix

\section{Relation between Fubini-Study and Wigner-Yanase metrics}

In this appendix, we show that the Fubini--Study metric $g_{\mathrm{FS}}$ is
proportional to the quantum Fisher information metric $g_{\mathrm{QFI}}$ (with
$g_{\mathrm{QFI}}=4g_{\mathrm{FS}}$ ). Indeed, the geometry of the Bloch
sphere, when utilizing the Fubini-Study metric \cite{provost80}, can be viewed
as a quantum information geometric framework for pure quantum states that are
associated with the Wigner-Yanase metric \cite{luo03,luo06}. The Wigner-Yanase
metric serves as a quantum information-geometric metric that is defined within
the realm of quantum states \cite{karol}. In the case of diagonal density
matrices, the Wigner-Yanase metric simplifies to the classical Fisher-Rao
information metric \cite{amari}. Conversely, for pure states, the
Wigner-Yanase metric aligns (up to a multiplicative factor) with the
Fubini-Study metric. In this appendix, we will illustrate this assertion.

\subsection{Fubini-Study metric}

We begin by presenting the Fubini-Study metric tensor components
$g_{ab}^{\mathrm{FS}}\left(  \xi\right)  $. We will consider a collection of
quantum state vectors $\left\{  \left\vert \psi\left(  \xi\right)
\right\rangle \right\}  $ that are parameterized by the parameters
$\xi\overset{\text{def}}{=}\left(  \xi^{1}\text{,..., }\xi^{m}\right)  $.
Here, $m$\textbf{\ }represents the number of real-valued parameters that are
assumed to characterize a quantum state\ $\left\vert \psi\left(  \xi\right)
\right\rangle $\textbf{\ }in\textbf{\ }$%
\mathbb{C}
P^{n-1}$\textbf{. }For the sake of clarity, we will assume that\textbf{\ \ }%
$\mathcal{H}$\textbf{\ }is the\textbf{\ }$n$\textbf{-}dimensional complex
Hilbert space\textbf{\ }$\mathcal{H}_{2}^{N}$\textbf{\ }of\textbf{\ }%
$N$\textbf{-}qubit quantum states with\textbf{\ }$n=2^{N}$\textbf{, }and we
concentrate on the straightforward case where\textbf{\ }$n=2$\textbf{.}
Consequently, irrespective of the selected definition of finite distance, the
infinitesimal line element $ds_{\text{\textrm{FS}}}^{2}$ which measures the
distance between two neighboring states $\left\vert \psi\left(  \xi\right)
\right\rangle $ and $\left\vert \psi\left(  \xi+d\xi\right)  \right\rangle $
can be expressed as%
\begin{equation}
ds_{\text{\textrm{FS}}}^{2}=g_{ab}\left(  \xi\right)  d\xi^{a}d\xi^{b}\text{.}
\label{oggi}%
\end{equation}
The metric tensor components $g_{ab}\left(  \xi\right)  $ in Eq. (\ref{oggi})
are given by \cite{provost80},
\begin{equation}
g_{ab}\left(  \xi\right)  \overset{\text{def}}{=}\left[  \gamma_{ab}\left(
\xi\right)  -\beta_{a}\left(  \xi\right)  \beta_{b}\left(  \xi\right)
\right]  \text{,} \label{metric}%
\end{equation}
where, assuming $\partial_{a}\overset{\text{def}}{=}\partial/\partial\xi^{a}$,
we define%
\begin{equation}
\gamma_{ab}\left(  \xi\right)  \overset{\text{def}}{=}\operatorname{Re}\left[
\left\langle \partial_{a}\psi\left(  \xi\right)  |\partial_{b}\psi\left(
\xi\right)  \right\rangle \right]  \text{, and }\beta_{a}\left(  \xi\right)
\overset{\text{def}}{=}-i\left\langle \psi\left(  \xi\right)  |\partial
_{a}\psi\left(  \xi\right)  \right\rangle \text{.} \label{metrica}%
\end{equation}
From Eq. (\ref{metrica}), we observe that%
\begin{equation}
\beta_{a}\left(  \xi\right)  \beta_{b}\left(  \xi\right)  =\left\langle
\partial_{a}\psi\left(  \xi\right)  |\psi\left(  \xi\right)  \right\rangle
\left\langle \psi\left(  \xi\right)  |\partial_{b}\psi\left(  \xi\right)
\right\rangle \text{,} \label{betaA}%
\end{equation}
given that $\partial_{a}\left[  \left\langle \psi\left(  \xi\right)
|\psi\left(  \xi\right)  \right\rangle \right]  =0$ leads to $\left\langle
\psi\left(  \xi\right)  |\partial_{a}\psi\left(  \xi\right)  \right\rangle
=-\left\langle \partial_{a}\psi\left(  \xi\right)  |\psi\left(  \xi\right)
\right\rangle $. Therefore, making use of Eqs. (\ref{metrica}) and
(\ref{betaA}), $g_{ab}\left(  \xi\right)  $ in Eq. (\ref{metric}) reduces to%
\begin{equation}
g_{ab}\left(  \xi\right)  =\left\{  \operatorname{Re}\left[  \left\langle
\partial_{a}\psi\left(  \xi\right)  |\partial_{b}\psi\left(  \xi\right)
\right\rangle \right]  -\left\langle \partial_{a}\psi\left(  \xi\right)
|\psi\left(  \xi\right)  \right\rangle \left\langle \psi\left(  \xi\right)
|\partial_{b}\psi\left(  \xi\right)  \right\rangle \right\}  \text{.}%
\end{equation}
For simplicity, let us introduce%
\begin{equation}
A_{ab}\left(  \xi\right)  \overset{\text{def}}{=}\left\langle \partial_{a}%
\psi\left(  \xi\right)  |\psi\left(  \xi\right)  \right\rangle \left\langle
\psi\left(  \xi\right)  |\partial_{b}\psi\left(  \xi\right)  \right\rangle
\text{.} \label{oggi1}%
\end{equation}
We shall verify that $A_{ab}\left(  \xi\right)  d\xi^{a}d\xi^{b}%
=\operatorname{Re}\left[  A_{ab}\left(  \xi\right)  \right]  d\xi^{a}d\xi^{b}%
$. Take into consideration,%
\begin{equation}
A_{ab}\left(  \xi\right)  =\operatorname{Re}\left[  A_{ab}\left(  \xi\right)
\right]  +i\operatorname{Im}\left[  A_{ab}\left(  \xi\right)  \right]
=A_{ab}^{\left(  1\right)  }\left(  \xi\right)  +iA_{ab}^{\left(  2\right)
}\left(  \xi\right)  \text{.}%
\end{equation}
Then, from Eq. (\ref{oggi1}), we arrive at%
\begin{align}
A_{ab}^{\left(  1\right)  }\left(  \xi\right)   &  =\operatorname{Re}\left[
A_{ab}\left(  \xi\right)  \right] \nonumber\\
&  =\operatorname{Re}\left[  \left\langle \partial_{a}\psi\left(  \xi\right)
|\psi\left(  \xi\right)  \right\rangle \left\langle \psi\left(  \xi\right)
|\partial_{b}\psi\left(  \xi\right)  \right\rangle \right] \nonumber\\
&  =\operatorname{Re}\left[  \left\langle \partial_{a}\psi\left(  \xi\right)
|\psi\left(  \xi\right)  \right\rangle ^{\ast}\left\langle \psi\left(
\xi\right)  |\partial_{b}\psi\left(  \xi\right)  \right\rangle ^{\ast}\right]
\nonumber\\
&  =\operatorname{Re}\left[  \left\langle \psi\left(  \xi\right)
|\partial_{a}\psi\left(  \xi\right)  \right\rangle \left\langle \partial
_{b}\psi\left(  \xi\right)  |\psi\left(  \xi\right)  \right\rangle \right]
\nonumber\\
&  =\operatorname{Re}\left[  \left\langle \partial_{b}\psi\left(  \xi\right)
|\psi\left(  \xi\right)  \right\rangle \left\langle \psi\left(  \xi\right)
|\partial_{a}\psi\left(  \xi\right)  \right\rangle \right] \nonumber\\
&  =\operatorname{Re}\left[  A_{ba}\left(  \xi\right)  \right] \nonumber\\
&  =A_{ba}^{\left(  1\right)  }\left(  \xi\right)  \text{,} \label{sym}%
\end{align}
and, moreover,
\begin{align}
A_{ab}^{\left(  2\right)  }\left(  \xi\right)   &  =\operatorname{Im}\left[
A_{ab}\left(  \xi\right)  \right] \nonumber\\
&  =\operatorname{Im}\left[  \left\langle \partial_{a}\psi\left(  \xi\right)
|\psi\left(  \xi\right)  \right\rangle \left\langle \psi\left(  \xi\right)
|\partial_{b}\psi\left(  \xi\right)  \right\rangle \right] \nonumber\\
&  =-\operatorname{Im}\left[  \left\langle \partial_{a}\psi\left(  \xi\right)
|\psi\left(  \xi\right)  \right\rangle ^{\ast}\left\langle \psi\left(
\xi\right)  |\partial_{b}\psi\left(  \xi\right)  \right\rangle ^{\ast}\right]
\nonumber\\
&  =-\operatorname{Im}\left[  \left\langle \partial_{b}\psi\left(  \xi\right)
|\psi\left(  \xi\right)  \right\rangle \left\langle \psi\left(  \xi\right)
|\partial_{a}\psi\left(  \xi\right)  \right\rangle \right] \nonumber\\
&  =-\operatorname{Im}\left[  A_{ba}\left(  \xi\right)  \right] \nonumber\\
&  =-A_{ba}^{\left(  2\right)  }\left(  \xi\right)  \text{.} \label{asym}%
\end{align}
Eqs. (\ref{sym}) and (\ref{asym}) imply that $A_{ab}^{\left(  1\right)
}\left(  \xi\right)  $ is symmetric under exchange of indices, whereas
$A_{ab}^{\left(  2\right)  }\left(  \xi\right)  $ is antisymmetric. Therefore,
given the symmetry of $d\xi^{a}d\xi^{b}$, we obtain $A_{ab}\left(  \xi\right)
d\xi^{a}d\xi^{b}=\operatorname{Re}\left[  A_{ab}\left(  \xi\right)  \right]
d\xi^{a}d\xi^{b}$. Summing up, $g_{ab}\left(  \xi\right)  $ can be recast as%
\begin{equation}
g_{ab}\left(  \xi\right)  =\operatorname{Re}\left[  \left\langle \partial
_{a}\psi\left(  \xi\right)  |\partial_{b}\psi\left(  \xi\right)  \right\rangle
\right]  -\operatorname{Re}\left[  \left\langle \partial_{a}\psi\left(
\xi\right)  |\psi\left(  \xi\right)  \right\rangle \left\langle \psi\left(
\xi\right)  |\partial_{b}\psi\left(  \xi\right)  \right\rangle \right]
\text{,}%
\end{equation}
that is,%
\begin{equation}
g_{ab}^{\mathrm{FS}}\left(  \xi\right)  =\operatorname{Re}\left[  \left\langle
\partial_{a}\psi\left(  \xi\right)  |\partial_{b}\psi\left(  \xi\right)
\right\rangle -\left\langle \partial_{a}\psi\left(  \xi\right)  |\psi\left(
\xi\right)  \right\rangle \left\langle \psi\left(  \xi\right)  |\partial
_{b}\psi\left(  \xi\right)  \right\rangle \right]  \text{.} \label{FS1}%
\end{equation}
Eq. (\ref{FS1}) specifies the Fubini-Study metric tensor on the manifold of
pure quantum states.

\subsection{Wigner-Yanase metric}

We now present the Wigner-Yanase metric tensor components $g_{ab}%
^{\mathrm{WY}}\left(  \xi\right)  $. We begin by recalling that the so-called
Wigner-Yanase metric $g_{ab}^{\mathrm{WY}}\left(  \xi\right)  $ is defined as
\cite{luo03,luo06},
\begin{equation}
g_{ab}^{\mathrm{WY}}\left(  \xi\right)  \overset{\text{def}}{=}%
4\text{\textrm{tr}}\left[  \left(  \partial_{a}\sqrt{\rho_{\xi}}\right)
\left(  \partial_{b}\sqrt{\rho_{\xi}}\right)  \right]  =4\text{\textrm{tr}%
}\left[  \left(  \partial_{a}\rho_{\xi}\right)  \left(  \partial_{b}\rho_{\xi
}\right)  \right]  \text{,} \label{WY}%
\end{equation}
since $\rho_{\xi}=\rho_{\xi}^{2}$ with $\rho_{\xi}\overset{\text{def}}%
{=}\left\vert \psi\left(  \xi\right)  \right\rangle \left\langle \psi\left(
\xi\right)  \right\vert =\left\vert \psi_{\xi}\right\rangle \left\langle
\psi_{\xi}\right\vert $ denoting a pure state. Note that $\partial_{a}%
\rho_{\xi}$ in Eq. (\ref{WY}) can be recast as,%
\begin{equation}
\partial_{a}\rho_{\xi}=\partial_{a}\left(  \left\vert \psi_{\xi}\right\rangle
\left\langle \psi_{\xi}\right\vert \right)  =\left\vert \partial_{a}\psi_{\xi
}\right\rangle \left\langle \psi_{\xi}\right\vert +\left\vert \psi_{\xi
}\right\rangle \left\langle \partial_{a}\psi_{\xi}\right\vert \text{.}%
\end{equation}
Therefore, after some simple algebraic manipulations, we arrive at
\begin{align}
\left(  \partial_{a}\rho_{\xi}\right)  \left(  \partial_{b}\rho_{\xi}\right)
&  =\left\langle \psi_{\xi}|\partial_{b}\psi_{\xi}\right\rangle \left\vert
\partial_{a}\psi_{\xi}\right\rangle \left\langle \psi_{\xi}\right\vert
+\left\vert \partial_{a}\psi_{\xi}\right\rangle \left\langle \partial_{b}%
\psi_{\xi}\right\vert +\left\langle \partial_{a}\psi_{\xi}|\partial_{b}%
\psi_{\xi}\right\rangle \left\vert \psi_{\xi}\right\rangle \left\langle
\psi_{\xi}\right\vert +\nonumber\\
& \nonumber\\
&  +\left\langle \partial_{a}\psi_{\xi}|\psi_{\xi}\right\rangle \left\vert
\psi_{\xi}\right\rangle \left\langle \partial_{b}\psi_{\xi}\right\vert
\text{.} \label{doppio}%
\end{align}
Employing Eq. (\ref{doppio}), \textrm{tr}$\left[  \left(  \partial_{a}%
\rho_{\xi}\right)  \left(  \partial_{b}\rho_{\xi}\right)  \right]  $ in Eq.
(\ref{WY}) can be rewritten as%
\begin{align}
\text{\textrm{tr}}\left[  \left(  \partial_{a}\rho_{\xi}\right)  \left(
\partial_{b}\rho_{\xi}\right)  \right]   &  =\left\langle \psi_{\xi}|\left(
\partial_{a}\rho_{\xi}\right)  \left(  \partial_{b}\rho_{\xi}\right)
|\psi_{\xi}\right\rangle \nonumber\\
& \nonumber\\
&  =\left\langle \psi_{\xi}|\partial_{b}\psi_{\xi}\right\rangle \left\langle
\psi_{\xi}|\partial_{a}\psi_{\xi}\right\rangle +\left\langle \psi_{\xi
}|\partial_{a}\psi_{\xi}\right\rangle \left\langle \partial_{b}\psi_{\xi}%
|\psi_{\xi}\right\rangle +\nonumber\\
& \nonumber\\
&  +\left\langle \partial_{a}\psi_{\xi}|\partial_{b}\psi_{\xi}\right\rangle
+\left\langle \partial_{a}\psi_{\xi}|\psi_{\xi}\right\rangle \left\langle
\partial_{b}\psi_{\xi}|\psi_{\xi}\right\rangle \text{.} \label{quisopra}%
\end{align}
We observe that the normalization condition $\left\langle \psi_{\xi}|\psi
_{\xi}\right\rangle =1$ implies that $\left\langle \partial_{b}\psi_{\xi}%
|\psi_{\xi}\right\rangle =-\left\langle \psi_{\xi}|\partial_{b}\psi_{\xi
}\right\rangle $. Therefore, \textrm{tr}$\left[  \left(  \partial_{a}\rho
_{\xi}\right)  \left(  \partial_{b}\rho_{\xi}\right)  \right]  $ in Eq.
(\ref{quisopra}) reduces to%
\begin{equation}
\text{\textrm{tr}}\left[  \left(  \partial_{a}\rho_{\xi}\right)  \left(
\partial_{b}\rho_{\xi}\right)  \right]  =\left\langle \partial_{a}\psi_{\xi
}|\partial_{b}\psi_{\xi}\right\rangle +\left\langle \partial_{a}\psi_{\xi
}|\psi_{\xi}\right\rangle \left\langle \partial_{b}\psi_{\xi}|\psi_{\xi
}\right\rangle \text{.} \label{1}%
\end{equation}
Following Ref. \cite{provost80}, the inner product $\left\langle \partial
_{a}\psi_{\theta}|\partial_{b}\psi_{\theta}\right\rangle $ can be recast as
\begin{equation}
\left\langle \partial_{a}\psi_{\xi}|\partial_{b}\psi_{\xi}\right\rangle
=\gamma_{ab}+i\sigma_{ab}\text{,} \label{2}%
\end{equation}
where $\gamma_{ab}$ and $\sigma_{ab}$ are defined as%
\begin{equation}
\gamma_{ab}\overset{\text{def}}{=}\operatorname{Re}\left[  \left\langle
\partial_{a}\psi_{\xi}|\partial_{b}\psi_{\xi}\right\rangle \right]  \text{,
and }\sigma_{ab}\overset{\text{def}}{=}\operatorname{Im}\left[  \left\langle
\partial_{a}\psi_{\xi}|\partial_{b}\psi_{\xi}\right\rangle \right]  \text{, }
\label{3}%
\end{equation}
respectively. Observe that $\operatorname{Re}\left(  z\right)  $ and
$\operatorname{Im}\left(  z\right)  $ represent the real and imaginary parts
of a complex quantity $z$, respectively. Note that $\gamma_{ab}$ and
$\sigma_{ab}$ in Eq. (\ref{3}) are symmetric and antisymmetric quantities,
respectively. Indeed, we have
\begin{equation}
\gamma_{ba}=\operatorname{Re}\left[  \left\langle \partial_{b}\psi_{\xi
}|\partial_{a}\psi_{\xi}\right\rangle \right]  =\operatorname{Re}\left[
\left\langle \partial_{a}\psi_{\xi}|\partial_{b}\psi_{\xi}\right\rangle
^{\ast}\right]  =\operatorname{Re}\left[  \left\langle \partial_{a}\psi_{\xi
}|\partial_{b}\psi_{\xi}\right\rangle \right]  =\gamma_{ab}\text{,}%
\end{equation}
and,%
\begin{equation}
\sigma_{ba}=\operatorname{Im}\left[  \left\langle \partial_{b}\psi_{\xi
}|\partial_{a}\psi_{\xi}\right\rangle \right]  =\operatorname{Im}\left[
\left\langle \partial_{a}\psi_{\xi}|\partial_{b}\psi_{\xi}\right\rangle
^{\ast}\right]  =-\operatorname{Im}\left[  \left\langle \partial_{a}\psi_{\xi
}|\partial_{b}\psi_{\xi}\right\rangle \right]  =-\sigma_{ab}\text{.}%
\end{equation}
Since $\sigma_{ab}=-\sigma_{ba}$, $\sigma_{ab}d\xi^{a}d\xi^{b}=0$.
By\textbf{\ }using Eqs. (\ref{1}), (\ref{2}) and (\ref{3}), $g_{ab}%
^{\mathrm{WY}}\left(  \xi\right)  $ in Eq. (\ref{WY}) becomes
\begin{equation}
g_{ab}^{\mathrm{WY}}\left(  \xi\right)  =4\operatorname{Re}\left[
\left\langle \partial_{a}\psi_{\xi}|\partial_{b}\psi_{\xi}\right\rangle
-\left\langle \partial_{a}\psi_{\xi}|\psi_{\xi}\right\rangle \left\langle
\psi_{\xi}|\partial_{b}\psi_{\xi}\right\rangle \right]  \text{,} \label{WY1}%
\end{equation}
since $\left\langle \partial_{a}\psi_{\xi}|\psi_{\xi}\right\rangle
\left\langle \psi_{\xi}|\partial_{b}\psi_{\xi}\right\rangle $ is a real-valued
quantity. Comparing Eqs. (\ref{FS1}) and (\ref{WY1}), we finally arrive at the
relation $g_{ab}^{\mathrm{WY}}\left(  \xi\right)  =4g_{ab}^{\mathrm{FS}%
}\left(  \xi\right)  $. For further details, we suggest Refs.
\cite{cafaro12,cafaro18}. With this statement, we conclude our discussion on
the link between Fubini-Study and Wigner-Yanase metrics at this point.

\section{Rotation matrix $\mathcal{R}\left(  t\right)  $}

In this appendix, we find the explicit expression of the rotation matrix
$\mathcal{R}\left(  t\right)  $ such that $\mathbf{a}\left(  t\right)
=\mathcal{R}\left(  t\right)  \mathbf{a}\left(  0\right)  $, when
$\mathbf{\dot{a}=}\left(  2/\hslash\right)  \mathbf{h\times a}$, the
Hamiltonian \textrm{H}$\overset{\text{def}}{=}\mathbf{h\cdot\boldsymbol{\sigma
}}$ is time-independent, and $\mathbf{h}\overset{\text{def}}{\mathbf{=}%
}h\mathbf{n}$ with $h\overset{\text{def}}{=}\left\Vert \mathbf{h}\right\Vert $.

We start by noting that the equation $\mathbf{\dot{a}=}\left(  2/\hslash
\right)  \mathbf{h\times a}$ can be recast as $\mathbf{\dot{a}=}\left(
2/\hslash\right)  H_{\mathbf{h}}\mathbf{a}$, with $H_{\mathbf{h}}$ being a
$\left(  3\times3\right)  $-matrix defined as%
\begin{equation}
H_{\mathbf{h}}\overset{\text{def}}{=}\left(
\begin{array}
[c]{ccc}%
0 & -h_{z} & h_{y}\\
h_{z} & 0 & -h_{x}\\
-h_{y} & h_{x} & 0
\end{array}
\right)  \text{.} \label{mary1}%
\end{equation}
Therefore, after a brief inspection, one realizes that integration of
$\mathbf{\dot{a}=}\left(  2/\hslash\right)  H_{\mathbf{h}}\mathbf{a}$ yields
\begin{equation}
\mathbf{a}\left(  t\right)  =e^{\frac{2}{\hslash}H_{\mathbf{h}}t}%
\mathbf{a}\left(  0\right)  \text{,}%
\end{equation}
with $\mathcal{R}\left(  t\right)  =e^{\frac{2}{\hslash}H_{\mathbf{h}}%
t}=e^{\frac{2h}{\hslash}\frac{H_{\mathbf{h}}}{h}t}=e^{\frac{2h}{\hslash
}H_{\mathbf{n}}t}=\mathcal{R}_{\mathbf{n}}(\frac{2h}{\hslash}t)$ being a
rotation with axis of rotation $\mathbf{n}$ and angle of rotation $\frac
{2h}{\hslash}t$. Recalling the Rodrigues rotation formula in matrix form
\cite{park17}, we observe that $\mathcal{R}_{\mathbf{n}}(\frac{2h}{\hslash}t)$
can be rewritten as%
\begin{equation}
\mathcal{R}_{\mathbf{n}}(\frac{2h}{\hslash}t)=\cos\left(  \frac{2h}{\hslash
}t\right)  I_{3\times3}\mathbf{+}\left[  1-\cos\left(  \frac{2h}{\hslash
}t\right)  \right]  \left(  \mathbf{n\cdot n}^{T}\right)  +\sin\left(
\frac{2h}{\hslash}t\right)  H_{\mathbf{n}}\text{,} \label{mary2}%
\end{equation}
where $I_{3\times3}$ is the $\left(  3\times3\right)  $-identity matrix,
$\left(  \mathbf{n\cdot n}^{T}\right)  $ is the $\left(  3\times3\right)
$-matrix obtained from the matrix algebra multiplication between the column
vector $\mathbf{n}$ and the row vector $\mathbf{n}^{T}$ (with
\textquotedblleft$T$\textquotedblright\ denoting the transposition operation),
and $H_{\mathbf{n}}\overset{\text{def}}{=}(1/h)H_{\mathbf{h}}$ with
$H_{\mathbf{h}}$ in Eq. (\ref{mary1}). We can explicitly verify from Eq.
(\ref{mary2}) that $\mathcal{R}_{\mathbf{n}}(0)=I_{3\times3}$ and
$\mathcal{R}_{\mathbf{n}}(\frac{2h}{\hslash}t)\in\mathrm{SO}\left(  3;%
\mathbb{R}
\right)  $ with $\det\left[  \mathcal{R}_{\mathbf{n}}(\frac{2h}{\hslash
}t)\right]  =1$. From Eq. (\ref{mary2}), we notice that $\mathbf{a}%
_{t}=\mathcal{R}_{\mathbf{n}}(\frac{2h}{\hslash}t)\mathbf{a}_{0}$ reduces to
\begin{equation}
\mathbf{a}_{t}=\cos\left(  \frac{2h}{\hslash}t\right)  \mathbf{a}%
_{0}\mathbf{+}\left[  1-\cos\left(  \frac{2h}{\hslash}t\right)  \right]
\left(  \mathbf{n\cdot a}_{0}\right)  \mathbf{n}+\sin\left(  \frac{2h}%
{\hslash}t\right)  \left(  \mathbf{n\times a}_{0}\right)  \text{,}
\label{mary3}%
\end{equation}
since $\left(  \mathbf{n\cdot n}^{T}\right)  \mathbf{a}_{0}=\left(
\mathbf{n\cdot a}_{0}\right)  \mathbf{n}$ is the projection of $\mathbf{a}%
_{0}$ onto $\mathbf{n}$ and $H_{\mathbf{n}}\mathbf{a}_{0}=\mathbf{n\times
a}_{0}$. Clearly, $\mathbf{a}_{t}$ and $\mathbf{a}_{0}$ denote $\mathbf{a}%
\left(  t\right)  $ and $\mathbf{a}\left(  0\right)  $, respectively. To
understand Eq. (\ref{mary3}), one can decompose $\mathbf{a}_{0}$ relative to
the rotation axis $\mathbf{n}$ as $\mathbf{a}_{0}^{\parallel}=\left(
\mathbf{n\cdot a}_{0}\right)  \mathbf{n}$ (i.e., a part parallel to
$\mathbf{n}$) and $\mathbf{a}_{0}^{\perp}=\mathbf{a}_{0}-\mathbf{a}%
_{0}^{\parallel}$ (i.e., a part orthogonal to $\mathbf{n}$). Since
$\mathbf{a}_{0}^{\parallel}$ lies along the axis $\mathbf{n}$, it does not
change under rotation and $\mathbf{a}_{0}^{\parallel}\left(  t\right)
=\mathbf{a}_{0}^{\parallel}\left(  0\right)  $. The orthogonal component
$\mathbf{a}_{0}^{\perp}$ rotates in the plane orthogonal to $\mathbf{n}$,
$\mathbf{a}_{0}^{\perp}\left(  t\right)  =\cos(\frac{2h}{\hslash}%
t)\mathbf{a}_{0}^{\perp}\left(  0\right)  +\sin(\frac{2h}{\hslash
}t)\mathbf{n\times a}_{0}^{\perp}\left(  0\right)  $. Therefore, we have%
\begin{align}
\mathbf{a}\left(  t\right)   &  =\mathbf{a}_{0}^{\parallel}\left(  0\right)
+\mathbf{a}_{0}^{\perp}\left(  0\right) \nonumber\\
&  =\mathbf{a}_{0}^{\parallel}\left(  0\right)  +\cos(\frac{2h}{\hslash
}t)\mathbf{a}_{0}^{\perp}\left(  0\right)  +\sin(\frac{2h}{\hslash}t)\left[
\mathbf{n\times a}_{0}^{\perp}\left(  0\right)  \right] \nonumber\\
&  =\left(  \mathbf{n\cdot a}_{0}\right)  \mathbf{n+}\cos(\frac{2h}{\hslash
}t)\left[  \mathbf{a}_{0}-\left(  \mathbf{n\cdot a}_{0}\right)  \mathbf{n}%
\right]  +\sin(\frac{2h}{\hslash}t)\left(  \mathbf{n\times a}_{0}\right)
\nonumber\\
&  =\cos(\frac{2h}{\hslash}t)\mathbf{a}_{0}+\left[  1-\cos(\frac{2h}{\hslash
}t)\right]  \left(  \mathbf{n\cdot a}_{0}\right)  \mathbf{n+}\sin(\frac
{2h}{\hslash}t)\left(  \mathbf{n\times a}_{0}\right)  \text{,} \label{mary3B}%
\end{align}
where in the second to last line we used the fact that $\mathbf{n\times a}%
_{0}^{\perp}\left(  0\right)  =\mathbf{n\times a}_{0}$. Clearly, since
$\mathbf{a}_{t}=\mathbf{a}\left(  t\right)  $, Eqs. (\ref{mary3}) and
(\ref{mary3B}) are identical. Finally, given that the vector triple product
$\mathbf{n\times}\left(  \mathbf{n\times a}_{0}\right)  $ is equal to $\left(
\mathbf{n\cdot a}_{0}\right)  \mathbf{n-}\left(  \mathbf{n\cdot n}\right)
\mathbf{a}_{0}=\left(  \mathbf{n\cdot a}_{0}\right)  \mathbf{n-a}_{0}$, Eq.
(\ref{mary3}) can be recast as%
\begin{equation}
\mathbf{a}_{t}=\mathbf{a}_{0}+\sin\left(  \frac{2h}{\hslash}t\right)  \left(
\mathbf{n\times a}_{0}\right)  \mathbf{+}\left[  1-\cos\left(  \frac
{2h}{\hslash}t\right)  \right]  \left[  \mathbf{n\times}\left(
\mathbf{n\times a}_{0}\right)  \right]  \text{.} \label{mary4}%
\end{equation}
Eq. (\ref{mary4}) is the vectorial form of the so-called Rodrigues rotation
formula \cite{park17}.

For completeness, we stress that the interested reader can verify that Eq.
(\ref{mary2}) works properly. For example, one can recover $\mathbf{a}_{t}$ in
Eq. (\ref{giorgia}) as $\mathbf{a}_{t}=\mathcal{R}_{\mathbf{n}}(\frac
{2h}{\hslash}t)\mathbf{a}_{0}$ by setting $\mathbf{a}_{0}=\left(  0\text{,
}0\text{, }1\right)  $, $\mathbf{n=}\left(  0,1,0\right)  $, and $h=\left\Vert
\mathbf{h}\right\Vert =\left(  \hslash\omega\right)  /\sqrt{6}$. Indeed, a
simple calculation yields%
\begin{equation}
\mathbf{a}_{t}=\mathcal{R}_{\mathbf{n}}(\frac{2h}{\hslash}t)\mathbf{a}%
_{0}=\left(
\begin{array}
[c]{ccc}%
\cos(\frac{2\omega t}{\sqrt{6}}) & 0 & \sin(\frac{2\omega t}{\sqrt{6}})\\
0 & 1 & 0\\
-\sin(\frac{2\omega t}{\sqrt{6}}) & 0 & \cos(\frac{2\omega t}{\sqrt{6}})
\end{array}
\right)  \left(
\begin{array}
[c]{c}%
0\\
0\\
1
\end{array}
\right)  =\left(
\begin{array}
[c]{c}%
\sin(\frac{2\omega t}{\sqrt{6}})\\
0\\
\cos(\frac{2\omega t}{\sqrt{6}})
\end{array}
\right)  \text{.}%
\end{equation}
Following this verification, we conclude our discussion on the rotation matrix
$\mathcal{R}\left(  t\right)  $ at this point.

\section{Basis independence of quantum IG\ complexity}

In the first part of this appendix, we discuss in an explicit illustrative
example the fact that our complexity measure \textrm{C} does not depend on the
orthonormal basis chosen to express the time-evolved quantum state of the
system. However, although $\overline{\mathrm{V}}$ and \textrm{V}$_{\max}$ do
not change, the temporal dependence of the instantaneous volume $V\left(
t\right)  $ can depend on the particular basis chosen. In the second part, we
present some remarks for future investigations.

\subsection{Explicit example}

For clarity of exposition, assume the evolution from $\frac{\left\vert
0\right\rangle +i\left\vert 1\right\rangle }{\sqrt{2}}$ to $\left\vert
0\right\rangle $ under \textrm{H}$\overset{\text{def}}{\mathbb{=}%
}E\mathbb{\sigma}_{x}$. With respect to the computational basis $\left\{
\left\vert 0\right\rangle \text{, }\left\vert 1\right\rangle \right\}  $ and
the eigenbasis $\left\{  \left\vert E_{0}\right\rangle \overset{\text{def}}%
{=}\frac{\left\vert 0\right\rangle +i\left\vert 1\right\rangle }{\sqrt{2}%
}\text{, }\left\vert E_{1}\right\rangle \overset{\text{def}}{=}\frac
{\left\vert 0\right\rangle -i\left\vert 1\right\rangle }{\sqrt{2}}\right\}  $
(which, in this particular case, coincides with the Krylov basis $\left\{
\left\vert K_{0}\right\rangle =\left\vert E_{0}\right\rangle \text{,
}\left\vert K_{1}\right\rangle =\left\vert E_{1}\right\rangle \right\}  $),
the time-evolved state $\left\vert \psi\left(  t\right)  \right\rangle $ is
given by%
\begin{equation}
\left\vert \psi\left(  t\right)  \right\rangle =\frac{\cos(\frac{E}{\hslash
}t)+\sin(\frac{E}{\hslash}t)}{\sqrt{2}}\left\vert 0\right\rangle +i\frac
{\cos(\frac{E}{\hslash}t)-\sin(\frac{E}{\hslash}t)}{\sqrt{2}}\left\vert
1\right\rangle \text{,} \label{luca1}%
\end{equation}
and,%
\begin{equation}
\left\vert \psi\left(  t\right)  \right\rangle =\cos(\frac{E}{\hslash
}t)\left\vert K_{0}\right\rangle +\sin(\frac{E}{\hslash}t)\left\vert
K_{1}\right\rangle \text{,} \label{luca2}%
\end{equation}
respectively. The spherical angles that parametrize the qubit in the states in
Eqs. (\ref{luca1}) and (\ref{luca2}) are given by%
\begin{equation}
\theta\left(  t\right)  =2\arctan\left(  \sqrt{\left(  \frac{\cos(\frac
{E}{\hslash}t)-\sin(\frac{E}{\hslash}t)}{\cos(\frac{E}{\hslash}t)+\sin
(\frac{E}{\hslash}t)}\right)  ^{2}}\right)  \text{, }\varphi\left(  t\right)
=\frac{\pi}{2}\text{,} \label{luca3}%
\end{equation}
and,%
\begin{equation}
\theta\left(  t\right)  =\frac{2E}{\hslash}t\text{, }\varphi\left(  t\right)
=0\text{,} \label{luca4}%
\end{equation}
respectively, with $0\leq t\leq\left(  \pi\hslash\right)  /\left(  4E\right)
$. Although we use\textbf{ }the same notation for the spherical angles in Eqs.
(\ref{luca3}) and (\ref{luca4}), the $z$-axis is specified by the state
$\left\vert 0\right\rangle $ in Eq. (\ref{luca3}), while $\left\vert
K_{0}\right\rangle $ defines the $z$-axis in Eq.\ (\ref{luca4}). Setting
$E=1=\hslash$ for simplicity, we note that although $V(t)$ for Eq.
(\ref{luca3}) differs from $V(t)$ for Eq. (\ref{luca4}), we have%
\begin{equation}
\int_{0}^{\frac{\pi}{4}}tdt=\frac{\pi^{2}}{32}=\int_{0}^{\frac{\pi}{4}}\left[
\frac{\pi}{4}-\arctan\left(  \sqrt{\left(  \frac{\cos(t)-\sin(t)}{\cos
(t)+\sin(t)}\right)  ^{2}}\right)  \right]  dt\text{.} \label{luca5}%
\end{equation}
Therefore, after some inspection, we have $\left(  \overline{\mathrm{V}%
}\right)  _{\text{Eq. (\ref{luca3})}}=\pi/8=\left(  \overline{\mathrm{V}%
}\right)  _{\text{Eq. (\ref{luca4})}}$ and $($\textrm{V}$_{\max})_{\text{Eq.
(\ref{luca3})}}=\pi/4=($\textrm{V}$_{\max})_{\text{Eq. (\ref{luca4})}}$.
Finally, we show that $($\textrm{C}$)_{\text{Eq. (\ref{luca1})}}=($%
\textrm{C}$)_{\text{Eq. (\ref{luca2})}}$.

\subsection{Exploratory remarks}

The definition of $V(t)$ suggests a volume that does not correspond to a
geometrically invariant area on the Bloch sphere. This arises from the
integration over a rectangle in coordinate space $\left(  \theta\text{,
}\varphi\right)  $, which, when subjected to a rotation (i.e., a change of
basis), does not transform into a rectangle in the new coordinate system.
Typically, this rectangle becomes distorted in a complex manner. Nevertheless,
we have observed empirically that the averages and maxima of $V(t)$,
calculated over the quantum evolutions defined by stationary Hamiltonians,
appear to be independent of the specific single-qubit space basis selected.
However, we cannot include a formal mathematical proof of such
basis-independence. We intend to address this matter in our future research.

\smallskip

It is noteworthy to mention that an alternative method for defining the
concept of IG complexity on the Bloch sphere, utilizing natural geometric or
coordinate-free quantities, can be articulated as follows. Consider the
trajectory's image on the Bloch sphere as the collection of all points on the
Bloch sphere that are traversed by the state $\left\vert \psi\left(  t\right)
\right\rangle $\textbf{ }for\textbf{ }$t_{A}\leq t\leq t_{B}$. This quantity
is represented by \textrm{Image}$\left(  \mathbf{a}\left(  t\right)  \right)
\overset{\text{def}}{=}\left\{  \mathbf{a}\left(  t\right)  \in S^{2}\subset%
\mathbb{R}
^{3}:t\in\left[  t_{A}\text{, }t_{B}\right]  \right\}  $. It is evident that
$\mathbf{a}\left(  t\right)  \overset{\text{def}}{=}\left\langle \psi\left(
t\right)  \left\vert \mathbf{\boldsymbol{\sigma}}\right\vert \psi\left(
t\right)  \right\rangle \in%
\mathbb{R}
^{3}$ denotes the Bloch vector of the pure qubit state, with $\left\Vert
\mathbf{a}\left(  t\right)  \right\Vert =1$, while $S^{2}$ signifies the Bloch
sphere. Although \textrm{Image}$\left(  \mathbf{a}\left(  t\right)  \right)  $
represents a geometric quantity that is independent of parametrization, it
does not possess an intrinsic area. Nevertheless, it is possible to associate
an invariant area with it by defining a geometrically determined region on the
sphere through the solid angle encompassed by the path. Indeed, on a unit
sphere, the surface area of a region is precisely equivalent to the solid
angle it subtends. The solid angle $\Omega$ quantifies the apparent size of a
surface as viewed from the center of the sphere. For a surface patch
$\tilde{S}$ on a sphere with radius $r_{0}$, $\Omega=\mathrm{Area}(\tilde
{S})/r_{0}^{2}=\mathrm{Area}(\tilde{S})$ (when $r_{0}=1$). Consequently, the
solid angle is clearly invariant under rotations, and the area of the region
corresponds to the solid angle subtended by the boundary curve. In cases where
the dynamical path is closed, the resulting solid angle is unique. Conversely,
if the path is open, one can adopt a closure prescription to guarantee
uniqueness. The geodesic closure is the standard and geometrically natural
one. Considering the image \textrm{Image}$\left(  \gamma\left(  t\right)
\right)  \equiv$\textrm{Image}$\left(  \mathbf{a}\left(  t\right)  \right)  $
of the trajectory $\gamma:\left[  t_{A}\text{, }t_{B}\right]  \ni t\mapsto$
$\gamma\left(  t\right)  =\mathbf{a}\left(  t\right)  \in S^{2}\subset%
\mathbb{R}
^{3}$, along with the initial and final points $(\mathbf{a}\left(
t_{A}\right)  $\textbf{ }and $\mathbf{a}\left(  t_{B}\right)  $,
respectively), the region\textbf{ }$\mathcal{\tilde{R}}$ swept by the
trajectory $\gamma\left(  t\right)  $ is defined as the area on $S^{2}$ that
is enclosed by the path $\gamma\left(  t\right)  $\textbf{ }(which is
generally open) in addition to its geodesic closure $\gamma_{\mathrm{geo}}$,
which signifies the shortest geodesic path from $\mathbf{a}\left(
t_{B}\right)  $\textbf{ }to $\mathbf{a}\left(  t_{A}\right)  $. Consequently,
$\mathcal{\tilde{R}}$ is the region on\textbf{ }$S^{2}$ that is\textbf{
}enclosed by the closed curve $\gamma_{\mathrm{tot}}\overset{\text{def}}%
{=}\gamma\circ\gamma_{\mathrm{geo}}$. The region\textbf{ }$\mathcal{\tilde{R}%
}$\textbf{ }represents a geometrically well-defined intrinsic quantity on the
Bloch sphere, and its area, denoted as $\mathrm{Area}(\mathcal{\tilde{R}%
})\overset{\text{def}}{=}\int_{\mathcal{\tilde{R}}}d\Omega$,\textbf{ }where
$d\Omega\overset{\text{def}}{=}\sin\left(  \theta\right)  d\theta d\varphi$ is
the surface element of the Bloch sphere, remains invariant under rotations.
The rotational invariance of\textbf{ }$\mathrm{Area}(\mathcal{\tilde{R}})$ is
derived from the invariance of the surface element\textbf{ }$d\Omega$,\textbf{
i}n conjunction with the understanding that rotations serve as isometries of
the sphere, mapping regions to regions without any distortion. The
corresponding expressions for $V\left(  t\right)  $, $\overline{\mathrm{V}%
}\left(  t_{A}\text{, }t_{B}\right)  $, and $\mathrm{V}_{\max}\left(
t_{A}\text{, }t_{B}\right)  $ can be stated as $\mathrm{Area}(\mathcal{\tilde
{R}}(t))\overset{\text{def}}{=}\int_{\mathcal{\tilde{R}}\left(  t\right)
}d\Omega$, $\left\langle \mathrm{Area}(\mathcal{\tilde{R}}(t))\right\rangle
_{t_{A}\leq t\leq t_{B}}\overset{\text{def}}{=}\frac{1}{t_{B}-t_{A}}%
\int_{t_{A}}^{t_{B}}\mathrm{Area}(\mathcal{\tilde{R}}(t^{\prime}))dt^{\prime}%
$, and $\mathrm{Area}_{\max}\overset{\text{def}}{=}\underset{t_{A}\leq t\leq
t_{B}}{\max}\left[  \mathrm{Area}(\mathcal{\tilde{R}}(t))\right]  $, respectively.

\smallskip

We will examine an IG complexity measure that is defined by these
coordinates-free quantities as part of our construction in future scientific
endeavors\textbf{.}

\section{Links between $\mathcal{K}\left(  t\right)  $, $\left\Vert
\mathbf{a}\left(  t\right)  -\mathbf{a}\left(  0\right)  \right\Vert ^{2}$,
and $V_{\mathrm{FS}}\left(  t\right)  $}

In this appendix, to gain some understanding of the connections between
Krylov's state complexity $\mathcal{K}\left(  t\right)  $, the distance
squared $\left\Vert \mathbf{a}\left(  t\right)  -\mathbf{a}\left(  0\right)
\right\Vert ^{2}$ between the initial Bloch vector and the Bloch vector at an
instant $t$ with $0\leq t\leq t_{\mathrm{f}}$, and the Fubini-Study volume
element $V_{\mathrm{FS}}\left(  t\right)  $, we consider an arbitrary
stationary Hamiltonian \textrm{H} with a spectral decomposition given by,%
\begin{equation}
\mathrm{H}\overset{\text{def}}{=}E_{0}\left\vert E_{0}\right\rangle
\left\langle E_{0}\right\vert +E_{1}\left\vert E_{1}\right\rangle \left\langle
E_{1}\right\vert \text{.}%
\end{equation}
Assume the initial state of the system is $\left\vert \psi\left(  0\right)
\right\rangle \overset{\text{def}}{=}\cos(\frac{\theta_{0}}{2})\left\vert
E_{0}\right\rangle +e^{i\varphi_{0}}\sin(\frac{\theta_{0}}{2})\left\vert
E_{1}\right\rangle $. Then, the time-evolved state $\left\vert \psi\left(
t\right)  \right\rangle \overset{\text{def}}{=}e^{-\frac{i}{\hslash}%
\mathrm{H}t}\left\vert \psi\left(  0\right)  \right\rangle $ is physically
equivalent to the state
\begin{equation}
\left\vert \psi\left(  t\right)  \right\rangle \simeq\cos(\frac{\theta_{0}}%
{2})\left\vert E_{0}\right\rangle +e^{i\left(  \varphi_{0}+\Delta Et\right)
}\sin(\frac{\theta_{0}}{2})\left\vert E_{1}\right\rangle \text{,}
\label{campa1}%
\end{equation}
where $\Delta E\overset{\text{def}}{=}E_{0}-E_{1}$, and $\mathrm{H}\left\vert
E_{i}\right\rangle =E_{i}\left\vert E_{i}\right\rangle $ for $i\in\left\{
0\text{, }1\right\}  $. From Eq. (\ref{campa1}), the time evolution of the
spherical angles that parametrize the qubit $\left\vert \psi\left(  t\right)
\right\rangle $ on the Bloch sphere is specified by $\theta\left(  t\right)
=\theta_{0}$ and $\varphi\left(  t\right)  =\varphi_{0}+\Delta Et$. A simple
calculation yields%
\begin{equation}
\mathcal{K}\left(  t\right)  =\sin^{2}(\theta_{0})\sin^{2}(\frac{\Delta E}%
{2}t)\text{, }\left\Vert \mathbf{a}\left(  t\right)  -\mathbf{a}\left(
0\right)  \right\Vert ^{2}=4\mathcal{K}\left(  t\right)  \text{, and
}V_{\mathrm{FS}}\left(  t\right)  =\sin(\theta_{0})\frac{\Delta E}{2}t\text{.}
\label{campa2}%
\end{equation}
We recall that the Fubini-Study line element is given by $ds_{\mathrm{FS}}%
^{2}\overset{\text{def}}{=}\frac{1}{4}\left[  d\theta^{2}+\sin^{2}%
(\theta)d\varphi^{2}\right]  $ and, in general, $V_{\mathrm{FS}}\left(
t\right)  =\int dV_{\mathrm{FS}}\left(  t\right)  $ with $dV_{\mathrm{FS}%
}\left(  t\right)  =\frac{1}{4}\sqrt{\sin^{2}(\theta)}d\theta d\varphi$.
Clearly, when $\theta\left(  t\right)  =\theta_{0}$ is constant, we have
$dV_{\mathrm{FS}}\left(  t\right)  =\frac{1}{2}\sin(\theta_{0})d\varphi$.
Moreover, the Krylov basis used to calculate $\mathcal{K}\left(  t\right)  $
in Eq. (\ref{campa2}) is given by $\left\{  \left\vert K_{0}\right\rangle
\text{, }\left\vert K_{1}\right\rangle \right\}  $ where $\left\vert
K_{0}\right\rangle =\left\vert \psi\left(  0\right)  \right\rangle
\overset{\text{def}}{=}\left\vert \psi_{0}(\theta_{0}\text{, }\varphi
_{0})\right\rangle $ and $\left\vert K_{1}\right\rangle \overset{\text{def}%
}{=}\left\vert \psi_{0}(\pi-\theta_{0}\text{, }\pi+\varphi_{0})\right\rangle
=\sin(\frac{\theta_{0}}{2})\left\vert E_{0}\right\rangle -e^{i\varphi_{0}}%
\cos(\frac{\theta_{0}}{2})\left\vert E_{1}\right\rangle $. From Eq.
(\ref{campa2}), we observe that%
\begin{equation}
\mathcal{K}\left(  t\right)  =\frac{1}{4}\left\Vert \mathbf{a}\left(
t\right)  -\mathbf{a}\left(  0\right)  \right\Vert ^{2}\text{, and }%
\sqrt{\mathcal{K}\left(  t\right)  }\overset{t\ll1}{\approx}V_{\mathrm{FS}%
}\left(  t\right)  \text{.} \label{campa3}%
\end{equation}
Eq. (\ref{campa3}) implies that Krylov's state complexity $\mathcal{K}\left(
t\right)  $ in qubit dynamics is proportional to the Euclidean distance
between the initial and the final (i.e., at time $t$) Bloch vectors that
specify the qubit on the Bloch sphere. Moreover, Eq. (\ref{campa3}) leads to
the conclusion that the square root of Krylov's state complexity
$\mathcal{K}\left(  t\right)  $ is approximately equal to the instantaneous
volume $V_{\mathrm{FS}}\left(  t\right)  $ in the short-time limit. It is
noteworthy to mention that for short times, $V_{\mathrm{FS}}\left(  t\right)
$ (which, in this instance, represents the Fubini-Study length) increases
linearly, whereas Krylov's state complexity escalates quadratically. The
emergence of the quadratic behavior is universal and signifies that the
probability flow into new Krylov levels is second order in time. This
universality arises from the unitarity of quantum-mechanical time evolution,
the Taylor expansion of $e^{-\frac{i}{\hslash}\mathrm{H}t}$, and the Born rule
(i.e., \textrm{probability}$=\left\vert \text{\textrm{amplitude}}\right\vert
^{2}$). The flow of probability is characterized as second order in time. If
$\left\vert K_{0}\right\rangle \overset{\text{def}}{=}\left\vert \psi
_{0}\right\rangle $ represents the initial Krylov state, the time-evolved
state can be expressed as%
\begin{equation}
\left\vert \psi(t)\right\rangle =e^{-\frac{i}{\hslash}\mathrm{H}t}\left\vert
\psi_{0}\right\rangle =\left[  \mathbf{1-}i\frac{\mathrm{H}}{\hslash}%
t-\frac{1}{2}\frac{\mathrm{H}^{2}}{\hslash^{2}}t^{2}+\mathcal{O}\left(
t^{3}\right)  \right]  \left\vert \psi_{0}\right\rangle \text{,}%
\end{equation}
it follows that the amplitude for transitioning from $\left\vert
K_{0}\right\rangle $ is linear in $t$, specifically $\left\langle
K_{1}\left\vert \psi(t)\right.  \right\rangle \sim-\frac{i}{\hslash
}\left\langle K_{1}\left\vert \mathrm{H}\right\vert K_{0}\right\rangle t$.
Consequently, the probability of transitioning to a new Krylov level is second
order in time, as indicated by $\left\vert \left\langle K_{1}\left\vert
\psi(t)\right.  \right\rangle \right\vert ^{2}\sim t^{2}$. With these
considerations, we end our presentation here.

\bigskip

\bigskip

\bigskip

\bigskip

\bigskip

\bigskip

\bigskip

\bigskip

\bigskip

\bigskip

\end{document}